\documentclass{aa}  

\usepackage{graphicx}
\usepackage{multirow}
\usepackage{color}
\usepackage{natbib}
\usepackage{txfonts}
\usepackage{hyperref}
\usepackage[T1]{fontenc}

\begin{document}

   \title{Stellar population astrophysics (SPA) with the TNG. Abundance analysis of nearby red giants and red clump stars: combining high resolution spectroscopy and asteroseismology\thanks{Based on observations made with the Italian Telescopio Nazionale Galileo (TNG) operated on the island of La Palma by the Fundaci\'on Galileo Galilei of the INAF (Istituto Nazionale di Astrofisica) at the Spanish Observatorio del Roque de los Muchachos of the Instituto de Astrofisica de Canarias.}}

 \authorrunning{N. Vernekar et al.}
 \titlerunning{ Red giants, spectroscopy, and asteroseismology}

   \author{Nagaraj Vernekar\inst{1,2}
\and
Sara Lucatello\inst{2}
\and
Angela Bragaglia\inst{3}
\and
Andrea Miglio\inst{3,4}
\and
Nicoletta Sanna\inst{5}
\and
Gloria Andreuzzi\inst{6,7}
\and
Antonio Frasca \inst{8}
}

   \institute{
   Dipartimento di Fisica e Astronomia, Universit\'a di Padova, vicolo dell'Osservatorio 2, 35122 Padova, Italy \\
  \email{nagarajbadarinarayan.vernekar@phd.unipd.it, nagaraj.vernekar@inaf.it}
   \and
  INAF - Ossevatorio Astronomico di Padova, vicolo dell'Osservatorio 5, 35122 Padova, Italy
  \and
  INAF - Osservatorio di Astrofisica e Scienza dello Spazio, via P. Gobetti 93/3, 40129 Bologna. Italy
  \and
  Department of Physics and Astronomy,  Università di Bologna, Via Zamboni 33, 40126 Bologna, Italy
  \and
  INAF - Osservatorio Astrofisico di Arcetri, Largo Enrico Fermi, 5, 50125 Firenze, Italy
  \and
  Fundación Galileo Galilei - INAF, Rambla José Ana Fernández Pérez 7, 38712, Breña Baja, Tenerife, Spain
  \and
  INAF - Osservatorio Astronomico di Roma, Via Frascati 33, 00078, Monte Porzio Catone, Italy
  \and
  INAF – Osservatorio Astrofisico di Catania, via S. Sofia 78, 95123 Catania, Italy
 }

   \date{}

 
  \abstract
{Asteroseismology, a powerful approach for obtaining internal structure and stellar properties, requires surface temperature and chemical composition information to determine mass and age. High-resolution spectroscopy is a valuable technique for precise stellar parameters (including surface temperature) and chemical composition analysis.}
   {We aim to combine spectroscopic parameters with asteroseismology to test stellar models.}
   {Using high-resolution optical and near-IR spectra from GIARPS at the Telescopio Nazionale Galileo, we conducted a detailed spectroscopic analysis of 16 stars photometrically selected to be on the red giant and red clump. Stellar parameters and chemical abundances for light elements (Li, C, N, F), Fe peak, $\alpha$ and n-capture elements were derived using a combination of equivalent widths and spectral synthesis techniques, based on atomic and molecular features. Ages were determined through asteroseismic scaling relations and compared with ages based on chemical clocks, [Y/Mg] and [C/N].}
   {Spectroscopic parameters confirmed the stars as part of the red giant branch and red clump. Two objects, HD 22045 and HD 24680 exhibited relatively high Li abundances, with HD 24680 potentially being a Li-rich giant resulting from mass transfer with an intermediate-mass companion, which already underwent its AGB phase. Stellar parameters derived from scaling different sets of relations were consistent with each other. For what concerns ages, the values based on asteroseismology were in excellent agreement with those derived from theoretical evolutionary tracks, but did not align with ages derived from the chemical clocks [Y/Mg] and [C/N].}
   {}

   \maketitle
%

\section{Introduction}
Galactic archaeology is the study of present-time stellar populations in the Milky Way to probe the physical processes that resulted in the formation and subsequent evolution of our Galaxy and its components. 
The study of the properties of stars, position, kinematics, and chemical composition, provides key information on how and where the stars were formed. In this context, the upcoming high-resolution, high-multiplexing spectroscopic surveys (e.g. WEAVE, 4MOST, SDSSV-MWM), building upon previous surveys, such as APOGEE, Gaia-ESO, and GALAH combined with Gaia parallaxes and proper motions are expected to enable a major step ahead in the next few years. 

The determination of ages is a critical point in galactic archaeology.  In fact, their derivation is rather challenging due to the lack of observables uniquely sensitive to age alone \citep{soderblom2010}. Traditionally, stellar ages are obtained through isochrone fitting, on the basis of their atmospheric parameters (T$_{\rm eff}$, log(g),  and chemical composition). While this approach proves highly effective for members of stellar clusters, it encounters limitations when applied to individual field stars due to inherent degeneracies in the properties. 

An alternative approach is to rely on information resulting from asteroseismological studies.
Asteroseismology, the study of stellar oscillations, is a powerful technique for probing the internal structure and properties of stars. Through the analysis of the observed frequencies, amplitudes, and mode patterns of stellar oscillations, it allows to unravel fundamental information about stellar interiors \citep[see][and references therein]{hekker2016}.
Scaling relations, that link asteroseismic observables and the fundamental properties of stars are used to estimate stellar masses, radii, and surface gravity, requiring as input also information about the stellar surface temperature, and, in some formulations of said relations, about the chemical composition \citep[e.g.][]{1995A&A...293...87K,guggen2016}. 

Due to this, complementing asteroseismology with high-resolution spectroscopy proves to be a powerful tool to obtain stellar properties, ages and internal structures of stars. This combination of techniques could also be used to test the reliability of theoretical models and probe the reliability of alternative approaches to age measurements (e.g. chemical clocks).

In this paper, we present the analysis of high-resolution optical and near infra-red spectroscopy for a sample of bright, local red giant stars for which K2 data \citep{2014PASP..126..398H} are available, comparing quantities inferred from the spectra to those derived thanks to asteroseimological data. In Section 2, we describe the sample selection along with the data used for the analysis. Section 3 describes the procedure followed for the determination of stellar parameters and elemental abundances, including obtaining ages from asteroseismology and stellar tracks. Discussion of the results is given in Section 4 with summary and conclusion in Section 5.




\section{Sample and observations} \label{sample}
As we wanted to test the stellar evolutionary and asteroseismological models, it was necessary to obtain good-quality photometric and spectroscopic data. We selected 16 bright red giant stars that were nearby and within the field-of-view of the K2 mission \citep{2014PASP..126..398H}, to make it easy to obtain high-resolution spectroscopic data with good signal-to-noise. We also gathered high-precision photometric data for all the stars through the K2 mission. The fact that these were field stars and not part of any cluster, made it difficult to accurately determine their evolutionary stage. We used color indices and {\em Gaia} information \citep{2022yCat.1357....0G} to estimate their evolutionary stage and select the ones that were most likely to be red clump or lower red giant branch (RGB) stars. We mainly concentrated on these because they are homogeneous type stars, which are warm enough to obtain a good estimation of the stellar parameters and abundances without having to worry about line crowding as seen in brighter giants at solar metallicity. 

All the spectroscopic data used in this work were obtained using the 3.5m Telescopio Nazionale Galileo (TNG) located in La Palma, Spain. Out of multiple instruments, TNG has two high-resolution spectrographs: HARPS-N \citep{2012SPIE.8446E..1VC} and GIANO \citep{2012SPIE.8446E..3TO,2014SPIE.9147E..1EO}. HARPS-N works in the optical, with a resolution of 115000 and a wavelength coverage of 3800 - 6900 $\AA$, whereas GIANO works in the infrared, with a resolution of 50000 and wavelength coverage of 0.97 - 2.4 $\mu$m, i.e. YJHK. At TNG, one can use both these instruments at the same time using the GIARPS mode \citep{2017EPJP..132..364C}. In this mode, every observation covers the optical as well as the infrared wavelength regions. The log of the observations, together with information on the S/N reached with HARPS-N and GIANO, can be found in Table~\ref{tab:log}.

\begin{table*}[]
\centering
\caption{Observation log of the sample containing the date and time of the observation, exposure time used, and signal-to-noise of the spectrum at
a specific wavelength.}
\label{tab:log}
\begin{tabular}{llrrccccc}
\hline
\multicolumn{1}{c}{\multirow{2}{*}{StarID}} & \multicolumn{1}{c}{\multirow{2}{*}{EPIC}} & \multicolumn{1}{c}{\multirow{2}{*}{RA}} & \multicolumn{1}{c}{\multirow{2}{*}{DEC}} & \multicolumn{1}{c}{\multirow{2}{*}{Date-obs}} & \multicolumn{2}{c}{Optical} & \multicolumn{2}{c}{Infrared}\\ \cline{6-9} 
\multicolumn{1}{c}{}& \multicolumn{1}{c}{}& \multicolumn{1}{c}{}  & \multicolumn{1}{c}{}   & \multicolumn{1}{c}{}  & \multicolumn{1}{c}{\begin{tabular}[c]{@{}c@{}}Exposure\\ time (s)\end{tabular}} & \multicolumn{1}{c}{\begin{tabular}[c]{@{}c@{}}S/N at\\ 5700 $\AA$\end{tabular}} & \multicolumn{1}{c}{\begin{tabular}[c]{@{}c@{}}Exposure\\ time (s)\end{tabular}} & \multicolumn{1}{c}{\begin{tabular}[c]{@{}c@{}}S/N at \\ 1.5 $\mu$m\end{tabular}} \\ \hline \hline
HD 218330   & 251773509& 346.761& 3.052   & 2019-12-08   & 690& 186& 600& 368\\
HD 4313& 220548055& 11.418 & 7.845   & 2019-12-08   & 690& 155& 600& 282\\
HD 5214& 220456348& 13.493 & 5.809   & 2019-12-08   & 690& 146& 600& 283\\
HD 6432& 220568655& 16.351 & 8.332   & 2019-12-08   & 690& 137& 600& 307\\
HD 22045& 210690537& 53.452 & 18.228  & 2019-12-08   & 690& 137& 600& 272\\
HD 24680& 210859667& 59.027 & 20.772  & 2019-12-08   & 690& 217& 1200& 179\\
HD 76445& 211811597& 134.192& 17.481  & 2019-12-08   & 1400& 84 & 600& 200\\
HD 77776& 211514553& 136.242& 13.344  & 2019-12-08   & 1400& 97 & 1200& 299\\
HD 78419& 211756677& 137.142& 16.702  & 2019-12-08   & 1400& 90 &  -  & - \\
78 Cnc & 211810753& 137.259& 17.470  & 2019-12-08   & 1400& 107& 1200& 336\\
HD 99596& 201528051& 171.892& 0.956   & 2019-01-14   & 900& 213& 400& 116\\
HD 100872   & 201839927& 174.142& 6.106   & 2019-01-14   & 600& 202& 240 & 232\\
HD 97716& 201379481& 168.657& -1.271  & 2019-01-14   & 600& 216& 240 & 267\\
HD 97491& 201370145& 168.286& -1.410  & 2019-01-15   & 900& 186& 400& 145\\
HD 97197& 201371239& 167.830& -1.395  & 2019-01-15   & 1400& 114& 800& 178 \\
p04 Leo& 201594287& 166.724& 1.955   & 2019-01-15   & 600& 169& 400 & 272 \\ \hline
\end{tabular}%

\end{table*}

\begin{figure}
   \centering
   \includegraphics[width=1 \linewidth]{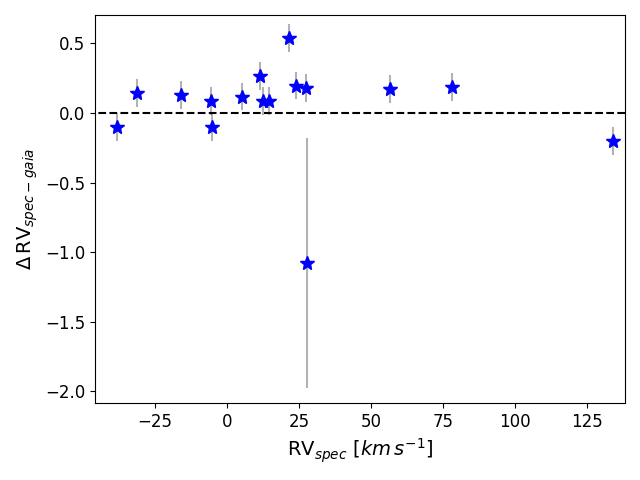}
   \caption{Comparison of RV values obtained in this work (indicated by the suffix spec) and {\em Gaia}. Errors plotted are the propagated uncertainties on both RV$_{\rm spec}$ and RV$_{\rm Gaia}$.}
   \label{fig:RV}
\end{figure}

Data reduction of the optical spectra from HARPS-N which includes flat-field and bias correction, spectral extraction, and wavelength calibration, was performed by the dedicated pipeline. For GIANO, the reduction was also performed with the offline version of the GOFIO reduction software \citep{Rainer2018}, and the telluric correction was performed using the spectra of telluric standards. Details of the procedure are described in \citep{origlia2019}.
Once the reduced spectra were obtained, we used the iSpec tool \citep{2014ascl.soft09006B} to carry out continuum normalisation and radial velocity (RV) correction. For continuum normalisation, we divided the observed spectrum by a fitted spline function.
iSpec determines the RV by computing the cross-correlation function 
between the observed and synthetic spectra. Table \ref{tab:sample} provides an overview of all the stars along with the estimated RV. Upon comparing our RV values with {\em Gaia}'s,  we see an excellent agreement between the two with a mean difference of 0.2 km\,s$^{-1}$ and no trends, as shown in Fig.~\ref{fig:RV}. Only one star (HD24680) shows a large discrepancy in the RV along with a large error in the {\em Gaia} RV; This star is 
classified as an SB1 system \citep{2022yCat.1357....0G}. 

For the purpose of validation of our results, we also performed analysis on Arcturus. We obtained a high-resolution spectrum of Arcturus in the optical range from \cite{2000vnia.book.....H} with a resolution R = 150000 and SNR $\sim$ 1000. The spectrum had already been corrected for tellurics as well as continuum normalised. The IR spectrum was obtained from \cite{1995PASP..107.1042H} with a wavelength coverage of 0.9 - 5.3 $\mu$m and a resolution R = 100000.

\begin{table*}[]
\centering
\caption{Properties of the stars used in this work. Photometric colors were taken from the following: ($B, V$) from \cite{2000A&A...355L..27H}, ($J, H, K$) from \cite{2003yCat.2246....0C}, and ($G, G_{BP}, G_{RP}$) from \cite{2021A&A...649A...1G}.}
\label{tab:sample}
\begin{tabular}{llccccccccr}
\hline 
\multicolumn{1}{c}{\multirow{1}{*}{StarID}}   & \multicolumn{1}{c}{\multirow{1}{*}{GAIA DR3 ID}}& \begin{tabular}[c]{@{}c@{}}\textit{B}\\ (mag)\end{tabular} & \begin{tabular}[c]{@{}c@{}}\textit{V}\\ (mag)\end{tabular} & \begin{tabular}[c]{@{}c@{}}\textit{J}\\ (mag)\end{tabular} & \begin{tabular}[c]{@{}c@{}}\textit{H}\\ (mag)\end{tabular} & \begin{tabular}[c]{@{}c@{}}\textit{K}\\ (mag)\end{tabular} & \begin{tabular}[c]{@{}c@{}}\textit{G}\\ (mag)\end{tabular} & \begin{tabular}[c]{@{}c@{}}\textit{G}$_{\rm BP}$\\ (mag)\end{tabular} & \begin{tabular}[c]{@{}c@{}}\textit{G}$_{\rm RP}$\\ (mag)\end{tabular} & \begin{tabular}[c]{@{}c@{}}RV$_{\rm spec}$\\ (km\,s$^{-1}$)\end{tabular} \\ \hline  
HD 218330 & 2658767863565504512 & 8.09  & 7.00  & 5.05  & 4.54  & 4.33  & 6.66  & 7.23& 5.95& -5.56(3)   \\
HD 4313   & 2557541493057378048 & 8.78  & 7.82  & 6.15  & 5.77  & 5.60  & 7.59  & 8.06& 6.95& 14.56(4)   \\
HD 5214   & 2553288689455891072 & 8.97  & 7.92  & 6.11  & 5.58  & 5.47  & 7.64  & 8.17& 6.95& 5.08(3)\\
HD 6432   & 2578083844893292672 & 9.04  & 7.92  & 5.99  & 5.44  & 5.29  & 7.60  & 8.16& 6.89& 23.92(3)   \\
HD 22045  & 56717246464212736   & 9.06  & 8.01  & 6.14  & 5.65  & 5.54  & 7.71  & 8.23& 7.03& 27.26(3)   \\
HD 24680  & 51557169676123008   & 8.88  & 7.87  & 6.01  & 5.60  & 5.42  & 7.60  & 8.12& 6.92& 27.56(3)   \\
HD 76445  & 611642860944510848  & 8.59  & 7.64  & 5.85  & 5.38  & 5.28  & 7.36  & 7.85& 6.71& -16.17(3)  \\
HD 77776  & 605364202873870336  & 8.37  & 7.37  & 5.60  & 5.14  & 4.97  & 7.10  & 7.60& 6.44& -5.33(3)   \\
HD 78419  & 610968448000722432  & 8.63  & 7.58  & 5.77  & 5.32  & 5.13  & 7.30  & 7.82& 6.62& 12.26(3)   \\
78 Cnc& 635095890642370560  & 8.38  & 7.17  & 5.20  & 4.75  & 4.48  & 6.86  & 7.45& 6.13& 77.91(3)   \\
HD 99596  & 3798435683212668288 & 8.16  & 7.17  & 5.43  & 5.04  & 4.77  & 6.89  & 7.39& 6.22& -31.42(3)  \\
HD 100872 & 3909470345980014592 & 7.79  & 6.83  & 5.12  & 4.67  & 4.47  & 6.57  & 7.07& 5.91& -38.33(5)  \\
HD 97716  & 3791794117584128640 & 7.77  & 6.62  & 4.82  & 4.43  & 4.09  & 6.33  & 6.88& 5.64& 21.44(4)   \\
HD 97491  & 3791765599002094464 & 8.53  & 7.62  & 5.86  & 5.36  & 5.22  & 7.33  & 7.81& 6.68& 134.04(4)  \\
HD 97197  & 3791775219728772480 & 8.60  & 7.48  & 5.64  & 5.15  & 4.92  & 7.19  & 7.74& 6.49& 11.39(3)   \\
p04 Leo   & 3811456580944840704 & 6.49  & -& 4.03  & 3.35  & 3.33  & 5.25  & 5.76& 4.59& 56.63(3)   \\ \hline
\end{tabular}%

\end{table*}

\section{Analysis}
\subsection{Stellar parameters} \label{stellarparameters}
In order to perform a uniform analysis of all the stars, we started with the determination of the stellar parameters (T$_{\rm eff}$, log(g), [Fe/H] and $\xi_{\rm mic}$) for all the stars in the sample and Arcturus, which is used to assess the accuracy of our measurements. 
Stellar parameters were determined spectroscopically, with the standard approach of excitation equilibrium of iron lines, where T$_{\rm  eff}$ and $\xi_{\rm mic}$ were determined by minimising the trends of Fe {\sc i} abundance with respect to excitation potential (EP) and reduced equivalent widths (EWs), respectively. The log(g) were derived using the ionisation equilibrium, where the difference in the abundances of Fe {\sc i} and Fe {\sc ii} should be less than 1$\sigma$. 

The initial guesses for temperature and log(g) were derived using a combination of photometry and isochrone fitting.
We determined the color temperature by averaging those based on combination of photometric colors ($B, ~V, ~J, ~K, ~G, ~G_{BP}$ and $G_{RP}$)  using calibrations from \cite{2020RNAAS...4...52M} for {\em Gaia DR2}. 
For log(g), an initial guess was obtained by fitting an isochorone obtained from PARSEC\footnote{http://stev.oapd.inaf.it/cgi-bin/cmd}\footnote{Note: The choice of isochrones is not crucial in this process, as these were used only for the initial guess of the parameters.}\citep{2012MNRAS.427..127B} with age = 0.5 Gyr 
and metallicity of -0.1 dex. For microturbulance, a constant value of 1.5 km\,s$^{-1}$ was taken as the initial estimate. Naturally, this is a rough approximation, however the reader should keep in mind that the parameters are determined spectroscopically and the result is rather robust with respect to 
variations of the first guess values. 

The line list adopted for Fe transitions is reported in Table \ref{tab:ironelements} and is based on that in \citet{2015MNRAS.449.4038D}.
Equivalent widths were measured using ARES \citep{ARES2007}. We removed any line with an EW of less than 4$\sigma$ of the fitting error, and lines with EW > 150 m\AA. The assumptions adopted by ARES in line fitting become less appropriate in strong lines, moreover strong saturated lines are not good indicators of the chemical abundance of the species they arise from.
Finally, we adopted the grid of Kurucz stellar atmospheres \citep{kuruz1992}. 

\begin{figure}
   \centering
   \includegraphics[width=0.9 \linewidth]{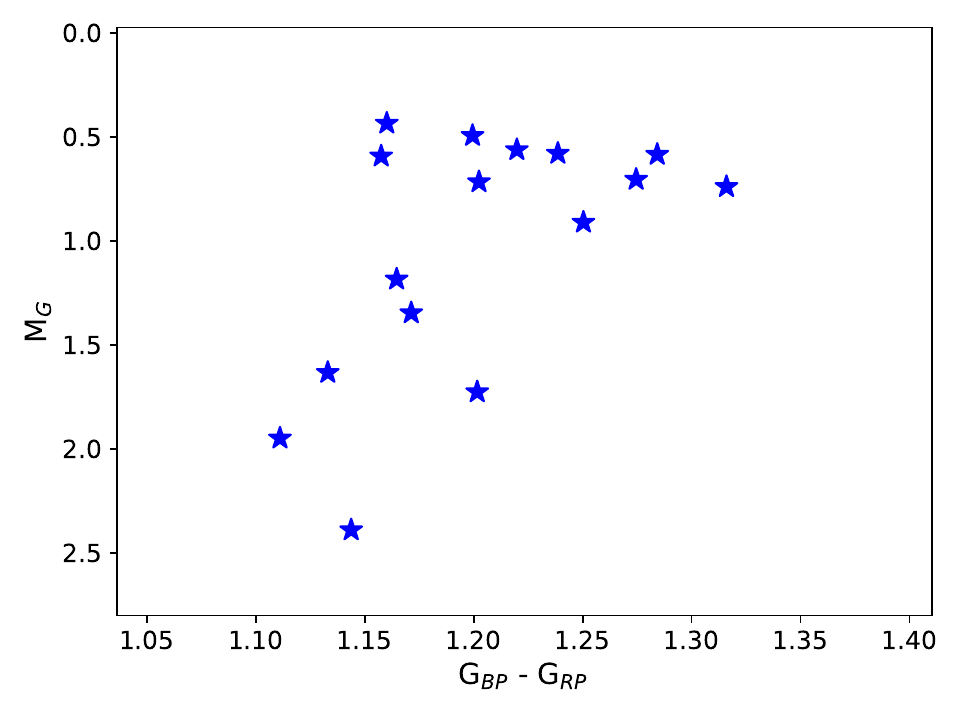}
   \caption{Color magnitude diagram of the stars studied using $Gaia$ magnitudes and parallaxes to compute the absolute magnitude $M_G$.}
   \label{fig:cmd}
\end{figure}

\begin{figure}
   \centering
   \includegraphics[width=1 \linewidth]{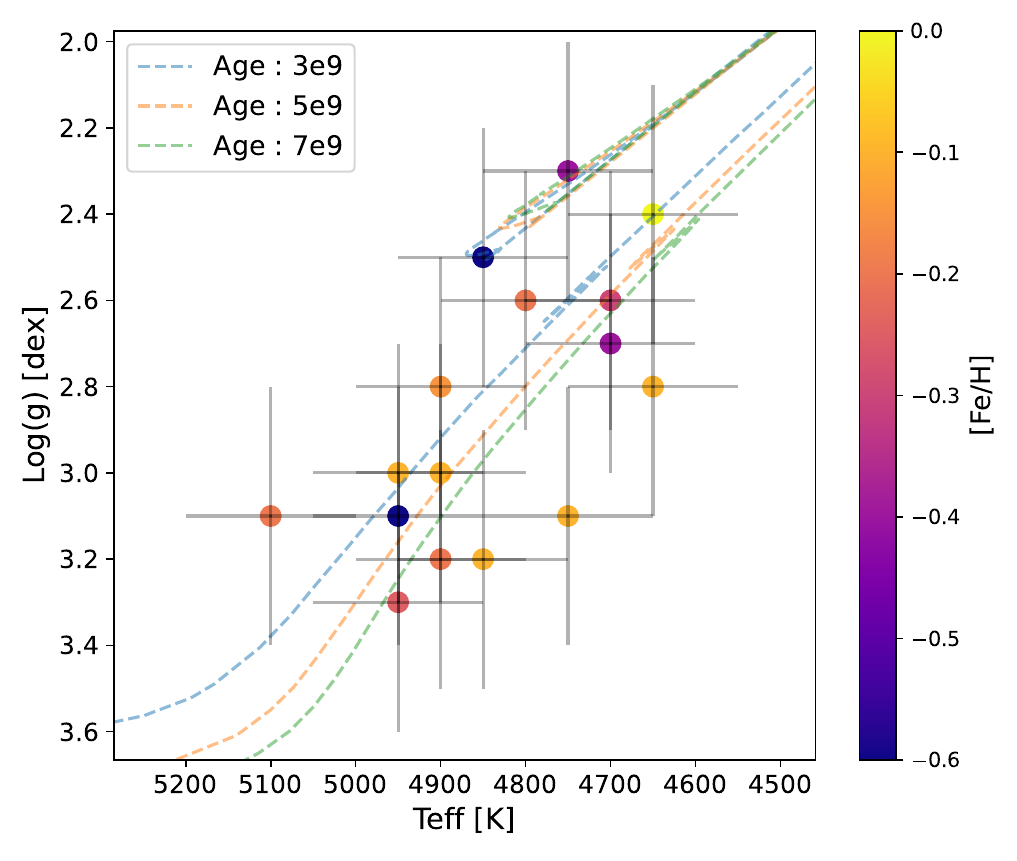}
   \caption{Kiel diagram (T$_{\rm eff}$ vs log(g)) of the stars using stellar parameters from spectroscopic analysis with an overlay of metallicity. The three dashed lines represent the stellar isochrones with ages of 3 Gyr (blue), 5 Gyr (orange), and 7 Gyr (green), and metallicity of -0.1 dex.}
   \label{fig:kiel}
\end{figure}

We performed the analysis using the {\tt abfind} driver of MOOG \citep{MOOG1973}, in its python-wrapped version,  pyMOOGi\footnote{https://github.com/madamow/pymoogi}). 
The values of the input stellar parameters were changed until the trends of Fe abundance with excitation potential, reduced EWs were flattened (within the errors), and the differences between Fe II and Fe I abundances were within the observational uncertainties. Once the parameters were estimated using Fe, we refined them by repeating the whole process with Ti lines given in Table \ref{tab:Tielements}. The difference in parameters estimated using Fe and Ti was smaller than the uncertainties on the parameters ($\delta\,T_{\rm eff}$ = 50 K, $\delta\,log(g)$ = 0.1 dex and $\delta\,\xi_{\rm mic}$ = 0.05 km\,s$^{-1}$). The parameters obtained for our sample are given in Table \ref{tab:stellarparameter} with the color-magnitude diagram (CMD) and Kiel diagram in Fig. \ref{fig:cmd} and \ref{fig:kiel}, respectively.

Along with our sample, we performed an analysis of Arcturus.
For Arcturus, we obtained an effective temperature of 4300 $\pm$ 70 K, log(g) of 1.66 $\pm$ 0.20 dex, [Fe/H] of -0.50 $\pm$ 0.10 dex and $\xi_{\rm mic}$ of 1.74 $\pm$ 0.08 km\,s$^{-1}$. These are consistent with the values obtained by \cite{2011ApJ...743..135R}.

Given the homogeneous nature of our sample, the uncertainties on each of the parameters were very similar for all the stars. We adopt uniform uncertainties as follows: 100 K on T$_{\rm eff}$, 0.30 dex on log(g), 0.13 dex on [Fe/H]\footnote{In this paper, we used the standard notations to express the abundances: for an element X, [X] = log $\epsilon(X)_{star}$ - log $\epsilon(X)_{\odot}$, [X/Fe] = [X/H] - [Fe/H] and A($\epsilon$) = log($N_{x}/N_{H}$) + 12.} and 0.10 km\,s$^{-1}$ on $\xi_{\rm mic}$. 

\begin{figure}
   \centering
   \includegraphics[width=1 \linewidth]{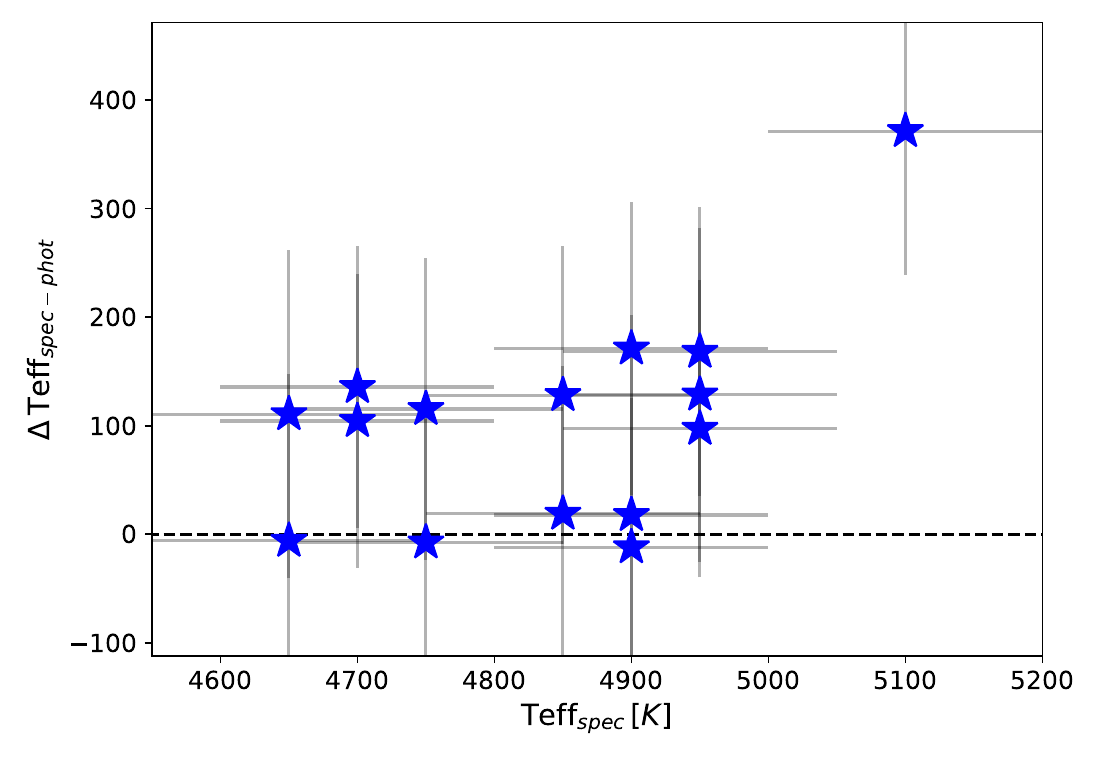}
   \caption{Comparison of effective temperature obtained from spectroscopy and photometry.}
   \label{figure:teff_phot_vs_teff_spec}
\end{figure}

\begin{table}[]
\centering
\caption{Stellar properties (effective temperature, surface gravity, metallicity and microturbulant velocity) for all the stars in the sample and Arcturus. The uncertainties on the parameters for the sample stars are as follows: 100 K for T$_{\rm eff}$, 0.30 dex for log(g), 0.13 dex for [Fe/H] and 0.10  km\,s$^{-1}$ for $\xi_{\rm mic}$. 
}
\label{tab:stellarparameter}
\resizebox{\linewidth}{!}{%
\begin{tabular}{lcccc}
\hline
\multicolumn{1}{c}{\multirow{1}{*}{StarID}} & \begin{tabular}[c]{@{}c@{}}T$_{\rm eff}$\\ (K)\end{tabular} & \begin{tabular}[c]{@{}c@{}}log(g)\\ (dex)\end{tabular} & \begin{tabular}[c]{@{}c@{}}{[}Fe/H{]}\\ dex\end{tabular} & \begin{tabular}[c]{@{}c@{}}$\xi_{\rm mic}$\\ (km\,s$^{-1}$)\end{tabular} \\ \hline
HD 218330 & 4700  & 2.70& -0.40 & 1.60   \\
HD 4313   & 4900  & 3.20& -0.20 & 1.20   \\
HD 5214   & 4900  & 2.80& -0.15 & 1.70   \\
HD 6432   & 4700  & 2.60& -0.30 & 1.77   \\
HD 22045  & 4850  & 3.20& -0.10 & 1.40\\
HD 24680  & 5100  & 3.10& -0.20 & 1.40   \\
HD 76445  & 4950  & 3.30 & -0.25 & 1.20\\
HD 77776  & 4750  & 2.30& -0.40 & 1.50   \\
HD 78419  & 4800  & 2.60 & -0.20 & 1.65   \\
78 Cnc& 4650  & 2.90& -0.10 & 1.90   \\
HD 99596  & 4950  & 3.00& -0.10 & 1.45   \\
HD 100872 & 4850  & 2.50& -0.60 & 1.70   \\
HD 97716  & 4650  & 2.40& 0.00  & 1.68   \\
HD 97491  & 4950  & 3.10   & -0.60 & 1.40   \\
HD 97197  & 4750  & 3.10   & -0.10 & 1.60   \\
p04 Leo   & 4900  & 3.00& -0.10 & 1.40   \\ \hline \hline
Arcturus  & 4300 $\pm$ 70  & 1.66 $\pm$ 0.20 & -0.50 $\pm$ 0.10 & 1.74 $\pm$ 0.08   \\ \hline
\end{tabular}
}
\end{table}

\begin{figure}
   \centering
   \includegraphics[width= 1\linewidth]{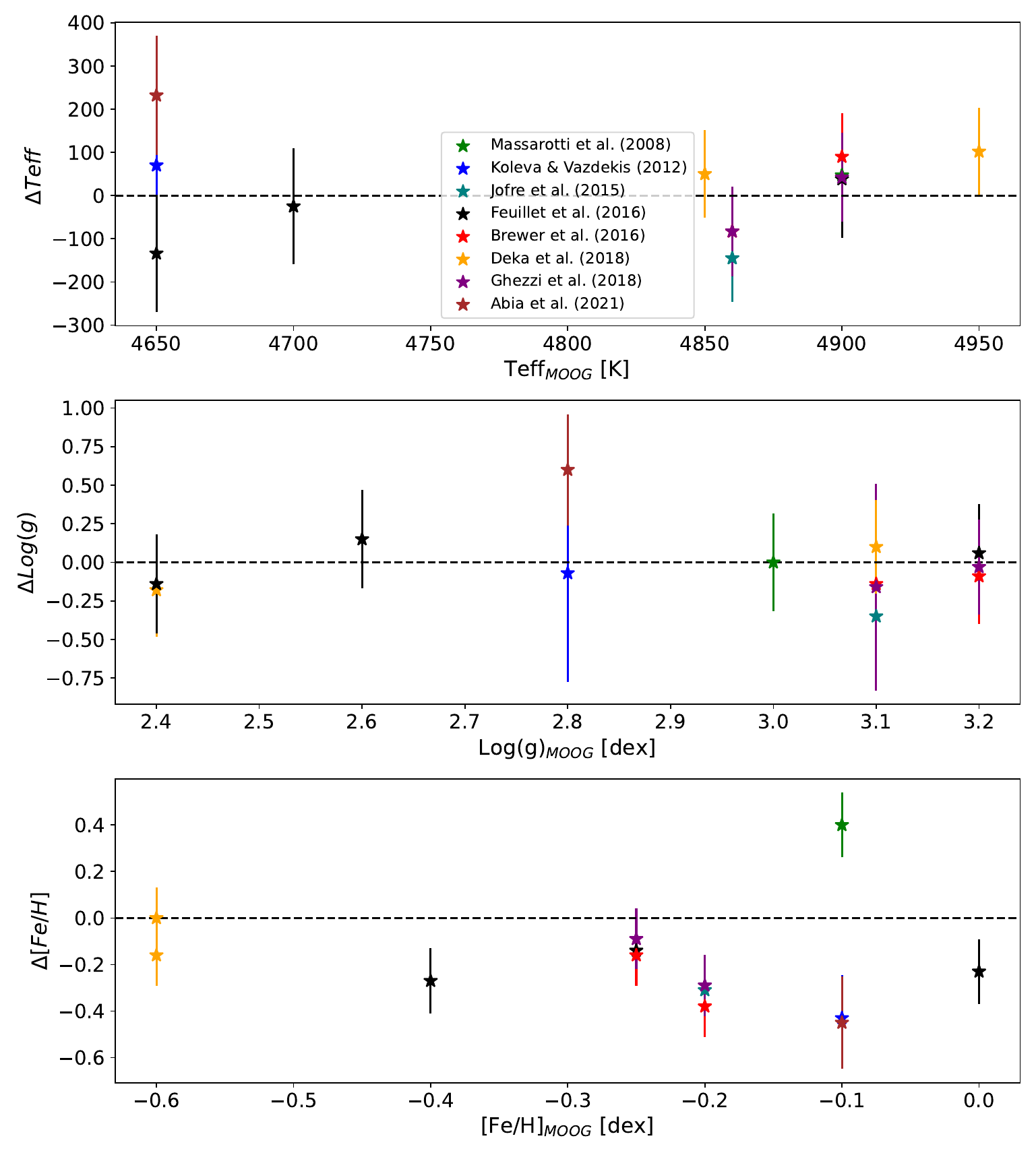}
   \caption{Comparison of stellar parameters (T$_{\rm eff}$, log(g) and [Fe/H]) obtained in the study with literature sources from \cite{teff7,teff5,teff4,teff1,teff3,teff2,teff6,teff8}. The black line represents the $y=0$ line. In all three panel, each color represents a specific study as described in the legend.}
   \label{Figure:MOOG_vs_GAIA}
\end{figure}

A comparison between the effective temperature obtained through photometric magnitudes and from spectroscopic analysis is shown in Fig. \ref{figure:teff_phot_vs_teff_spec}. Even though the two temperatures agree within the errors for some stars, as a whole sample, we see a mean difference of about 90 K and a standard deviation of 85 K, with spectroscopic temperatures being higher. This offset could be due to the metallicity value of -0.1 dex that was assumed for calculating the photometric temperature, or to unaccounted reddening, which however seems unlikely given the spatial distribution of the objects. In Fig. \ref{Figure:MOOG_vs_GAIA} and \ref{fig:MOOGcomparison}, we compare our stellar parameters (labelled as MOOG) with those available in the literature, APOGEE, LAMOST and {\em Gaia}. The literature values were taken from different studies such as \cite{teff7}, \cite{teff5}, \cite{teff4}, \cite{teff1}, \cite{teff3}, \cite{teff2}, \cite{teff6} and \cite{teff8}, whereas the APOGEE and LAMOST values were taken from \cite{HD218330HD97716} and \cite{2018ApJ...858L...7T}, respectively. 

There is generally very good agreement with data from large surveys, that is Gaia, APOGEE, and LAMOST, as can be seen in Figure \ref{fig:MOOGcomparison}. On the other hand, the agreement is rather poor for what concerns \cite{2020ApJS..247...28H}, who report on the parameters derived with LAMOST spectra for four stars, especially for what concerns temperature. We note that, however, their values are quite discrepant from the corresponding Gaia ones.

 Fig. \ref{Figure:MOOG_vs_GAIA}, shows the comparison with atmospheric parameters reported in small-scale studies. Temperature is overall a good agreement, with the exception of 78Cnc, where the value from  \cite{teff8} shows a significantly higher temperature than the one derived in the present study. We have no explanation for this discrepancy, and we can only speculate that it might have to do with the different approach to the analysis adopted by the authors (automatic parameter determination based on the MATISSE algorithm) but we note, however, that the value reported for the same object by \cite{teff2} is is good agreement with ours. 

We note, however, that the presence of some offsets between different studies is not surprising, due to different choices made in analysis (e.g. classes of model atmosphere, solar composition adopted, line transition parameters etc). 

Given the uncertainties on the MOOG parameters, these differences are not significant. The only significant difference obtained was in the effective temperature of MOOG and \cite{2020ApJS..247...28H}. This difference is not reflected in the surface gravity or the metallicity. Assuming the T$_{\rm eff}$ from \cite{2020ApJS..247...28H} is correct, that would imply that both the spectroscopic and photometric T$_{\rm eff}$ are wrong. Having used different color indices to measure the photometric T$_{\rm eff}$, only extinction can introduce such a large deviation. But as the stars are located in the solar neighbourhood, the extinction is likely to be low, which was verified using {\em Gaia} DR3. This is further supported by the small deviations in derived temperatures of different color indices. Due to this, we assume our parameters to be more reliable than \cite{2020ApJS..247...28H}, which is supported by {\em Gaia} and other literature values.  

\subsection{Elemental abundances}
In order to test the stellar evolutionary models, we focused our abundance analysis on elements particularly affected by evolution such as carbon, nitrogen, oxygen, and lithium. Along with these, we also performed analysis on some of the $\alpha$- and Fe-peak elements to obtain a better understanding of the chemical signatures. Additionally, we derived the abundance of fluorine, rarely measured, and of yttrium, to check on the relations between age and $\alpha$ to neutron capture element ratio. For the analysis, we used two methods: spectral synthesis and equivalent width. We used the {\tt abfind} driver of pyMOOGi for the equivalent width method and {\tt synth} driver for spectral synthesis with interpolated Kurucz atmosphere models.  

\subsubsection{Carbon, Nitrogen, and Oxygen (CNO)} \label{sec:CNO}

Due to the presence of GIARPS data for all the stars except one, we conducted separate analysis on the optical and the IR spectrum using the spectral synthesis method. The linelist used for the entire CNO analysis was generated using the Linemake\footnote{https://github.com/vmplacco/linemake} tool \citep{2021RNAAS...5...92P}; the interested reader is referred to the paper for details on the sources of the parameters of the atomic and molecular transitions that are taken into account. 

\begin{figure}
   \centering
   \includegraphics[width=1 \linewidth]{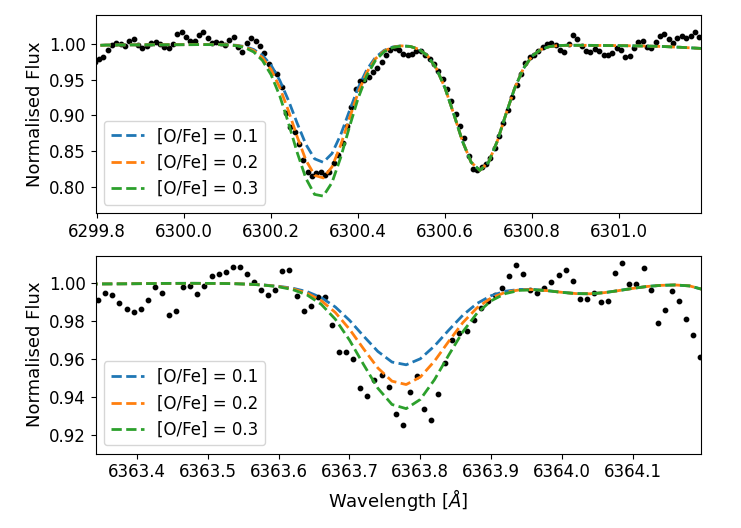}
   \caption{Fitting of two oxygen lines at 6300 and 6363 $\AA$ in HD 5214. The observed spectrum is shown in black points with synthetic spectra of different oxygen abundance shown in blue ([O/Fe] = 0.1 dex), orange ([O/Fe] = 0.2 dex) and green ([O/Fe] = 0.3 dex). A line-by-line scatter of 0.1 dex can be seen between the two. }
   \label{fig:oxygen}
\end{figure}

\begin{figure*}
   \centering
   \includegraphics[width=0.49 \linewidth]{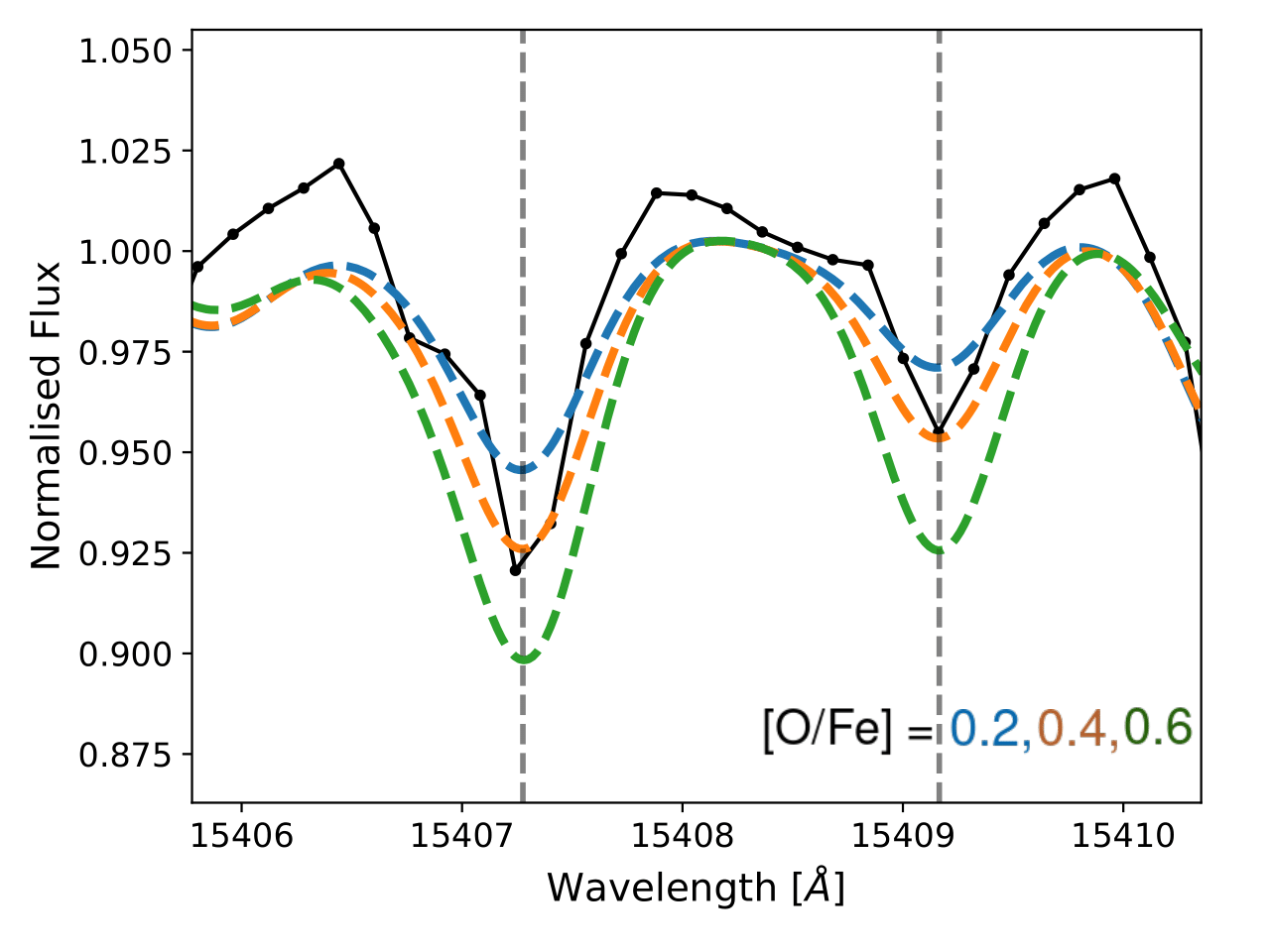}
   \includegraphics[width=0.49 \linewidth]{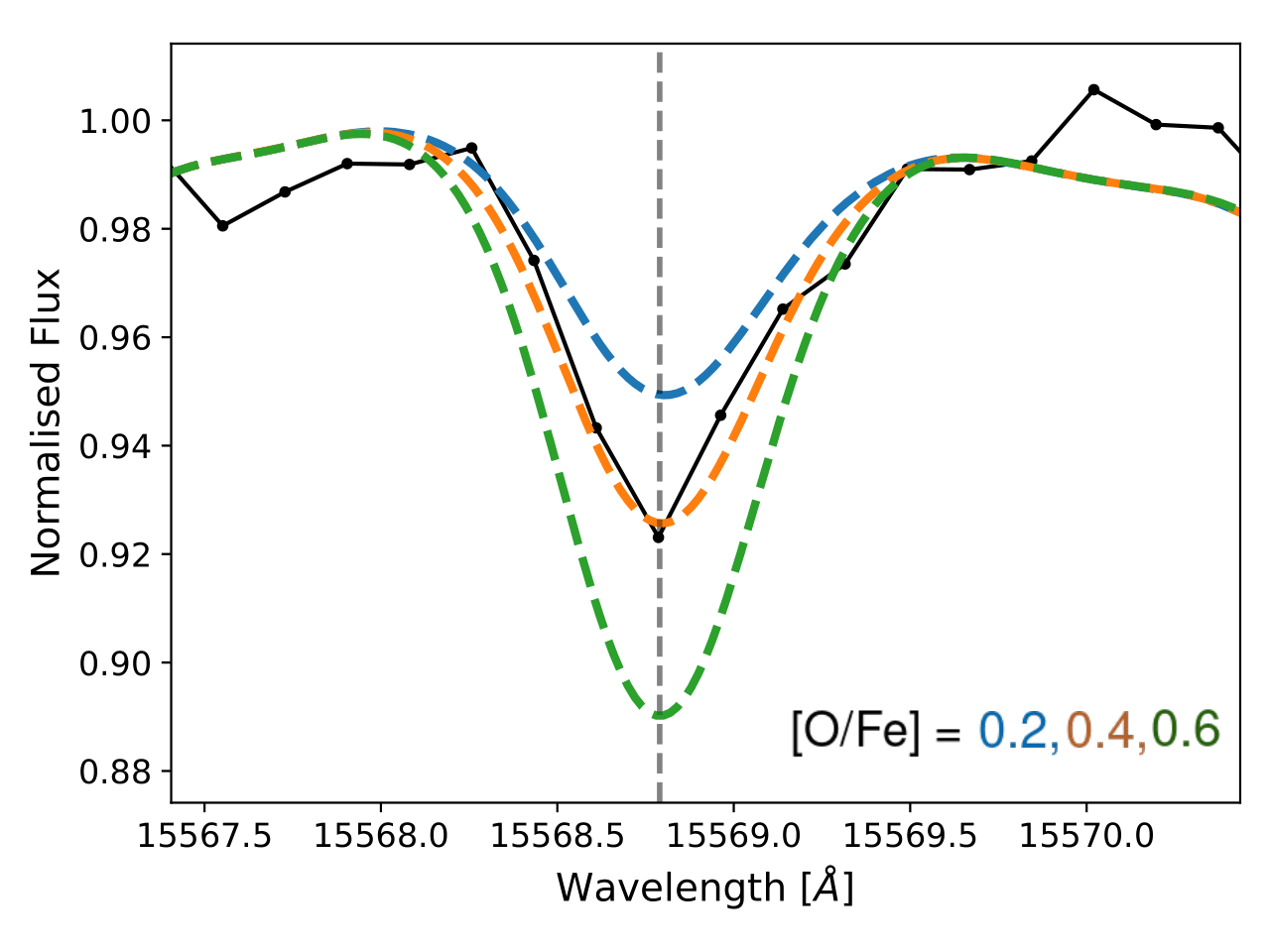}
   \includegraphics[width=0.49 \linewidth]{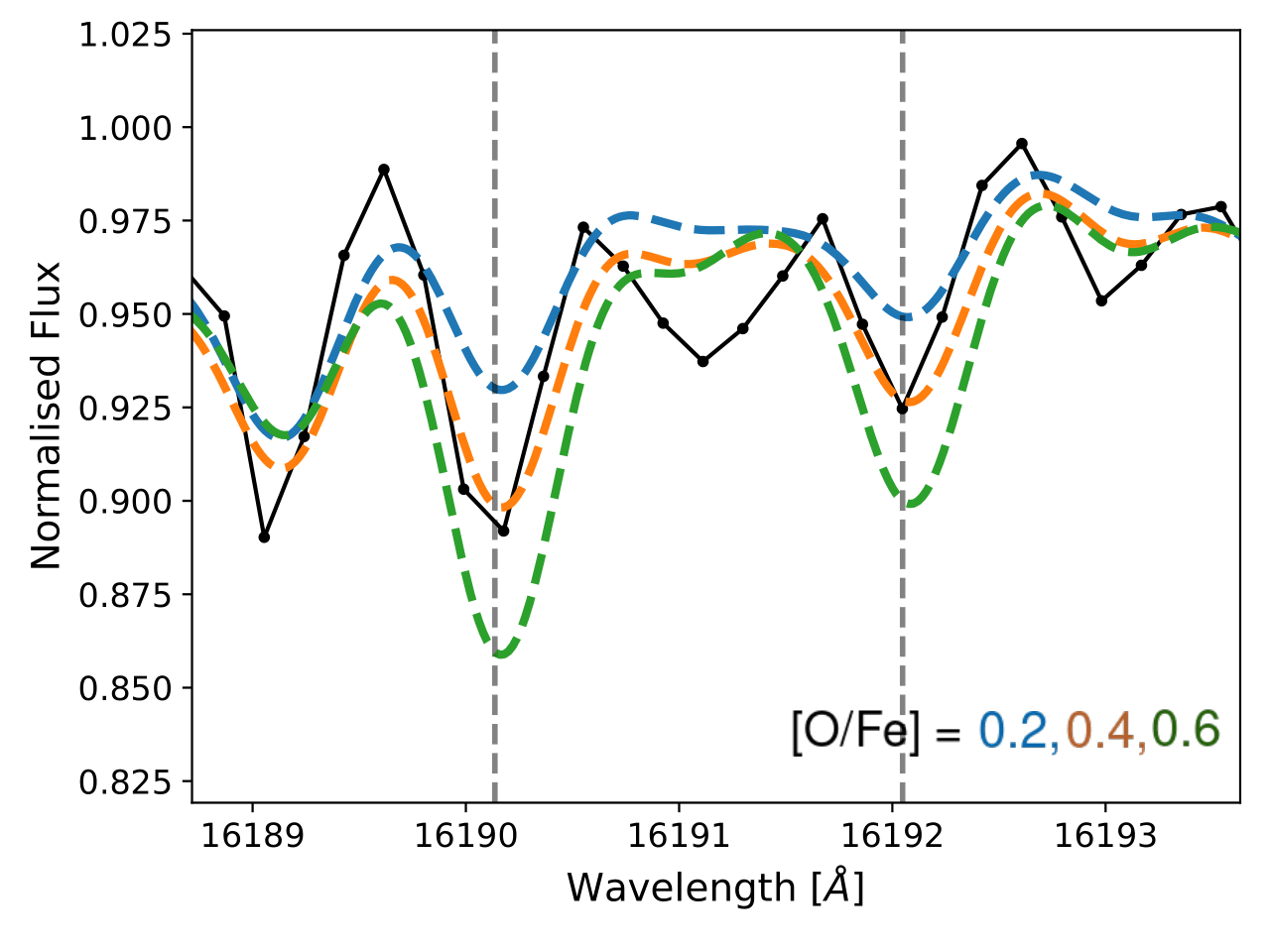}
   \includegraphics[width=0.49 \linewidth]{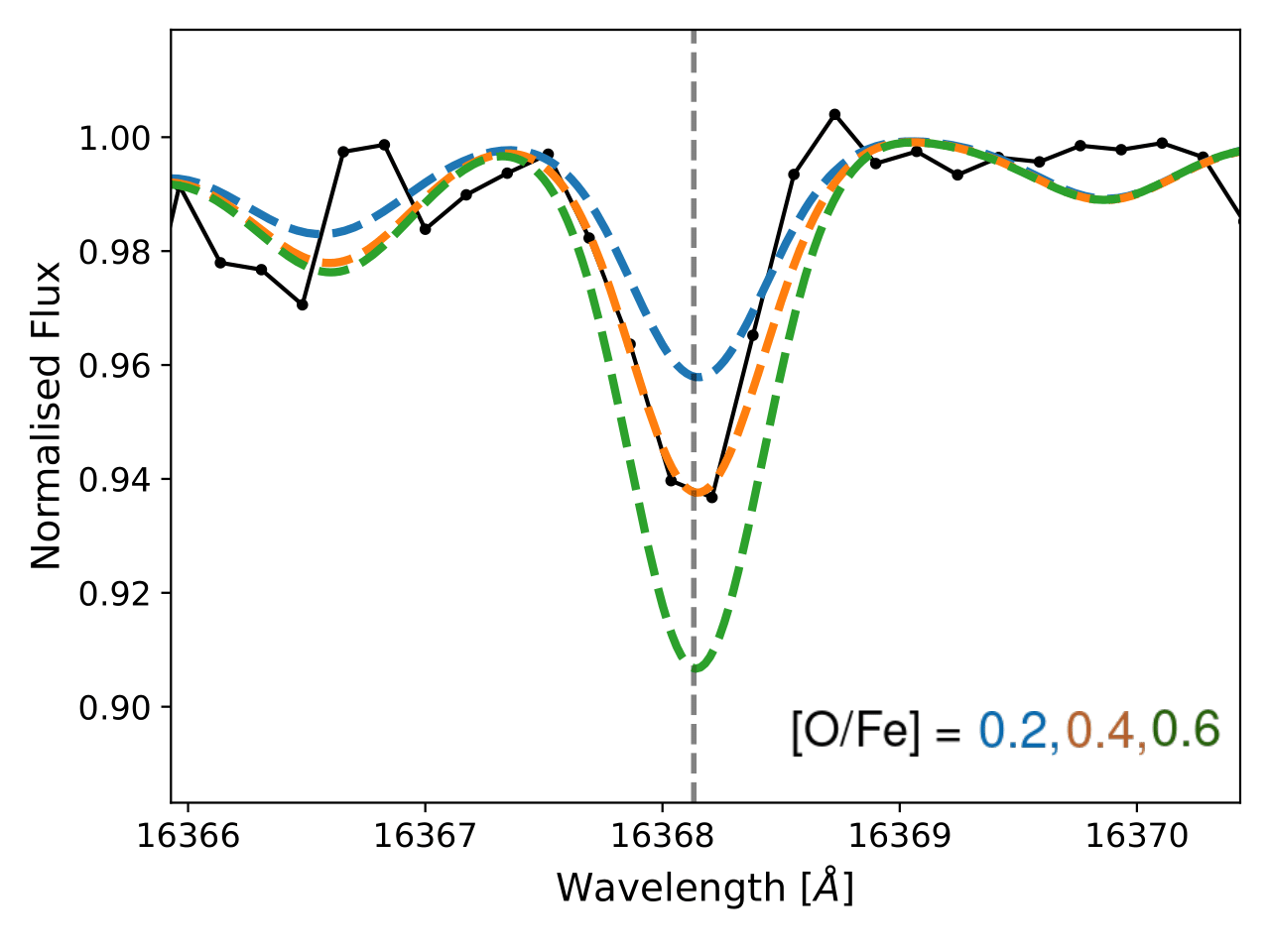}

   \caption{Fitting of OH band at 1.5 and 1.6 $\mu$m in HD 218330. The observed spectrum is shown in black with synthetic spectra of different oxygen abundance shown in blue ([O/Fe] = 0.2 dex), orange ([O/Fe] = 0.4 dex) and green ([O/Fe] = 0.6 dex). The vertical black dashed line shows the OH lines considered for the abundance analysis. The best-fit line is shown in orange with O = 0.40 dex.}
   \label{fig:OH}
\end{figure*}

For Carbon, we used the CH band at 4300 $\AA$ and one carbon high-excitation line at 5380 $\AA$ in the optical, and the CO band at 2.30 $\mu$m in the IR. Nitrogen was determined using the CN bands at 4100 $\AA$ and 1.5 $\mu$m in optical and IR region, respectively. We estimated oxygen abundance from the two forbidden lines at 6300 and 6363 $\AA$ in the optical and two OH bands at 1.5 and 1.6 $\mu$m in the IR region. We note that for one star (HD 78419) only the optical spectrum was available. 
For HD 97716, we could not measure the oxygen abundance from the OH molecular band as the band was exceedingly weak. Usually, telluric contamination becomes prominent at longer wavelength (about 6000 $\AA$), with large contamination in the IR region. As the two oxygen lines were located in the region affected by the tellurics, we made a visual inspection of the two lines in all the stars and found no contamination. For confirmation, we looked at the line-by-line scatter in the sample, with the understanding that a contaminated line would result in a vastly different oxygen abundance compared to the other line. We found the line-by-scatter to be between 0.07 to 0.13 dex with no drastic difference between the two lines in any of the stars, thus concluding that the lines were not affected by telluric contamination. 
The telluric contamination in the IR spectrum was modelled (as shown in Fig. \ref{fig:telluric}) and subtracted using the TelFit code \citep{2014AJ....148...53G}. Fig. \ref{fig:oxygen} and \ref{fig:OH} show the fitting of the two oxygen lines in the optical and IR, respectively. While fitting of the CH- (optical) and CO- (IR) bands is shown in Fig. \ref{fig:c} and fitting of the CN-band (both optical and IR) is shown in Fig. \ref{fig:n}.

From the abundances reported in Table \ref{tab:cno}, we found that the oxygen abundances obtained from the IR region were larger than those obtained from the optical atomic lines (see Fig.~\ref{fig:CNO}). Such a offset is not seen in Arcturus. However, all the stars in our sample are considerably hotter and younger than Arcturus, therefore the absence of an offset in Arcturus might not exclude the presence of systematic differences from measurements based on IR vs optical spectrum, intrinsic to the features analysed. As the OH band was weak and suffered from crowding, which is seen in Fig. \ref{fig:oxygen_weak}, we gave higher weight to the optical abundance because the atomic lines were relatively strong and well isolated. If the discrepancy between the two were within the errors of each other, we took an average but if the discrepancy was larger, the optical abundance was assumed to be a better estimate and used for further analysis of carbon and nitrogen. 

For carbon, we again found that the abundance from the IR molecular band was larger than optical in the majority of the sample. However, the difference between the two is not as large as in case of oxygen. This discrepancy between the abundance of optical and IR also extends to nitrogen where the IR abundances are larger than the optical once more. High excitation carbon lines also do not agree with the abundance from the CH and CO bands. This could be due to the availability of only one high excitation line within our spectral coverage. A similar offset is found also in Arcturus, where the high excitation line returns an abundance +0.3 dex larger than those obtained from the same CH and CO bands. 

Arcturus shows a good agreement in abundances for CNO, with the difference between optical and IR being $\delta$[O/Fe] = 0.01 dex, $\delta$[C/Fe] = 0.01 dex and $\delta$[N/Fe] = 0.05 dex, which are consistent within the errors. 
 
\begin{figure}
   \centering
   \includegraphics[width=1 \linewidth]{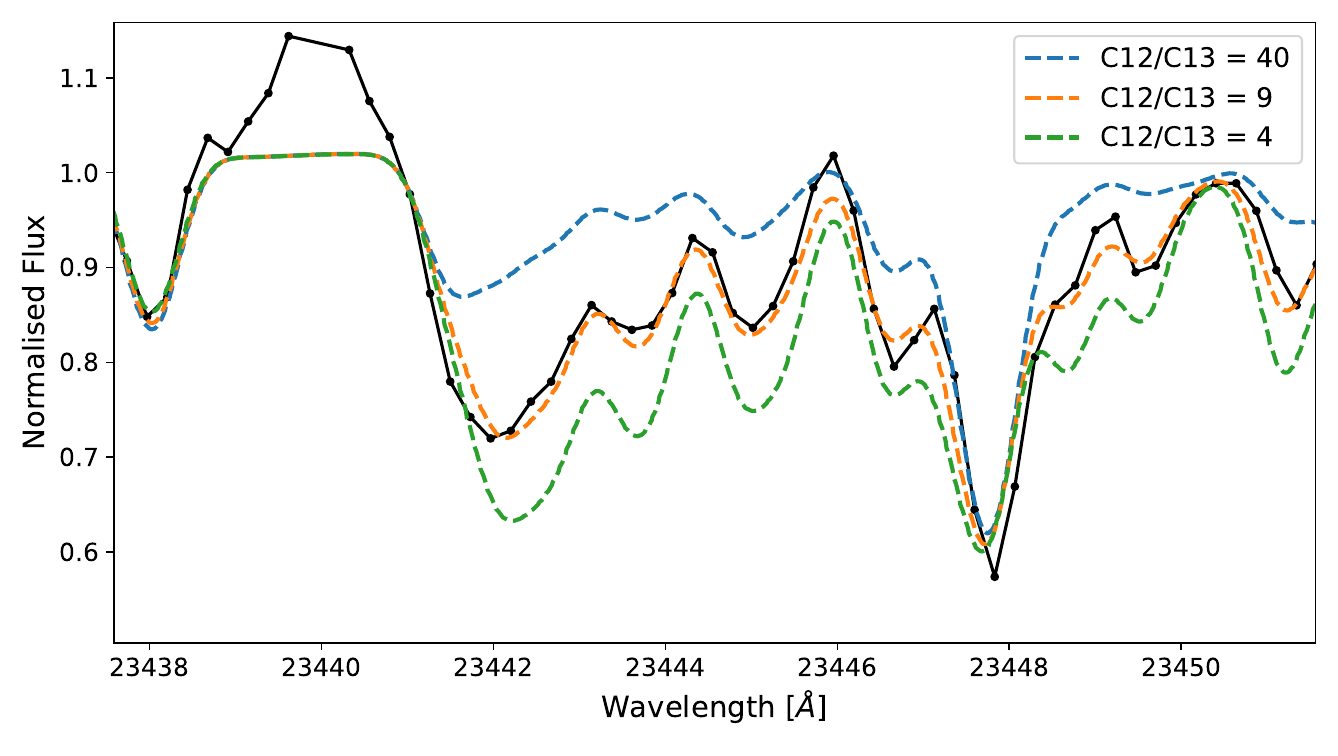}
   \caption{Fitting of $^{13}$CO (2-0) band at 23442 $\AA$ in HD6432 for the estimation of carbon isotopic ratio. Observed spectrum is shown in black with synthetic spectra of varying isotopic ratios shown in different color. The best-fit carbon isotopic ratio is 9 (dashed orange line). }
   \label{fig:CO1213}
\end{figure}

After the estimation of carbon abundance in optical, we pursued the estimate of the carbon isotopic ratio, i.e. $^{12}$C/$^{13}$C using the CH and $^{13}$CO (2-0) band-head in the optical and IR, respectively. Given the stars were observed in sub-optimal weather conditions, the $^{13}$CO molecular band at 2.34 $\mu$m was suffering from a large telluric contamination. Even after modelling and subtracting these telluric lines using Telfit, many spectra had significant residuals which increased the difficulty of obtaining a reliable measure of the isotopic ratios in majority of the stars. Due to this, we first measured the isotopic ratio from the CH band in the optical and then used the $^{13}$CO band to refine the ratio. Figure \ref{fig:CO1213} shows the fitting of the CO band in HD6432. We find values ranging between 15 and 6, as seen from Table \ref{tab:cno}.

\begin{table*}[]
\centering
\caption{Abundances of carbon, nitrogen, and oxygen obtained from the optical and IR region. The abundances are given in terms of [X/Fe]. Carbon isotopic ratio is given in the last column. The uncertainties in the optical and IR abundances of CNO elements are as follows: 0.16 and 0.15 dex for carbon, 0.18 and 0.16 dex for nitrogen, and 0.09 and 0.17 dex for oxygen. }
\label{tab:cno}
\begin{tabular}{lccccccc}
\hline
\multicolumn{1}{c}{\multirow{2}{*}{Star}} & \multicolumn{2}{c}{Carbon}& \multicolumn{2}{c}{Nitrogen}& \multicolumn{2}{c}{Oxygen}& \multirow{2}{*}{$^{12}$C / $^{13}$C} \\ \cline{2-7}
\multicolumn{1}{c}{} & \multicolumn{1}{c}{Optical} & \multicolumn{1}{c}{Infrared} & \multicolumn{1}{c}{Optical} & \multicolumn{1}{c}{Infrared} & \multicolumn{1}{c}{Optical} & \multicolumn{1}{c}{Infrared} &\\ \hline\hline
HD 218330& -0.06   & 0.01& 0.33& 0.50& 0.20& 0.40& $\sim$15\\
HD 4313 & -0.03   & -0.06& 0.34& 0.52& 0.10& 0.37& $\sim$15\\
HD 5214 & -0.11   & -0.04& 0.40& 0.55& 0.25& 0.35& $\sim$15\\
HD 6432 & -0.03& 0.05& 0.29& 0.65& 0.22& 0.20& 9 (3) \\
HD 22045& -0.14   & -0.01& 0.45& 0.55& 0.20& 0.20& $\sim$15\\
HD 24680& 0.02& -0.07& 0.55& 1.00& 0.19& 0.40& 6 (3) \\
HD 76445& 0.00&0.05& 0.48& 0.53& 0.27& 0.58& \textgreater 15\\
HD 77776& -0.15   & -0.20& 0.25& 0.50& -0.05& 0.18& 6 (3) \\
HD 78419& -0.08   & - & 0.30& - & 0.18& - & $\sim$15\\
78 Cnc  & 0.18& 0.29& 0.15& 0.70& 0.30& 0.20& 9 (3) \\
HD 99596& -0.10   & 0.05& 0.30& 0.45& 0.15& 0.10& $\sim$15\\
HD 100872& -0.03   & -0.04& 0.23& 0.43& 0.29& 0.50& 6 (3) \\
HD 97716& -0.23   & -0.02& 0.10& 0.35& -0.13   & - & $\sim$15\\
HD 97491& -0.05   & 0.25& 0.15& 0.16& 0.27& 0.50& $\sim$6\\
HD 97197& -0.07   & 0.12& 0.30& 0.70& 0.07& 0.22& $\sim$15\\
p04 Leo & -0.15   & -0.09& 0.12& 0.32& 0.21& 0.35& $\sim$15\\ \hline
Arcturus& 0.05 $\pm$ 0.10  & 0.05 $\pm$ 0.05 & 0.45 $\pm$ 0.11 & 0.40 $\pm$ 0.11 & 0.45 $\pm$ 0.06 & 0.49 $\pm$ 0.06 & 6 (3) \\ 
\hline
\end{tabular}
\end{table*}

\subsubsection{Lithium}
\begin{figure}
   \centering
   \includegraphics[width=1 \linewidth]{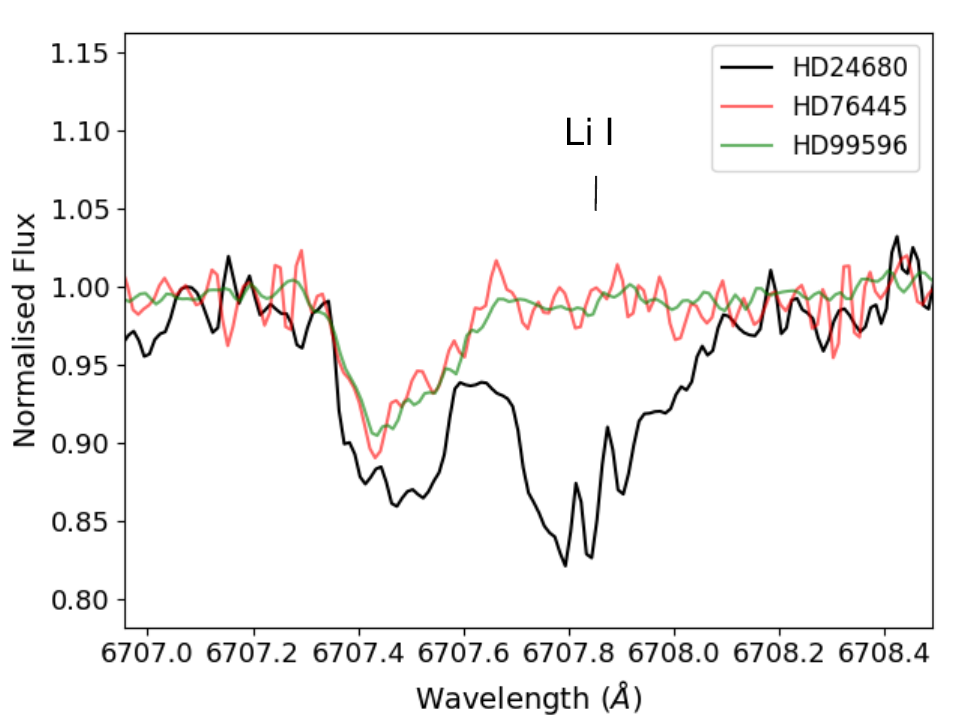}
   \caption{Comparison of Li line strength in HD24680 (black), HD76445 (red) and HD99596 (green). The line in HD24680 is slightly blended with a nearby CN line at 6707.45 $\AA$.}
   \label{fig:Licomparison}
\end{figure}

We used spectral synthesis to probe the Li abundance using the line at 6707.78 $\AA$. The linelist used for the synthesis was adopted from \cite{2015MNRAS.449.4038D}. With two exceptions (HD24680 and HD22045), we could only estimate upper limits on the Li abundance as the lines were weak. Both HD24680 and HD22045 show relatively strong Li lines with A(Li) of 1.46 $\pm$ 0.20 and 0.65 $\pm$ 0.20 dex, respectively. Considering the errors, HD 24680 can be classified as a Li-rich giant (A(Li) > 1.5 dex,  \cite{1989ApJS...71..293B}). The lithium line of HD24680 is shown in Fig. \ref{fig:Licomparison} along with two other stars of the sample with similar stellar properties. The line is slightly blended with the neighbouring CN line. For Arcturus, we could not measure the Li abundance as the line was too weak.

For the considered Li feature, LTE is a poor approximation \citep{2009A&A...503..541L}, and in order to address this we applied non-local thermodynamical equilibrium (NLTE) corrections, as prescribed by the INSPECT\footnote{Data obtained from the INSPECT database, version 1.0 (\hyperlink{www.inspect-stars.net}{www.inspect-stars.net})} tool which allows the user to estimate the NLTE correction on a line-by-line basis. In our sample, the correction ranged between 0.13 and 0.25 dex with NLTE abundance being higher than the LTE. We like to note that the correction was not available for 3 stars due to weak lines (small EWs). Li abundances along with the NLTE corrections are reported in Table \ref{tab:abundance}.

\subsubsection{$\alpha$ and iron peak elements}
\begin{figure}
   \centering
   \includegraphics[width=1 \linewidth]{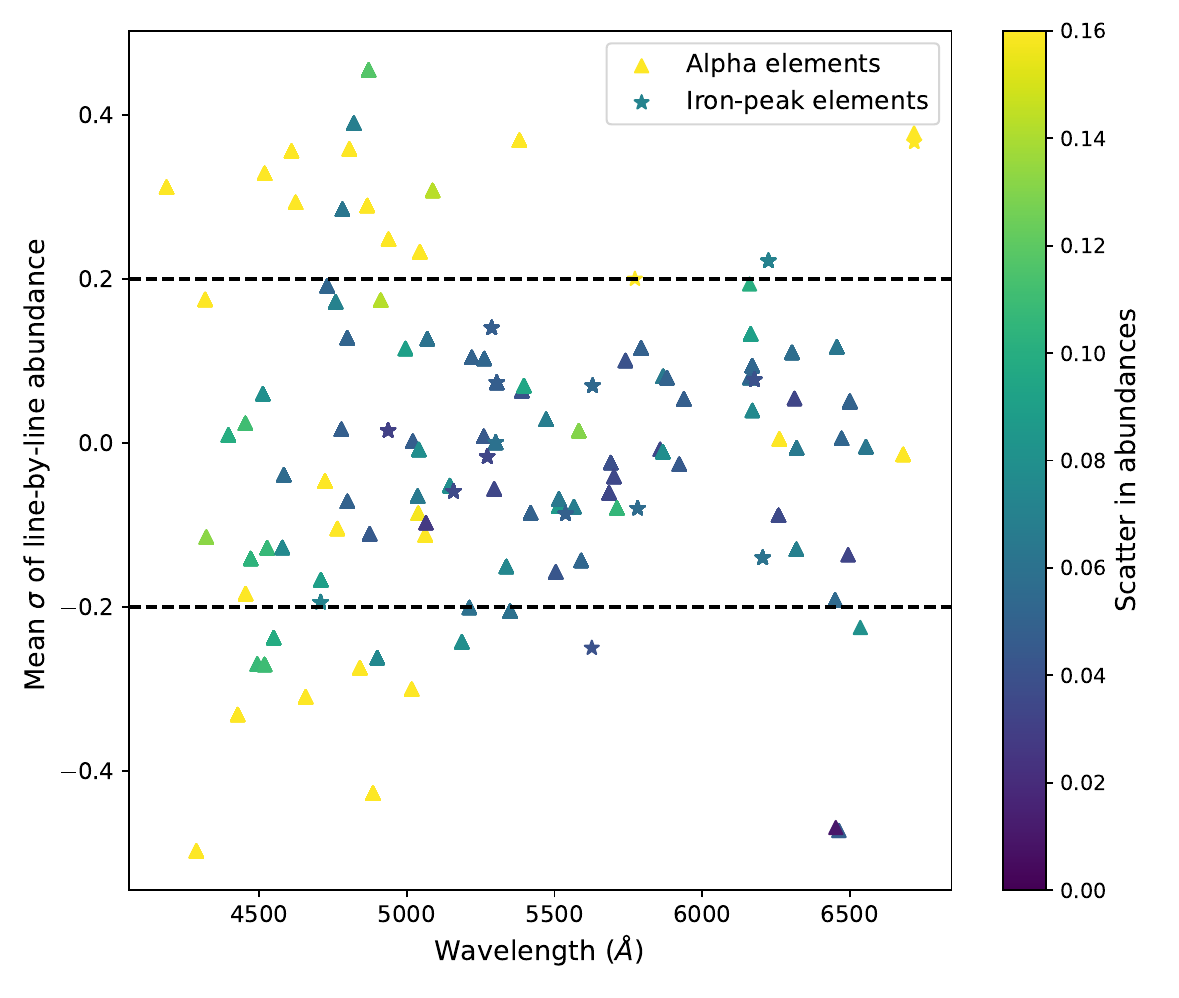}
   \caption{Offset and scatter in the line-by-line abundance of $\alpha$- and Fe-peak elements. The y-axis represents the mean of the standard deviation of line-by-line abundance (average abundance from all the lines - abundance from the line). The color represents the scatter of the standard deviation in different stars. The black line represents the offset limit of the mean, beyond which the line is discarded.}
   \label{fig:lines}
\end{figure}

We also derived the chemical abundances of a few $\alpha$- and iron peak elements, namely: Mg, Si, Ti, Ca, Cr, and Ni. For these elements, we used the equivalent width method based on measurement derived using ARES. The linelist used for each of the elements is given in Table \ref{tab:alphaelements}. Lines that had large errors on the EWs or resulted in discrepant abundances were manually checked using IRAF. 
The reliability of the abundances from individual lines were checked on a star-by-star basis. This was done by looking at the difference in the line-by-line abundance with respect to the average abundance of that element in that particular star. By repeating this for all the stars, we can plot the mean of this difference along with its scatter as shown in Fig. \ref{fig:lines}. In the figure, it is evident that a number of lines show an offset (mean is systematically higher or lower than 0) such as the Ca {\sc i} lines at 6717 $\AA$ and 6462 $\AA$. Some of the lines also show large scatter, e.g. Cr {\sc ii} line at 5772 $\AA$ and Ca {\sc i} line at 6717 $\AA$, with a standard deviation of 0.19 and 0.16, respectively. Offsets can be due to differences in the parameters of the transitions used and/or an unaccounted for contamination (i.e. blending with another line).
When a line of interest is blended with a nearby line, ARES can find it difficult to accurately measure the EW that results in determination of wrong abundance. Therefore, we discarded any lines with a mean difference outside the range of -0.2 to 0.2 as the errors on the abundances range between 0.05 and 0.2  dex for different elements. On average, we measured 4, 4, 18, 26, 8, 8 and 5 lines for Mg, Si, Ca, \ion{Ti}{i}, \ion{Ti}{ii}, Cr and Ni, respectively. The abundances are listed in Table \ref{tab:abundance}. Along with our sample, we also measure these elements in Arcturus. The abundances of all the elements except Ni are in agreement with \cite{2011ApJ...743..135R} within their respective errors. The Ni abundance is lower by 0.1 dex in our analysis when compared to \cite{2011ApJ...743..135R}. 

\subsubsection{Fluorine}
\begin{figure}
   \centering
   \includegraphics[width=1 \linewidth]{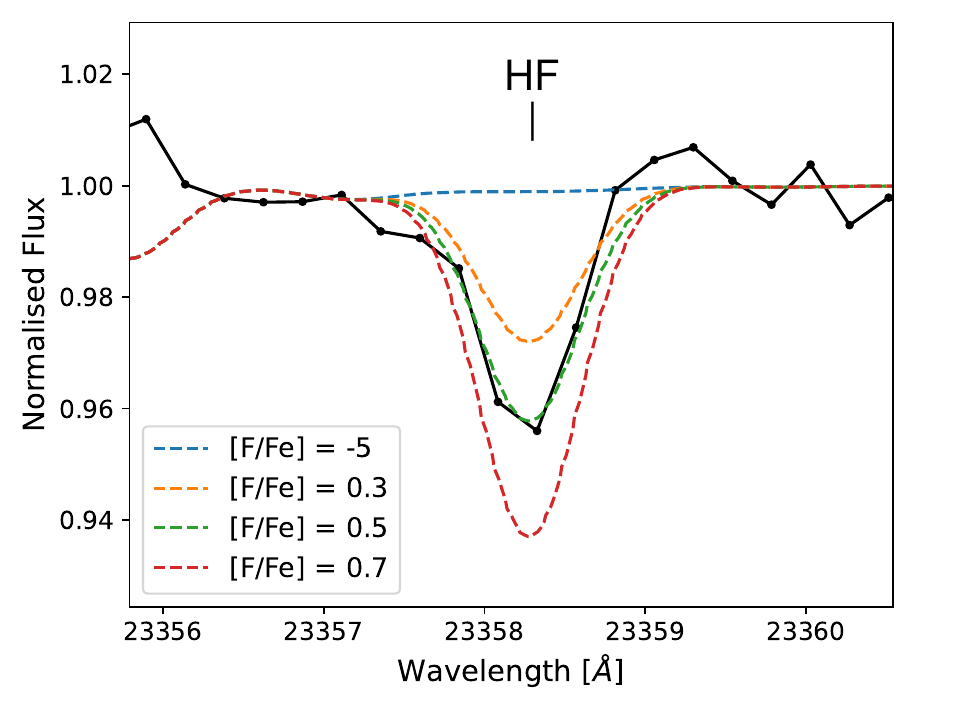}
   \caption{Fitting of HF line at 23358 $\AA$ in 78 Cnc. The observed spectrum is shown in black with synthetic spectra of varying fluorine abundance shown in different colors. The best-fit abundance for fluorine is 0.5 dex (green dashed line). }
   \label{fig:HF}
\end{figure}

Fluorine is a rather light element (Z = 9) in the periodic table neighbouring nitrogen, oxygen and neon. It is also quite a poorly understood element in the context of stellar nucleosynthesis. This is the result of its relatively low abundance when compared to its neighbouring elements (about four orders-of-magnitudes smaller than elements such as nitrogen and oxygen, \citealt{2022ApJ...929...24G}), which hampers the detection of its lines in stellar spectra. Its content in stellar atmospheres is in fact measurable only through a limited number of very weak and often telluric-contaminated transitions of the molecule HF. In addition to this, the possibility of fluorine formation in a variety of astrophysical sites makes it difficult to understand its origin. Abundance of fluorine is usually studied using transitions of the HF molecule in the IR. In this work, we used the spectral synthesis of the 23358 $\AA$ line to derive the abundances. We obtained reliable abundances only in two stars, reported in Table \ref{tab:abundance} and fitting shown in Fig. \ref{fig:HF} for one of them. We could not derive it for other stars due to problems caused by telluric removal as it was blended with a telluric line. 

\subsubsection{Uncertainties}

There are two components to the uncertainties associated to the abundance measurements of chemical species. One is that related to the error associated with the best-fit (for abundances derived through spectral synthesis) or due to line-by-line scatter (abundances derived through EW). The other is due to the uncertainties associated to the derived stellar parameters. Therefore, the uncertainties on these parameters needs to be propagated into the abundances. To calculate the errors, we employed the methodology outlined in \cite{2017A&A...598A..19D}. Our approach involved evaluating the impact of minor variations (corresponding to 1$\sigma$ change) in each stellar parameter on the abundance values. We then combined these effects in quadrature to determine the overall uncertainty.

As our sample contains stars that are similar to each other, we constructed the sensitivity matrix for one representative star (HD\,218330) which is given in Table \ref{tab:sensitivity}. For CNO abundances, we calculated the sensitivity matrix for both optical and IR, whereas for others, only optical is calculated. 
We assume the errors on CNO abundances in both optical and IR of all the stars in our sample to be the same as calculated for HD\,218330, whereas the errors for other abundances are given in Table \ref{tab:abundance} and \ref{tab:YMg}. 


\begin{table*}[]
\centering
\caption{Abundances of alpha-, Iron-peak elements, lithium and flourine. For lithium, we provide A(Li) along with the NLTE correction obtained from INSPECT. The majority of A(Li) measurements are upper limits except for two. }
\label{tab:abundance}
\resizebox{\textwidth}{!}{%
\begin{tabular}{lcccccccccccccccccc}
\hline
Star& {[}Mg/Fe{]} & $\sigma$ & {[}Si/Fe{]} & $\sigma$& {[}Ca/Fe{]} &   $\sigma$   & {[}Ti1/Fe{]} &   $\sigma$   & {[}Ti2/Fe{]} &   $\sigma$   & {[}Cr/Fe{]} &  $\sigma$& {[}Ni/Fe{]} &   $\sigma$   & A(Li) & NLTE\_Li & {[}F/Fe{]} & $\sigma$ \\ \hline
HD 218330 & 0.35 & 0.06 & 0.21 & 0.09 & 0.18 & 0.07 & 0.20  & 0.09 & 0.14  & 0.09 & 0.12 & 0.09 & 0.09 & 0.06 & $\leq$\,-0.95  & 0.20 & -   & -\\
HD 4313   & 0.42 & 0.09 & 0.19 & 0.11 & 0.22 & 0.10 & 0.19  & 0.11 & 0.16  & 0.12 & 0.20 & 0.10 & 0.14 & 0.09 & $\leq$\,-0.02  & 0.40 & -   & -\\
HD 5214   & 0.26 & 0.06 & 0.15 & 0.09 & 0.14 & 0.07 & 0.20  & 0.10 & 0.08  & 0.10 & 0.16 & 0.07 & 0.05 & 0.07 & $\leq$\,\,0.10 & 0.18 & -   & -\\
HD 6432   & 0.51 & 0.04 & 0.29 & 0.08 & 0.18 & 0.06 & 0.26  & 0.08 & 0.13  & 0.09 & 0.21 & 0.05 & 0.18 & 0.05 & $\leq$ -0.25   & 0.30 & -   & -\\
HD 22045  & 0.32 & 0.06 & 0.15 & 0.09 & 0.07 & 0.07 & 0.11  & 0.09 & 0.08  & 0.10 & 0.04 & 0.07 & 0.10 & 0.06 & 0.65   & 0.17 & 0.25& 0.18 \\
HD 24680  & 0.35 & 0.06 & 0.29 & 0.09 & 0.34 & 0.08 & 0.35  & 0.10 & 0.21  & 0.10 & 0.34 & 0.07 & 0.26 & 0.07 & 1.46   & 0.13 & -   & -\\
HD 76445  & 0.46 & 0.08 & 0.16 & 0.10 & 0.29 & 0.09 & 0.46  & 0.11 & 0.27  & 0.11 & 0.16 & 0.09 & 0.03 & 0.08 & $\leq$ -0.50   & 0.15 & -   & -\\
HD 77776  & 0.29 & 0.07 & 0.19 & 0.10 & 0.16 & 0.08 & 0.08  & 0.10 & 0.11  & 0.11 & 0.06 & 0.08 & -0.05& 0.08 & $\leq$ -0.25   & -& -   & -\\
HD 78419  & 0.32 & 0.11 & 0.20 & 0.13 & 0.18 & 0.12 & 0.20  & 0.13 & 0.09  & 0.14 & 0.17 & 0.12 & 0.10 & 0.12 & $\leq$ -0.15   & 0.20 & -   & -\\
78 Cnc    & 0.41 & 0.06 & 0.30 & 0.09 & 0.10 & 0.07 & 0.25  & 0.09 & 0.10  & 0.10 & 0.14 & 0.07 & 0.21 & 0.06 & $\leq$  0.25   & 0.23 & 0.50& 0.13 \\
HD 99596  & 0.19 & 0.07 & 0.00 & 0.11 & -0.03& 0.08 & 0.07  & 0.10 & -0.03 & 0.11 & -0.05& 0.08 & -0.12& 0.07 & $\leq$ -0.05   & 0.16 & -   & -\\
HD 100872 & 0.53 & 0.04 & 0.36 & 0.08 & 0.27 & 0.06 & 0.36  & 0.08 & 0.28  & 0.09 & 0.14 & 0.06 & 0.06 & 0.05 & $\leq$ -0.95   & -& -   & -\\
HD 97716  & 0.06 & 0.10 & 0.06 & 0.12 & -0.14& 0.11 & -0.13 & 0.13 & -0.14 & 0.13 & -0.13& 0.11 & -0.07& 0.11 & $\leq$ -0.15   & 0.25 & -   & -\\
HD 97491  & 0.48 & 0.04 & 0.27 & 0.08 & 0.30 & 0.06 & 0.38  & 0.09 & 0.32  & 0.09 & 0.10 & 0.06 & 0.00 & 0.05 & $\leq$   0.15  & -& -   & -\\
HD 97197  & 0.22 & 0.10 & 0.21 & 0.13 & 0.07 & 0.11 & 0.18  & 0.13 & 0.17  & 0.13 & 0.12 & 0.11 & 0.17 & 0.11 & $\leq$  0.00   & 0.19 & -   & -\\
p04 Leo   & 0.11 & 0.04 & -0.09& 0.08 & -0.18& 0.06 & -0.01 & 0.09 & -0.01 & 0.09 & -0.24& 0.06 & -0.28& 0.05 & $\leq$  0.20   & 0.17 & -   & -\\ \hline

Arcturus  & 0.37 & 0.05 & 0.27 & 0.08 & -0.02& 0.05 & 0.15  & 0.06 & 0.12  & 0.07 & -0.13& 0.05 & -0.14& 0.07 & -  & -& -   & -\\ \hline
\end{tabular}%
}
\end{table*}

\begin{table}[]
\centering
\caption{Sensitive matrix with different elements for the optical and infrared region of the representative star (HD 218330).}
\label{tab:sensitivity}
\resizebox{\columnwidth}{!}{%
\begin{tabular}{ccccc}
\hline
\multicolumn{1}{c}{} & \multicolumn{1}{c}{\begin{tabular}[c]{@{}c@{}}T$_{\rm eff}$\\ (+100 K)\end{tabular}} & \multicolumn{1}{c}{\begin{tabular}[c]{@{}c@{}}log(g)\\ (+0.30 dex)\end{tabular}} & \multicolumn{1}{c}{\begin{tabular}[c]{@{}c@{}}{[}Fe/H{]}\\ (+0.13 dex)\end{tabular}} & \begin{tabular}[c]{@{}c@{}}$\xi_{\rm mic}$\\ (+0.10 km\,s$^{-1}$)\end{tabular} \\ \hline
\multicolumn{5}{c}{Optical} \\ \hline
\multicolumn{1}{l}{A(C)}   &  \multicolumn{1}{c}{+0.01}& \multicolumn{1}{c}{+0.08}   & \multicolumn{1}{c}{+0.09}  & +0.01 \\
\multicolumn{1}{l}{A(N)}   & \multicolumn{1}{c}{+0.05}& \multicolumn{1}{c}{+0.10}   & \multicolumn{1}{c}{+0.09}  & +0.01 \\
\multicolumn{1}{l}{A(O)}   &  \multicolumn{1}{c}{+0.05}& \multicolumn{1}{c}{+0.05}   & \multicolumn{1}{c}{+0.03}  & +0.05 \\
\multicolumn{1}{l}{A(Mg)}  &  \multicolumn{1}{c}{+0.05}   & \multicolumn{1}{c}{+0.00}   & \multicolumn{1}{c}{+0.01}  & -0.02 \\
\multicolumn{1}{l}{A(Si)}  &  \multicolumn{1}{c}{-0.02}   & \multicolumn{1}{c}{+0.07}   & \multicolumn{1}{c}{+0.03}  & -0.02 \\
\multicolumn{1}{l}{A(Ca)} &  \multicolumn{1}{c}{+0.10}& \multicolumn{1}{c}{-0.01}   & \multicolumn{1}{c}{-0.01}  & -0.02 \\
\multicolumn{1}{l}{A(Ti I)} &  \multicolumn{1}{c}{+0.15}& \multicolumn{1}{c}{0.00}   & \multicolumn{1}{c}{-0.01}  & -0.02 \\
\multicolumn{1}{l}{A(Ti II)} & \multicolumn{1}{c}{-0.01}& \multicolumn{1}{c}{+0.13}   & \multicolumn{1}{c}{+0.04}  & -0.04 \\
\multicolumn{1}{l}{A(Ni)}  &\multicolumn{1}{c}{+0.02}& \multicolumn{1}{c}{+0.06}   & \multicolumn{1}{c}{+0.02}  & -0.02 \\
\multicolumn{1}{l}{A(Cr)}  & \multicolumn{1}{c}{+0.10}& \multicolumn{1}{c}{-0.01}   & \multicolumn{1}{c}{-0.01}  & -0.03 \\ 
\multicolumn{1}{l}{A(Y)}  & \multicolumn{1}{c}{+0.05}& \multicolumn{1}{c}{+0.10}   & \multicolumn{1}{c}{+0.06}  & +0.05 \\ 
\multicolumn{1}{l}{A(Fe I)}  & \multicolumn{1}{c}{+0.07}& \multicolumn{1}{c}{+0.03}   & \multicolumn{1}{c}{+0.02}  & -0.04 \\ 
\multicolumn{1}{l}{A(FeII)}  & \multicolumn{1}{c}{-0.08}& \multicolumn{1}{c}{+0.16}   & \multicolumn{1}{c}{+0.05}  & -0.04 \\ \hline
\multicolumn{5}{c}{IR}\\ \hline
\multicolumn{1}{l}{A(C)}   & \multicolumn{1}{c}{+0.15}& \multicolumn{1}{c}{+0.03}   & \multicolumn{1}{c}{+0.03}  & +0.01 \\
\multicolumn{1}{l}{A(N)}   &\multicolumn{1}{c}{+0.06}& \multicolumn{1}{c}{+0.06}   & \multicolumn{1}{c}{+0.09}  & +0.01 \\
\multicolumn{1}{l}{A(O)}   & \multicolumn{1}{c}{+0.10}& \multicolumn{1}{c}{+0.12}   & \multicolumn{1}{c}{+0.03}  & +0.01 \\ \hline
\end{tabular}%
}
\end{table}

\subsection{Asteroseismologic data}
\begin{figure}
   \centering
   \includegraphics[width=0.99 \linewidth]{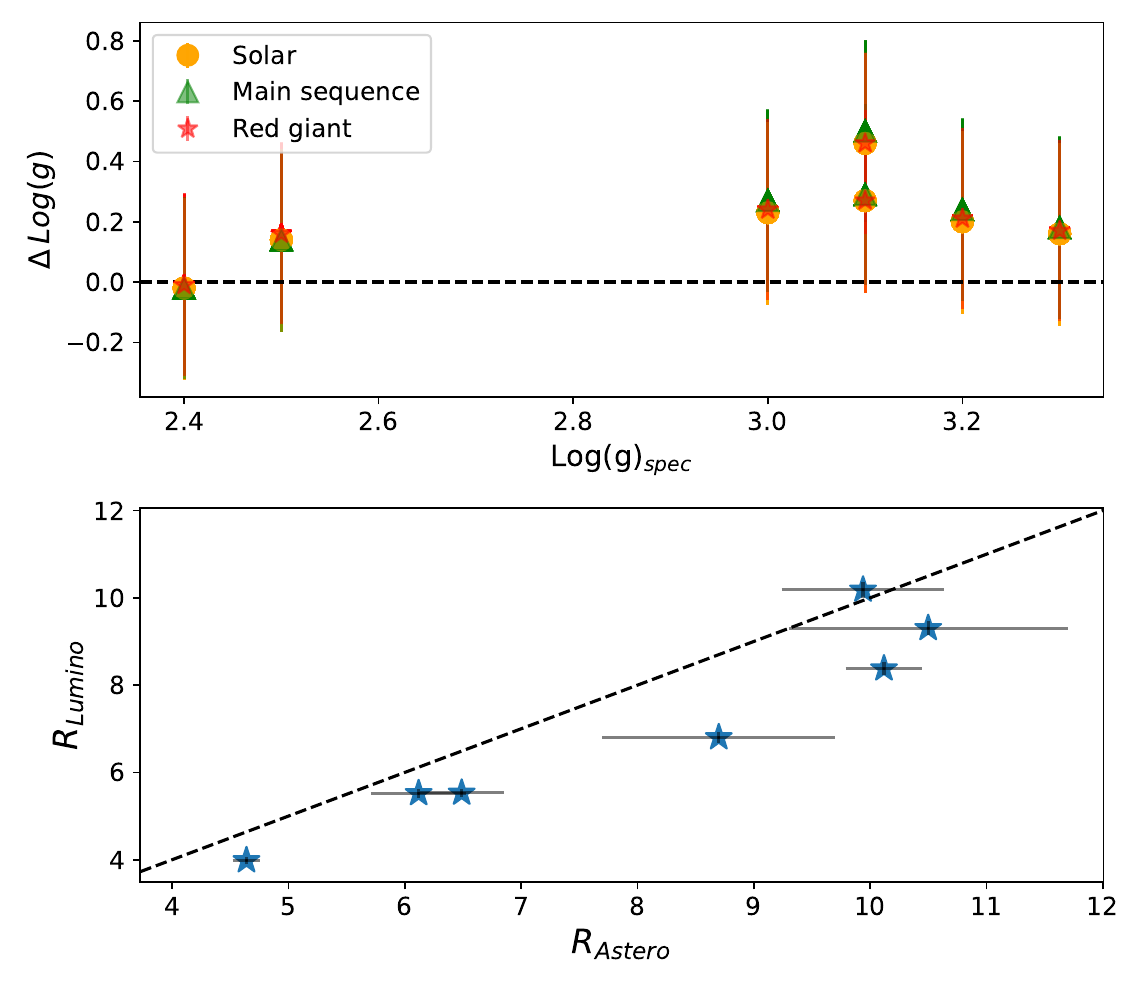}
\caption{\textit{Top: }Comparison of surface gravity obtained from spectroscopy and different asteroseimic scaling relations. Yellow circles represent log(g) from solar-type scaling relation \citep{1995A&A...293...87K}, green triangle from main-sequence scaling relation \citep{2019MNRAS.486.4612B} and red star from red giant scaling relation \citep{2020MNRAS.492L..50B}. \textit{Bottom:} Comparison of stellar radii obtained from asteroseismology and luminosity relation. Dashed Black line represents y=x.  }
   \label{parameters:aster_vs_spec}
\end{figure}

As mentioned in the introduction, the sample examined in this paper has been targeted by K2. \cite{2022MNRAS.511.5578R}
used K2 data and a machine-learning algorithm to extract $\nu_{\rm max}$ and $\Delta\nu$ for seven of the 16 stars in our sample. With these two quantities, we determined three sets of stellar parameters such as mass, radius, and surface gravity using three different scaling relations. Relations calibrated for solar-type stars, main sequence stars and red giant stars were taken from \cite{1995A&A...293...87K}, \cite{2019MNRAS.486.4612B}. and \cite{2020MNRAS.492L..50B}, respectively. For the Sun, we assumed T$_{\rm eff,\,\odot}$ = 5772 K \citep{2016AJ....152...41P}, $\nu_{\rm max,\,\odot}$ = 3090~$\mu$Hz, and $\Delta\nu_{\odot}$ = 135.1~$\mu$Hz \citep{2011ApJ...743..143H}. Table \ref{tab:astero} and Table \ref{tab:astero2} provide the asteroseismic parameters for different relations with the top panel of Fig. \ref{parameters:aster_vs_spec} showing the comparison between the spectroscopic and asteroseismic surface gravities. There is a good agreement between the spectroscopic and asteroseismic values with a mean difference of 0.18 dex. It is however worth noticing how the difference becomes larger with increasing gravity. The trend remains in all three scaling relations as the values are similar to each other. In a quick test of using the asteroseismic log(g) in MOOG, we find the ionic equilibrium worsens but remains with 1$\sigma$, with the effective temperature varying by 25 - 50 K and microturbulance showing no change.

The bottom panel of Fig. \ref{parameters:aster_vs_spec} shows the comparison between the stellar radii obtained from asteroseismology and from luminosity relation. The radii from luminosity relation were calculated using the apparent magnitude in G band and their parallaxes, taken from {\em Gaia}. For determining the bolometric magnitude of the stars, we used a python function \footnote{https://gitlab.oca.eu/ordenovic/gaiadr3{\_}bcg} developed by \cite{2023A&A...674A..26C} to obtain the bolometric correction. This correction is used to calculate the bolometric magnitude and the luminosity of the star which later results in measurement of the stellar radius. The two radii have a mean difference of  0.8 $\pm$ 0.1 R$_{\odot}$. 

Scaling relations in \cite{2020MNRAS.492L..50B} also provides an estimate for the stellar ages, and are reported in Table \ref{tab:astero2}. Along with this, we also derived ages by comparing stellar parameters with stellar evolutionary tracks. Given the mass and metallicity of a star, we obtained the corresponding stellar evolutionary track from MESA Isochrones \& Stellar Tracks (MIST)\footnote{https://waps.cfa.harvard.edu/MIST/index.html} \citep{2016ApJS..222....8D}. We then compared the T$_{\rm eff}$ and log(g) with the theoretical track to obtain the age. The uncertainties were calculated by randomly drawing 10000 values of T$_{\rm eff}$, log(g), mass and metallicity from four different Gaussian distributions (that were pre-defined using the spectroscopic and asteroseismic parameters) and comparing with a grid of MIST evolutionary tracks. Even after this, we were able to improve the uncertainties marginally. The ages obtained from the tracks are consistent with those obtained from scaling relation.

We further used PARAM \citep{2006A&A...458..609D}, a Bayesian estimator, to check the asteroseismic ages, masses and radii. As input, we provided the spectroscopic values for effective temperature, surface gravity and metallicity along with $\nu_{\rm max}$ and $\Delta\nu$ taken from \cite{2022MNRAS.511.5578R}. PARAM performs a comparison between the input observables and a grid of stellar evolutionary tracks through Bayesian analysis and outputs posteriors distributions for different parameters. The results from PARAM were consistent with the asteroseismic values for all seven stars. 

\begin{table*}[]
\centering
\caption{Surface gravity, radius and mass obtained using scaling relations for red giant stars \citep{2020MNRAS.492L..50B}. The asteroseismic parameters ($\nu_{\rm max}$ and $\Delta\nu$) were taken from \cite{2022MNRAS.511.5578R}. Age$_{\rm scaling}$ were obtained from scaling relations whereas Age$_{\rm tracks}$ were obtained by comparing the Masses and Radii (obtained from asteroseismology) with MIST evolutionary tracks.}
\label{tab:astero2}
\resizebox{\textwidth}{!}{%
\begin{tabular}{lcccccccccccc}
\hline
\multicolumn{1}{c}{\multirow{2}{*}{Star}} & \multicolumn{1}{c}{\multirow{2}{*}{$\nu_{\rm max}$}} & \multicolumn{1}{c}{\multirow{2}{*}{$\Delta\nu$}} & \multicolumn{8}{c}{Red giants scaling relation}  & \multicolumn{2}{c}{Evolutionary tracks}\\
\cline{4-13} 
\multicolumn{1}{c}{}   & \multicolumn{1}{c}{} &   & \multicolumn{1}{c}{Mass} & \multicolumn{1}{c}{$\sigma$} & \multicolumn{1}{c}{Radius} & \multicolumn{1}{c}{$\sigma$} & \multicolumn{1}{c}{log(g)} & \multicolumn{1}{c}{$\sigma$} & \multicolumn{1}{c}{Age$_{\rm scaling}$} & \multicolumn{1}{c}{$\sigma$} & \multicolumn{1}{c}{Age$_{\rm tracks}$} & \multicolumn{1}{c}{$\sigma$} \\ \hline
HD 22045  & 123.30 + 7.85   & 10.468 + 0.135  & 1.36   & 0.26   & 6.12   & 0.41   & 2.99  & 0.03  & 3.8  & 2.5  & 3.21 & 2.31\\
HD 76445  & 169.70 + 2.44   & 14.140 + 0.109  & 1.08   & 0.07   & 4.64   & 0.12   & 3.13  & 0.01  & 7.4  & 1.8  & 6.76 & 1.72 \\
HD 99596  & 72.61 + 8.44& 6.776 +0.122& 1.63   & 0.56   & 8.70   & 1.00   & 2.76  & 0.05  & 1.9  & 2.3  & 1.38 & 1.13 \\
HD 100872 & 28.29 + 2.92& 3.868 + 0.118   & 0.90   & 0.29   & 10.5   & 1.20   & 2.34  & 0.04  & 11.0 & 12.0 & 10.34 & 10.16   \\
HD 97716  & 33.16 + 2.26& 4.236 + 0.046   & 0.94   & 0.19   & 9.94   & 0.70   & 2.41  & 0.03  & 14.0 & 10.0 & 12.90 & 9.16 \\
HD 97491  & 83.28 + 3.79& 8.436 + 0.129   & 1.04   & 0.16   & 6.49   & 0.36   & 2.83  & 0.02  & 7.0  & 3.6  & 6.47 & 6.02\\
HD 97197  & 55.47 + 1.31& 5.437 + 0.055   & 1.64   & 0.14   & 10.12  & 0.33   & 2.64  & 0.01  & 2.0  & 0.7  & 1.84 & 0.72\\ \hline
\end{tabular}%
}
\end{table*}

\section{Discussion} \label{dis}

The composition of stellar atmospheres can provide useful information to probe the evolutionary status of our targets.

\subsection{Lithium}

Lithium, a fragile metal, exhibits high sensitivity to temperature variations. It undergoes destruction at low temperatures, approximately 2.5 million K, corresponding to the temperature at the base of the convective region in stars similar to the Sun. Consequently, during stellar evolution, as the outer envelope becomes mixed through convection, the destruction of lithium occurs readily, resulting in diminished lithium abundance in evolved stars. Given that the stars in our sample are evolved, we expect to obtain low lithium abundance, as supported by the values presented in Table \ref{tab:abundance}. A(Li) values below 0.25 dex must be considered as upper limits, because of the small intensity of the \ion{Li}{i}\,$\lambda6708$ line, whose EW is an upper limit. Only two stars exhibit sufficiently strong lines, enabling a reliable estimation of their lithium abundances (HD22045 and HD24680).

The analysis of HD24680 suggests that it is a unique star with an unusually high abundance of lithium, near the threshold for classification as a lithium-rich giant. The high lithium abundance, along with an elevated nitrogen abundance observed, particularly in the infrared region, encouraged us to perform a detailed investigation of its composition, with the aim of understanding the mechanisms behind the formation of such stars. HD24680 is classified as an SB1 binary, as also indicated by the large radial velocity error associated to {\em Gaia} average RV. 

Therefore, one plausible scenario to explain the high lithium and nitrogen abundances is mass transfer from an intermediate-mass asymptotic giant branch (AGB) companion that has now evolved into a white dwarf. In fact, these stars are known to be able to produce Na, Al, and, in a short period of their life, Li, through the Cameron-Fowler mechanism \citep{ventura08}. They also produce s-process elements, in amounts and relative proportions dependent on their mass and metallicity \citep[see e.g.][]{cseh2018}. 
By examining the abundance patterns of elements such as Na, Al, La, Zr, Y, Eu, and Sr (through synthesis), it is possible to gain further insights. 

The analysis reveals an increased abundance in s-process elements (given in Table \ref{tab:AGB}) with [Sr/Fe] = 0.25 $\pm$ 0.16 dex, [Y/Fe] = 0.30 $\pm$ 0.06 dex, [Zr/Fe] = 0.40 $\pm$ 0.10 dex and [La/Fe] = 0.46 $\pm$ 0.14 dex. Even though the uncertainty on Sr is high due to availability of only one line, the increase in the abundances of Y, Zr and La are significant. Along with this, we also observe an over-abundance of Na compared to Al ([Na/Fe] = 0.56 $\pm$ 0.07 dex and [Al/Fe] = 0.05 $\pm$ 0.10 dex) and marginal increase in Eu with [Eu/Fe] = 0.15 $\pm$ 0.07 dex. 
While a detailed analysis of the cause of the observed abundance anomalies in this star is beyond the scope of this paper, we note that the overall pattern supports the idea of the surface composition being the result of mass transfer from an low to intermediate mass AGB star (2-3\,M$_\odot$), capable of producing Na and also s-process elements.

\subsection{Carbon, Nitrogen and Oxygen}
As already introduced in Sect. \ref{sec:CNO}, the analysis of CNO elements in optical and IR spectra reveals a discrepancy between the abundances obtained from the two ranges. It is important to note in this context that the measurements from molecular bands are usually associated with larger uncertainties. This is due to the scatter in line-by-line abundance within the molecular band, but also to the fact that the strength of the molecular lines are strongly sensitive to atmospheric parameters. 
From the sensitivity matrix provided in Table \ref{tab:sensitivity}, it is shown that the errors associated with carbon are approximately 0.16 and 0.15 dex in optical and IR spectra, respectively. Similarly, for nitrogen, the uncertainties are found to be 0.18 and 0.16 dex for optical and IR abundances, respectively. Taking these uncertainties into account, we find that the abundances from the two regions for all three elements are consistent in the majority of the stars (as shown in Fig. \ref{fig:CNO}). 

A few stars show a discrepancy beyond those uncertainties. A total of five stars show discrepancy in nitrogen, whereas two and three stars show discrepancy in carbon and oxygen, respectively. Two reasons come to mind for the high IR abundance, especially for nitrogen: saturation of molecular bands in the optical and offsets introduced in the analysis by placing more weight on the optical abundance of oxygen. In the optical, carbon and nitrogen were analysed using CH and CN molecular bands at 4300 $\AA$ and 4100 $\AA$, respectively. These bands are crowded, blended with several atomic features, and are saturated. This makes it difficult to obtain an accurate determination of the abundances. In the IR region, the CO band is generally not saturated and rather clean, which makes the IR measurement of carbon more accurate and reliable compared to the optical. 
On the other hand, the CN is generally weaker, with some blending. This increases the uncertainties on the IR abundance. As the CNO elements are correlated with each other, a bias in one elements gets propagated to the other two. 
Oxygen abundance was adopted placing more weight on the optical value, which, even if more reliable than the IR one, is still based on a very small number of features, potentially introducing a bias that gets propagated into the abundances of carbon and nitrogen and results in a offset.

\begin{figure}
   \centering
   \includegraphics[width=0.99 \linewidth]{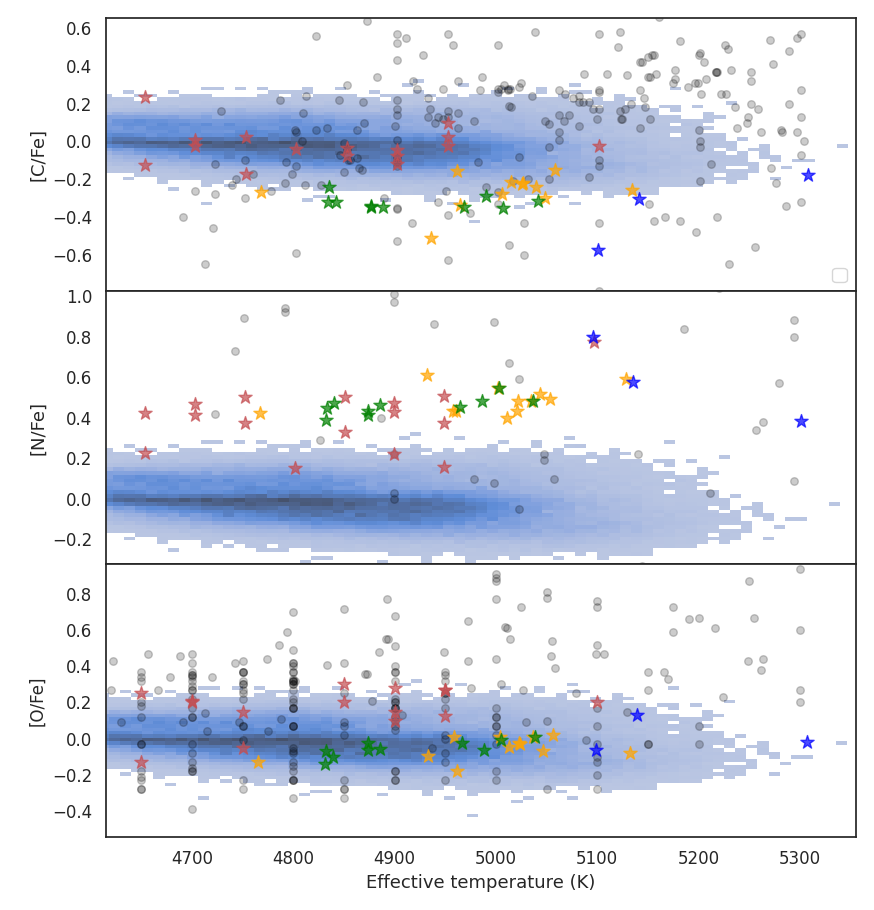}
\caption{Comparison of average CNO abundances obtained in this work (red stars) with other literature such as \cite{topco1} (orange stars), \cite{topco2} (green stars), \cite{afcsar} (blue stars), SAGA database \citep{2008PASJ...60.1159S} (grey circles) and APOGEE \citep{HD218330HD97716}. APOGEE is shows as the density plot in blue. }
   \label{fig:CNO_compa}
\end{figure}

From Table \ref{tab:cno}, considering the average abundances of carbon, nitrogen and oxygen, we find a majority of the stars have a carbon abundance between -0.1 and 0.0 dex, nitrogen abundance between 0.3 and 0.5 dex and oxygen abundance between 0.15 and 0.30 dex. As all the stars in the sample are giants, the carbon abundances are quite close to the expected value of -0.1 dex \citep{2000A&A...354..169G}. On the other side, the nitrogen abundance in majority of the stars is significantly higher than the expected value of 0.1 dex \citep{2000A&A...354..169G}. Two stars, 78 Cnc and HD 97197 show over-abundance in both nitrogen and carbon. In Fig. \ref{fig:CNO_compa}, We compared our results with the ones from \cite{topco1,topco2} and \cite{afcsar}, as these studies use the same atomic and molecular features to derive the CNO abundances in stars with similar parameters as our sample. We also plot the abundances from APOGEE \citep{HD218330HD97716} and SAGA database\footnote{Both APOGEE and SAGA data were cleaned to include stars with parameters similar to stars in our sample.} \citep{2008PASJ...60.1159S} as a density and scatter plot, respectively. We find our carbon and oxygen abundances to be higher than the three studies but within the scatter of APOGEE, whereas, for nitrogen, all three studies are consistent with our results but all of them have a significantly large abundance value relative to the expected values in RGB stars as well as APOGEE. The abundances from SAGA show a large scatter in all three elements which is expected as SAGA contains abundances taken from multiple different studies.

These large abundances could be attributed to an offset in the adopted temperatures or binaries in long-period orbits. If the estimated stellar temperatures had a deviation from the actual value, this can lead to either over- or under-estimation of the abundances. In case of CNO, this effect is amplified due to the usage of molecular band. For example, if the estimated temperature of a star is higher than the actual value, this results in an increase of carbon abundance. The reason being, at higher temperatures, it is harder to form molecules like CH and CN. Therefore, a larger carbon abundance is needed to reproduce the same line strength. Another possibility is that these stars could be part of binary systems with a long period orbit. Due to the large orbital period, it will be hard to identify their binarity. From previous studies, it has been established that the binary fraction of solar-type stars is about 33 $\pm$ 2 $\%$ with period in order of 1000 days \citep{2010ApJS..190....1R}. The stars could have interacted with their companions in the past resulting in accretion of large amounts of carbon or nitrogen.

To check the accuracy of the derived atmospheric parameters and look for offsets in estimated stellar temperatures, we made use of the q2 python package \citep{Ramirez2014} to conduct an independent determination of T$_{\rm eff}$ and log(g) along with a different analysis to estimate the metallicity of the stars. Even though q2 uses the same technique of minimisation of trends in Fe abundances as used in MOOG (see section \ref{stellarparameters}) to estimate T$_{\rm eff}$ and log(g), it performs the analysis in an automated manner and therefore reduces any biases and errors introduced by the user. On comparison with the spectroscopic parameters that were directly derived using MOOG (shown in Fig. \ref{fig:q2}), we found a difference of $\Delta$T$_{\rm eff}$ = 41 $\pm$ 23 K, $\Delta$log(g) = 0.13 $\pm$ 0.09 dex and $\Delta$[Fe/H] = 0.15 $\pm$ 0.08 dex. These differences are not significant given the errors on the spectroscopic parameters. 
As the q2 and MOOG values are consistent within the errors, we conclude that the stellar parameters obtained from MOOG are a good estimate and do not contain any offsets. 

As the q2 and MOOG values are consistent within the errors, we conclude that the stellar parameters obtained from MOOG are a good estimate and do not contain any offsets. 

Along with CNO abundances, we also determined the carbon isotopic ratios ($^{12}$C/$^{13}$C) for all the stars. Carbon isotopic ratio is an important quantity to probe the evolutionary phase of a star. This is because a typical solar neighborhood star is expected to form with a  C content that mostly consists of $^{12}$C with small traces of $^{13}$C, leading to a large value of the isotopic $^{12}$C$/ ^{13}$C ratio. As a star evolves, undergoes the first dredge-up and further mixing processes, the deepening of the convective zone brings $^{13}$C to the surface, lowering the isotopic ratio.

In our sample, we find stars with isotopic ratio ranging between 15 and \textbf{6}. For comparison, Sun has an isotopic ratio of $\sim$ 90 \citep{2013ApJ...765...46A}
and based on this work, Arcturus has a ratio of 6 $\pm$ 3 which is consistent with \cite{2012A&A...548A..55A}. Three of the 16 stars (HD24680, HD 77776 and HD 100872) are close to the equilibrium value, i.e. isotopic ratio equals 3. When a star reaches this equilibrium value, the ratio does not decrease any further as the rate of formation of $^{13}$C is equal to the rate of its destruction. The majority of giant stars have isotopic ratio of $\sim$10-15. These isotopic ratios can be explained by additional mixing in the stars such as thermohaline mixing \citep{2019A&A...621A..24L}. According to \cite{2019A&A...621A..24L}, thermohaline mixing is likely to have a large effect on the surface abundances of the stars used in this work, due to their stellar properties. This possibly also explains the high N abundance and low [C/N] ratio observed in our sample.

\subsection{$\alpha$ and Iron peak elements}
In 14 of the 16 stars, we observe a super solar ratio for the $\alpha$ and Fe-peak elements. 
On the other hand, HD 97716 and p04 Leo show sub-solar ratios, especially in elements like Ca, Ti {\sc ii}, Cr and Ni. 
This, combined with their metallicity of 0.0 and -0.1 dex, indicates that they are likely to be thin disk stars. 

\begin{figure}
   \centering
   \includegraphics[width=0.99 \linewidth]{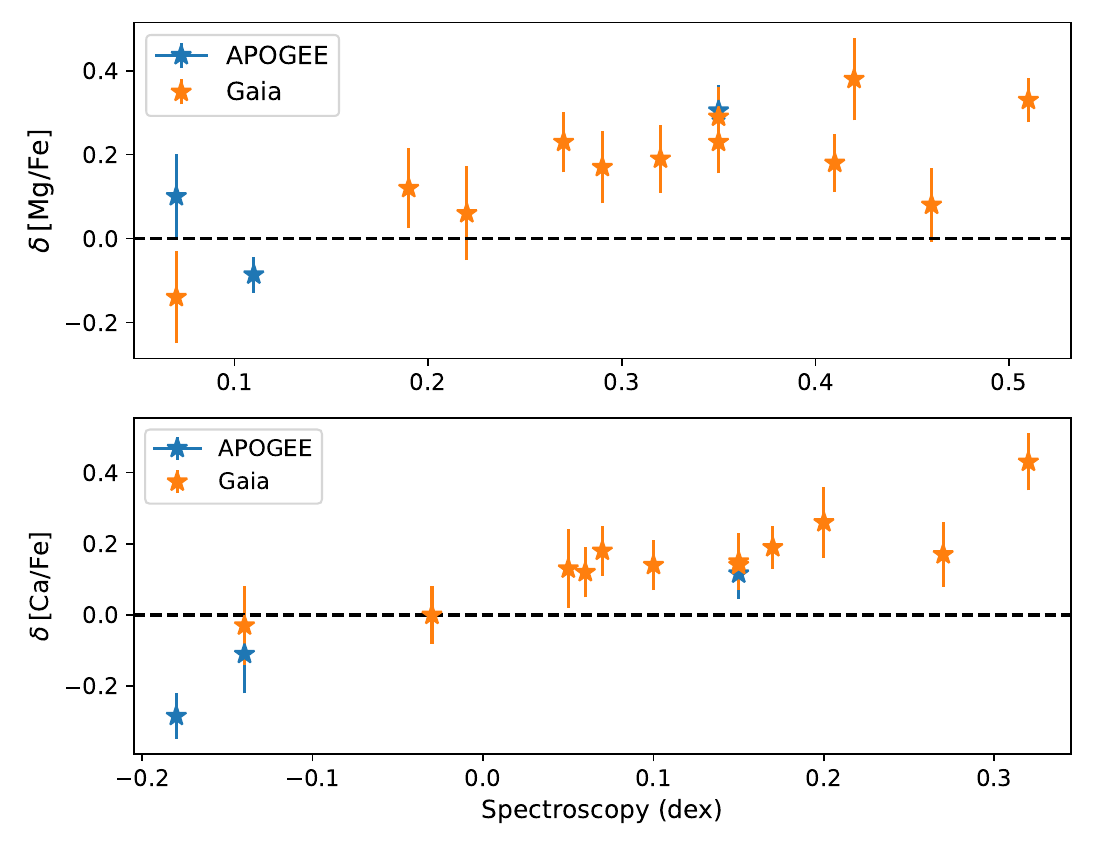}
   \caption{Comparison of Mg and Ca abundances obtained from this work, {\em Gaia} and APOGEE. Errors are propagations of individual uncertainties. }
   \label{fig:MgCa}
\end{figure}

\begin{figure*}
   \centering
   \includegraphics[width=0.99 \linewidth]{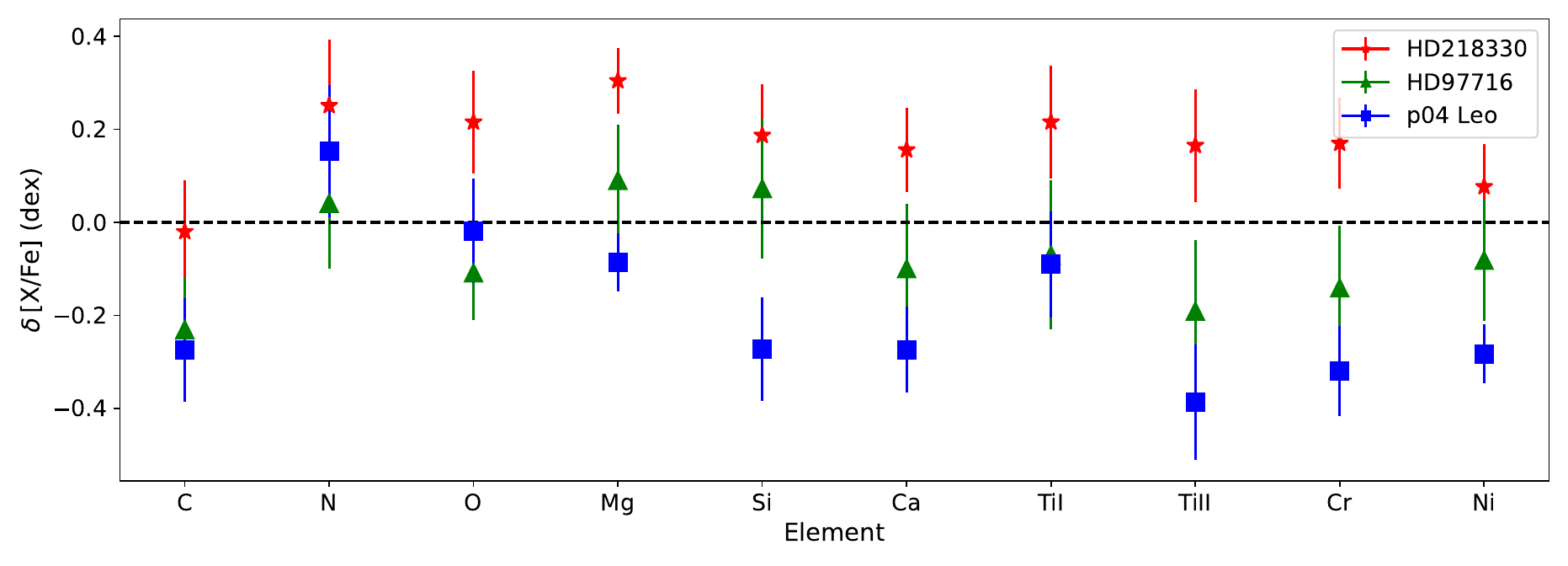}
   \caption{Comparison of abundances of CNO, $\alpha$ and Fe-peak elements obtained from this work and APOGEE.}
   \label{fig:apogee}
\end{figure*}

In Fig. \ref{fig:MgCa} and \ref{fig:apogee}, we make a comparison of abundances obtained in this work with the one from {\em Gaia} and APOGEE. For {\em Gaia}, we use Mg and Ca abundances as they are the most reliable from the RVS spectra whereas, from APOGEE, we obtain CNO, Mg, Si, Ca, Ti {\sc i}, Ti {\sc ii}, Ni and Cr. In Fig. \ref{fig:MgCa}, there is an offset of about 0.2 dex in both Mg and Ca with respect to {\em Gaia}, whereas in APOGEE, we see offsets in all the three stars, which can be attributed to the difference in the metallicity reported in our study and APOGEE. 

\subsection{Fluorine}

Out of the 16 stars, we obtained reliable fluorine measurements only in two stars, i.e. HD 22045 and 78 Cnc with [F/Fe] of 0.35 $\pm$ 0.18 and 0.50 $\pm$ 0.13 dex, respectively. Two main reasons for not having reliable detections in the other 14 stars are: higher temperatures and large telluric contamination. Higher temperatures make it easier to dissociate a molecule, due to which the line strength of the molecules decreases. \cite{Ryde2020} shows a clear signature of destruction of HF molecules in stars with T$_{\rm eff}$ > 4350 K. For stars warmer than 4350 K, they either obtain only upper limits or only measure stars with high fluorine abundance. In our sample, the coolest star has a T$_{\rm eff}$ = 4650 K and therefore, we expect the HF line to be exceedingly weak (which is what we observe). The situation is worsened by the heavy telluric contamination in the region around the HF line.

\begin{figure}
   \centering
   \includegraphics[width=0.99 \linewidth]{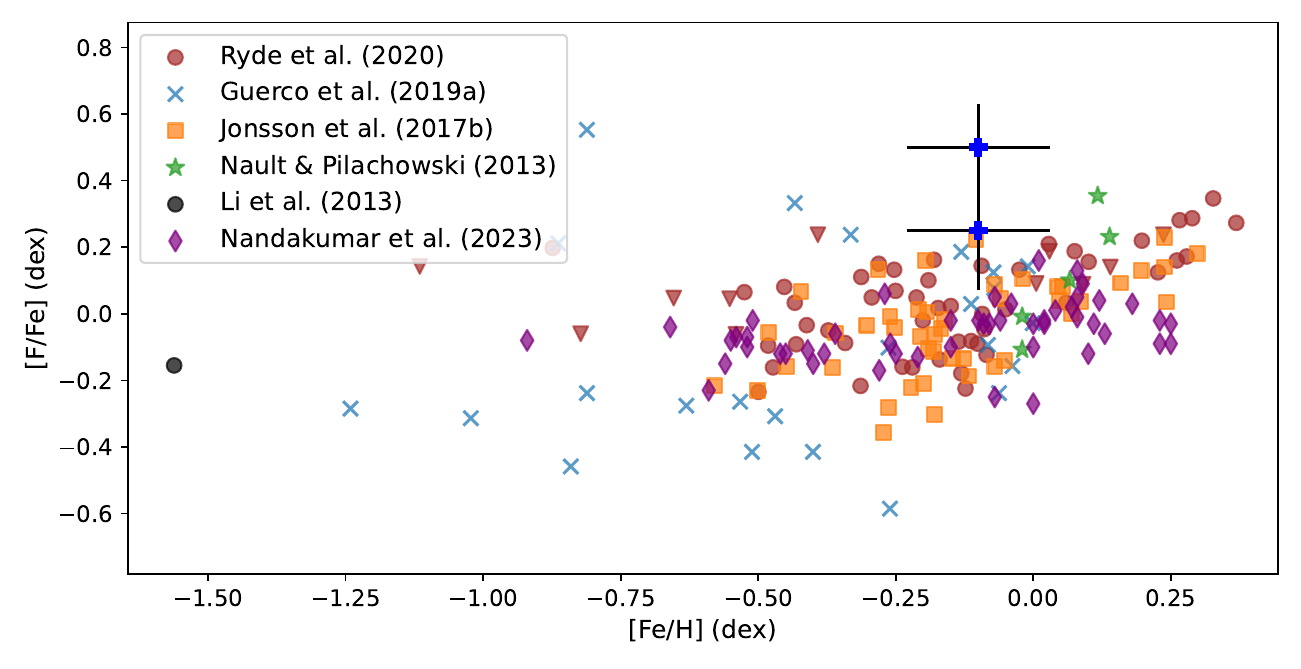}
   \caption{Comparison of Fluorine abundance from this work with literature. Literature abundances are taken from \cite{Nault2013} (green stars), \cite{Li2013} (black cicle), \cite{Jonsson2017b} (orange squares), \cite{Guerco2019} (blue crosses), \cite{Ryde2020} (brown circles and inverted triangles) and \cite{nanda2023} (purple diamonds). Two stars (HD 22045 and 78 Cnc) are shown as blue cross.}
   \label{fig:fluorine}
\end{figure}

To put the fluorine abundances into the context, we compared our results with the literature \citep{Nault2013,Li2013,Jonsson2017b,Guerco2019,Ryde2020,nanda2023} as shown in Fig. \ref{fig:fluorine}. We find HD 22045 to be on the upper end of the scatter whereas 78 Cnc is well above the flat distribution seen for sub-solar metallicity. \cite{abia2015} suggested that, an excess in fluorine abundance ([F/H] > 0.3 dex) was primarily found in stars with C/O close to unity (C/O less than 1.08). This is followed by 78 Cnc where the C/O ratio is equal to 0.98. 

Even though the two abundances do not allow us to obtain any conclusions on the abundance patterns, these measurements are reported here given the limited number of abundance analysis on fluorine. 

\subsection{Comparison between ages}
The stellar parameters obtained from all three scaling relations are consistent with each other. Even though asteroseismic ages obtained are consistent with the ones from theoretical evolutionary tracks, we found the methods to be equally uncertain in providing a precise measure of the age. With the agreement between the two proving the reliability of the theoretical models, we would like to point out that, these two methods are not fully independent of each other. The correlation between the two arises from the use of effective temperature and metallicity in the scaling relations and in determination of ages from evolutionary tracks.

We also tested the reliability of ages obtained from abundance ratios such as [Y/Mg] and [C/N]. The reasoning behind using the first of the two combinations is that magnesium is an $\alpha$-element while yttrium is a slow neutron-capture process element. The main process of Mg formation is the Type~II supernovae, i.e. massive stars (with mass $\ge$ 8 M$_{\odot}$). As massive stars live much shorter lives compared to low-mass stars, the abundance of Mg is expected to increase quickly. On the other side, Y is produced through slow neutron capture during the AGB phase of intermediate-mass stars. This process takes a long time as the evolutionary timescale of intermediate-mass star is much longer than for massive stars. We then expect [Y/Mg] to increase with time, therefore allowing us to probe the stellar ages. 
Carbon and nitrogen are important ingredients of the CNO-cycle. As a star enters the RGB, increasing depth of the convective zone carries a lot of material from the inner regions into the outer atmosphere. This causes variations in chemical abundances of many elements and specifically in the [C/N] ratio. Due to direct correlation between the dept of convective zone with the mass of the star and mass of the star with the age, we can use this ratio to estimate the age. For further details on [Y/Mg] and [C/N] as age estimators, refer to \cite{2022ApJ...936..100B} and \cite{2019A&A...629A..62C}, respectively.  

We derived yttrium abundance through spectral synthesis of 11 lines in the optical region (linelist given in Table \ref{tab:alphaelements}) as reported in Table \ref{tab:YMg}. The ages were estimated using the relations from \cite{2022ApJ...936..100B} for [Y/Mg] and \cite{2019A&A...629A..62C} for [C/N]. The uncertainties on the ages were determined using a similar method as the uncertainty on Age$_{\rm tracks}$, where we randomly draw 10000 abundance values from a Gaussian distribution giving a posterior distribution for the age. From Table \ref{tab:YMg}, we see that ages derived from [Y/Mg] are overestimated when compared to the ages from asteroseismology. For nine of the 16 stars, [Y/Mg] provides an age which is larger than the Hubble time and therefore is physically impossible (in some cases, even after considering the errors). On the other hand, ages from [C/N] are underestimated in comparison with asteroseismology. Most of the stars are reported to be younger than 5 Gys with two stars being reported to be larger than Hubble time (with large errors). Given the disagreement with the asteroseismic ages, which are considered to be quite precise and backed by the ages from evolutionary tracks, we conclude that the abundance-age relations from \cite{2022ApJ...936..100B} and \cite{2019A&A...629A..62C} are not a good estimate for the stars in our sample.

\begin{table*}[]
\centering
\caption{Abundances of Yttrium, Magnesium, Carbon and Nitrogen used to calculate the ages. Stellar ages from [Y/Mg] were calculated using relation from \cite{2022ApJ...936..100B} and ages from [C/N] were calculated using relation from \cite{2019A&A...629A..62C}. Mean abundance of carbon and nitrogen across optical and IR region is used to estimate the ages, except for HD 78419 (no IR spectrum).}
\label{tab:YMg}
\resizebox{\textwidth}{!}{%
\begin{tabular}{lcccccccc}
\hline
\multicolumn{1}{c}{Star} & \begin{tabular}[c]{@{}c@{}}[Mg/Fe]\\ (dex)\end{tabular} & \begin{tabular}[c]{@{}c@{}}[Y/Fe]\\ (dex)\end{tabular} & \begin{tabular}[c]{@{}c@{}}[Y/Mg]\\ (dex)\end{tabular} & \begin{tabular}[c]{@{}c@{}}Age\\ (Gyr)\end{tabular} & \begin{tabular}[c]{@{}c@{}}[C/Fe]\\ (dex)\end{tabular} & \begin{tabular}[c]{@{}c@{}}[N/Fe]\\ (dex)\end{tabular} & \begin{tabular}[c]{@{}c@{}}[C/N]\\ (dex)\end{tabular} & \begin{tabular}[c]{@{}c@{}}Age\\ (Gyr)\end{tabular} \\ \hline
HD 218330 & 0.35±0.06 & 0.05±0.08  & -0.30±0.10  & 14.10±2.93& -0.02±0.09   & 0.42±0.11& -0.44±0.14  & 3.31±5.23 \\
HD 4313 & 0.42±0.09   & -0.09±0.07  & -0.51±0.11  & 20.02±5.11& -0.05±0.09   & 0.43±0.11& -0.48±0.14  & 4.50±7.47 \\
HD 5214 & 0.26±0.06   & 0.05±0.08  & -0.21±0.10   & 12.00±4.76& -0.08±0.09   & 0.48±0.11& -0.56±0.14  & 2.02±3.63 \\
HD 6432 & 0.51±0.04   & -0.01±0.08  & -0.50±0.09   & 20.03±4.46& -0.01±0.09   & 0.47±0.11& -0.48±0.14  & 3.53±5.76 \\
HD 22045& 0.32±0.06   & 0.29±0.11  & -0.03±0.13   & 6.98±5.39 & -0.08±0.09   & 0.50±0.11& -0.58±0.14  & 1.94±3.65 \\
HD 24680& 0.35±0.06   & 0.34±0.06  & 0.01±0.08    & 5.69±4.76 & -0.03±0.09   & 0.78±0.11& -0.81±0.14  & 2.75±8.95 \\
HD 76445& 0.46±0.08   & -0.02±0.12 & -0.48±0.14   & 19.08±5.35&  0.03±0.09   & 0.50±0.11& -0.53±0.14  & 2.45±4.11 \\
HD 77776& 0.29±0.07   & -0.06±0.08 & -0.35±0.11   & 15.37±4.87& -0.18±0.09   & 0.38±0.11& -0.53±0.14  & 2.28±3.43 \\
HD 78419& 0.32±0.11   & 0.11±0.09  & -0.21±0.14   & 11.66±5.69& -0.04±0.16   & 0.30±0.18& -0.34±0.24  & 5.80±9.02 \\
78 Cnc  & 0.41±0.06   & -0.13±0.10  & -0.54±0.12   & 20.85±4.81&  0.24±0.09   & 0.43±0.11& -0.19±0.14  & 13.51±18.12 \\
HD 99596& 0.19±0.07   & -0.19±0.08 & -0.38±0.11   & 16.50±4.70& -0.03±0.09   & 0.38±0.11& -0.41±0.14  & 4.14±7.20 \\
HD 100872 & 0.53±0.04 & -0.28±0.11 & -0.81±0.12   & 28.21±4.72& -0.04±0.09   & 0.33±0.11& -0.37±0.14  & 4.63±6.94 \\
HD 97716&   0.06±0.10 & -0.18±0.10 & -0.24±0.14   & 12.67±5.32& -0.13±0.09   & 0.23±0.11& -0.36±0.14  & 5.69±9.74 \\
HD 97491&   0.48±0.04 & -0.13±0.08 & -0.61±0.09   & 22.72±4.52& 0.10±0.09& 0.16±0.11& -0.06±0.14  &18.89±24.00  \\
HD 97197&   0.22±0.10 & 0.11±0.08  & -0.11±0.13   & 9.18±5.50 & 0.03±0.09& 0.50±0.11& -0.47±0.14  & 2.93±5.69 \\
p04 Leo &   0.11±0.04 & -0.21±0.08 & -0.32±0.09   & 14.98±4.54& -0.12±0.09   & 0.22±0.11& -0.34±0.14  & 6.16±9.42 \\ \hline
\end{tabular}%
}
\end{table*}

\section{Summary and conclusion }

We examined a sample of 16 stars inhabiting the lower RGB and red clump regions. Initially, spectroscopic analysis was conducted to estimate the stellar parameters, followed by an abundance analysis of various elements such as CNO, $\alpha$, Fe-peak elements, Li, Y, and HF. Our focus was on CNO and Li elements to probe the stellar evolutionary phases, while Y was studied as it is one of the elements used in chemical clocks. Additionally, we expanded our carbon analysis to include the isotopic ratio, providing further insights into the stars' evolutionary stages.

All 16 stars were selected from the K2 field as a criterion, allowing us to leverage asteroseismology for additional insights. Asteroseismic parameters, including $\Delta\,\nu$ and $\nu_{\rm max}$, were obtained from \cite{2022MNRAS.511.5578R} for 7 out of the 16 stars. These parameters were then utilized to estimate stellar masses, radii, surface gravities, and ages using scaling relations. Furthermore, ages were also derived from stellar evolutionary tracks and abundance-age relations based on [C/N] and [Y/Mg].

The derived stellar parameters align well with values obtained from {\em Gaia} and previous literature, except for \cite{2020ApJS..247...28H}. These parameters confirm that the nature of these stars falls between the lower RGB and red clump evolutionary phases.

Regarding CNO abundances, we conducted analysis in both the optical and IR regions. Our analysis revealed that IR abundances for all three elements were higher than their optical counterparts, with N showing the largest disparity. Furthermore, the C and N abundances obtained for the stars were significantly higher than the expected values for evolved stars (approximately -0.1 and 0.1 dex, respectively). To rule out the possibility of incorrect stellar parameters causing these discrepancies, we performed a differential analysis using the q2 tool with three reference stars. The results from q2 were consistent with our analysis, ensuring the reliability of our stellar parameters.

In most cases, the Li line was weak, allowing us to determine only an upper limit on the Li abundance. This is consistent with expectations, as lithium is depleted during the ascent of the red giant branch. However, two stars, HD 22045 and HD 24680, exhibited relatively high lithium abundances, with HD 24680 being classified as a Li-rich giant. Considering HD 24680's classification as an SB1 binary, we suspected it to be a post-mass-transfer object, with an AGB star as the donor (which subsequently evolved into a white dwarf). To investigate this scenario, we analyzed elements such as Na, Al, and certain neutron capture elements (Sr, Y, Zr, La, and Eu). The overabundance of elements such as Na, Sr, Y, and La, along with the absence of an increase in Al, supports the hypothesis of mass transfer from a low- to intermediate-mass AGB star.

The derived mass, radii and surface gravities using the measurements from \cite{2022MNRAS.511.5578R} and three different asteroseismic scaling relations agree well with each other. The ages derived from asteroseismology were confirmed by comparing them with ages obtained from stellar evolutionary tracks. 

In recent years, studies such as \cite{2022ApJ...936..100B} and \cite{2019A&A...629A..62C} proposed relations between elemental abundances and stellar age. To test the reliability of these relations, we compared the ages derived from these relations with asteroseismic ages. However, the results either overestimated or underestimated the ages depending on the combination of elemental abundances used. Consequently, we concluded that these relations are not reliable estimators of age within our sample.

In conclusion, 
we found good agreement between the theory and observation, i.e. ages from theoretical tracks and asteroseismology. Along the way, we also identified some peculiarity in the abundances of elements, specifically Nitrogen and Li. We also have obtained additional evidence supporting the mass-transfer mechanism for the formation of Li-rich giants. But additional work, in terms of, constraining the nature of the companion and type of binary interaction, will help in obtaining a better understanding.




\begin{acknowledgements}
We thank Giada Casali and Massimiliano Matteuzzi for useful discussion on asteroseismology.
This work presents results from the European Space Agency (ESA) space mission Gaia. Gaia data are being processed by the Gaia Data Processing and Analysis Consortium (DPAC). Funding for the DPAC is provided by national institutions, in particular the institutions participating in the Gaia MultiLateral Agreement (MLA). The Gaia mission website is \hyperlink{https://www.cosmos.esa.int/gaia}{https://www.cosmos.esa.int/gaia}. The Gaia archive website is \hyperlink{https://archives.esac.esa.int/gaia}{https://archives.esac.esa.int/gaia}. This paper includes data collected by the Kepler mission and obtained from the MAST data archive at the Space Telescope Science Institute (STScI). Funding for the Kepler mission is provided by the NASA Science Mission Directorate. STScI is operated by the Association of Universities for Research in Astronomy, Inc., under NASA contract NAS 5–26555. This research used the facilities of the Italian Center for Astronomical Archive (IA2) operated by INAF at the Astronomical Observatory of Trieste. This work is partially funded by the PRIN INAF 2019 grant ObFu 1.05.01.85.14 (\emph{'Building up the halo: chemo-dynamical tagging in the age of large surveys'}, PI. S. Lucatello) and INAF Mini-Grants 2022 ({\em High resolution spectroscopy of open clusters}, PI Bragaglia).
\end{acknowledgements}

\addcontentsline{toc}{chapter}{Bibliography}
\bibliographystyle{aa}
\bibliography{reference.bib}

\begin{thebibliography}{71}
\expandafter\ifx\csname natexlab\endcsname\relax\def\natexlab#1{#1}\fi

\bibitem[{{Abia} {et~al.}(2015){Abia}, {Cunha}, {Cristallo}, \& {de
  Laverny}}]{abia2015}
{Abia}, C., {Cunha}, K., {Cristallo}, S., \& {de Laverny}, P. 2015, \aap, 581,
  A88

\bibitem[{{Abia} {et~al.}(2021){Abia}, {de Laverny}, {Korotin}, {Asensio
  Ramos}, {Recio-Blanco}, \& {Prantzos}}]{teff8}
{Abia}, C., {de Laverny}, P., {Korotin}, S., {et~al.} 2021, \aap, 648, A107

\bibitem[{{Abia} {et~al.}(2012){Abia}, {Palmerini}, {Busso}, \&
  {Cristallo}}]{2012A&A...548A..55A}
{Abia}, C., {Palmerini}, S., {Busso}, M., \& {Cristallo}, S. 2012, \aap, 548,
  A55

\bibitem[{{Af{\c{s}}ar} {et~al.}(2018){Af{\c{s}}ar}, {Sneden}, {Wood},
  {Lawler}, {Bozkurt}, {B{\"o}cek Topcu}, {Mace}, {Kim}, \& {Jaffe}}]{afcsar}
{Af{\c{s}}ar}, M., {Sneden}, C., {Wood}, M.~P., {et~al.} 2018, \apj, 865, 44

\bibitem[{{Ayres} {et~al.}(2013){Ayres}, {Lyons}, {Ludwig}, {Caffau}, \&
  {Wedemeyer-B{\"o}hm}}]{2013ApJ...765...46A}
{Ayres}, T.~R., {Lyons}, J.~R., {Ludwig}, H.~G., {Caffau}, E., \&
  {Wedemeyer-B{\"o}hm}, S. 2013, \apj, 765, 46

\bibitem[{{Bellinger}(2019)}]{2019MNRAS.486.4612B}
{Bellinger}, E.~P. 2019, \mnras, 486, 4612

\bibitem[{{Bellinger}(2020)}]{2020MNRAS.492L..50B}
{Bellinger}, E.~P. 2020, \mnras, 492, L50

\bibitem[{{Berger} {et~al.}(2022){Berger}, {van Saders}, {Huber}, \&
  {Gaidos}}]{2022ApJ...936..100B}
{Berger}, T.~A., {van Saders}, J.~L., {Huber}, D., \& {Gaidos}, E. e.~a. 2022,
  \apj, 936, 100

\bibitem[{{Blanco-Cuaresma} {et~al.}(2014){Blanco-Cuaresma}, {Soubiran},
  {Heiter}, \& {Jofr{\'e}}}]{2014ascl.soft09006B}
{Blanco-Cuaresma}, S., {Soubiran}, C., {Heiter}, U., \& {Jofr{\'e}}, P. 2014,
  {iSpec: Stellar atmospheric parameters and chemical abundances}, Astrophysics
  Source Code Library, record ascl:1409.006

\bibitem[{{B{\"o}cek Topcu} {et~al.}(2019){B{\"o}cek Topcu}, {Af{\c{s}}ar},
  {Sneden}, {Pilachowski}, {Denissenkov}, {VandenBerg}, {Strickland},
  {{\"O}zdemir}, {Mace}, {Kim}, \& {Jaffe}}]{topco1}
{B{\"o}cek Topcu}, G., {Af{\c{s}}ar}, M., {Sneden}, C., {et~al.} 2019, \mnras,
  485, 4625

\bibitem[{{B{\"o}cek Topcu} {et~al.}(2020){B{\"o}cek Topcu}, {Af{\c{s}}ar},
  {Sneden}, {Pilachowski}, {Denissenkov}, {VandenBerg}, {Wright}, {Mace},
  {Jaffe}, {Strickland}, {Kim}, \& {Sokal}}]{topco2}
{B{\"o}cek Topcu}, G., {Af{\c{s}}ar}, M., {Sneden}, C., {et~al.} 2020, \mnras,
  491, 544

\bibitem[{{Bressan} {et~al.}(2012){Bressan}, {Marigo}, {Girardi}, {Salasnich},
  {Dal Cero}, {Rubele}, \& {Nanni}}]{2012MNRAS.427..127B}
{Bressan}, A., {Marigo}, P., {Girardi}, L., {et~al.} 2012, \mnras, 427, 127

\bibitem[{{Brewer} {et~al.}(2016){Brewer}, {Fischer}, {Valenti}, \&
  {Piskunov}}]{teff3}
{Brewer}, J.~M., {Fischer}, D.~A., {Valenti}, J.~A., \& {Piskunov}, N. 2016,
  \apjs, 225, 32

\bibitem[{{Brown} {et~al.}(1989){Brown}, {Sneden}, {Lambert}, \&
  {Dutchover}}]{1989ApJS...71..293B}
{Brown}, J.~A., {Sneden}, C., {Lambert}, D.~L., \& {Dutchover}, Edward, J.
  1989, \apjs, 71, 293

\bibitem[{{Casali} {et~al.}(2019){Casali}, {Magrini}, {Tognelli}, \&
  {Jackson}}]{2019A&A...629A..62C}
{Casali}, G., {Magrini}, L., {Tognelli}, E., \& {Jackson}, R. e.~a. 2019, \aap,
  629, A62

\bibitem[{{Claudi} {et~al.}(2017){Claudi}, {Benatti}, {Carleo}, {Ghedina},
  {Guerra}, {Micela}, {Molinari}, \& {Oliva}}]{2017EPJP..132..364C}
{Claudi}, R., {Benatti}, S., {Carleo}, I., {et~al.} 2017, European Physical
  Journal Plus, 132, 364

\bibitem[{{Cosentino} {et~al.}(2012){Cosentino}, {Lovis}, {Pepe}, {Collier
  Cameron}, {Latham}, \& {Molinari}}]{2012SPIE.8446E..1VC}
{Cosentino}, R., {Lovis}, C., {Pepe}, F., {et~al.} 2012, in Society of
  Photo-Optical Instrumentation Engineers (SPIE) Conference Series, Vol. 8446,
  Ground-based and Airborne Instrumentation for Astronomy IV, ed. I.~S.
  {McLean}, S.~K. {Ramsay}, \& H.~{Takami}, 84461V

\bibitem[{{Creevey} {et~al.}(2023){Creevey}, {Sordo}, {Pailler}, {Fr{\'e}mat},
  {Heiter}, \& {Th{\'e}venin}}]{2023A&A...674A..26C}
{Creevey}, O.~L., {Sordo}, R., {Pailler}, F., {et~al.} 2023, \aap, 674, A26

\bibitem[{{Cseh} {et~al.}(2018){Cseh}, {Lugaro}, {D'Orazi}, {de Castro},
  {Pereira}, \& et~al.}]{cseh2018}
{Cseh}, B., {Lugaro}, M., {D'Orazi}, V., {et~al.} 2018, \aap, 620, A146

\bibitem[{{Cutri} {et~al.}(2003){Cutri}, {Skrutskie}, {van Dyk}, {Beichman},
  {Carpenter}, \& {Chester}}]{2003yCat.2246....0C}
{Cutri}, R.~M., {Skrutskie}, M.~F., {van Dyk}, S., {et~al.} 2003, VizieR Online
  Data Catalog, II/246

\bibitem[{{da Silva} {et~al.}(2006){da Silva}, {Girardi}, {Pasquini},
  {Setiawan}, \& {von der L{\"u}he}}]{2006A&A...458..609D}
{da Silva}, L., {Girardi}, L., {Pasquini}, L., {Setiawan}, J., \& {von der
  L{\"u}he}, O. e.~a. 2006, \aap, 458, 609

\bibitem[{{Deka-Szymankiewicz} {et~al.}(2018){Deka-Szymankiewicz},
  {Niedzielski}, {Adamczyk}, {Adam{\'o}w}, {Nowak}, \& {Wolszczan}}]{teff6}
{Deka-Szymankiewicz}, B., {Niedzielski}, A., {Adamczyk}, M., {et~al.} 2018,
  \aap, 615, A31

\bibitem[{{D'Orazi} {et~al.}(2017){D'Orazi}, {Desidera}, {Gratton}, {Lanza}, \&
  {Messina}}]{2017A&A...598A..19D}
{D'Orazi}, V., {Desidera}, S., {Gratton}, R.~G., {Lanza}, A.~F., \& {Messina},
  S. e.~a. 2017, \aap, 598, A19

\bibitem[{{D'Orazi} {et~al.}(2015){D'Orazi}, {Gratton}, {Angelou}, {Bragaglia},
  \& {Carretta}}]{2015MNRAS.449.4038D}
{D'Orazi}, V., {Gratton}, R.~G., {Angelou}, G.~C., {Bragaglia}, A., \&
  {Carretta}, E. e.~a. 2015, \mnras, 449, 4038

\bibitem[{{Dotter}(2016)}]{2016ApJS..222....8D}
{Dotter}, A. 2016, \apjs, 222, 8

\bibitem[{{Feuillet} {et~al.}(2016){Feuillet}, {Bovy}, {Holtzman}, {Girardi},
  {MacDonald}, {Majewski}, \& {Nidever}}]{teff1}
{Feuillet}, D.~K., {Bovy}, J., {Holtzman}, J., {et~al.} 2016, \apj, 817, 40

\bibitem[{{Gaia Collaboration}(2022)}]{2022yCat.1357....0G}
{Gaia Collaboration}. 2022, VizieR Online Data Catalog, I/357

\bibitem[{{Gaia Collaboration} {et~al.}(2021){Gaia Collaboration}, {Brown},
  {Vallenari}, {Prusti}, \& {de Bruijne}}]{2021A&A...649A...1G}
{Gaia Collaboration}, {Brown}, A.~G.~A., {Vallenari}, A., {Prusti}, T., \& {de
  Bruijne}, J.~H.~J. e.~a. 2021, \aap, 649, A1

\bibitem[{{Ghezzi} {et~al.}(2018){Ghezzi}, {Montet}, \& {Johnson}}]{teff2}
{Ghezzi}, L., {Montet}, B.~T., \& {Johnson}, J.~A. 2018, \apj, 860, 109

\bibitem[{{Gratton} {et~al.}(2000){Gratton}, {Sneden}, {Carretta}, \&
  {Bragaglia}}]{2000A&A...354..169G}
{Gratton}, R.~G., {Sneden}, C., {Carretta}, E., \& {Bragaglia}, A. 2000, \aap,
  354, 169

\bibitem[{{Guer{\c{c}}o} {et~al.}(2019){Guer{\c{c}}o}, {Cunha}, {Smith},
  {Hayes}, {Abia}, {Lambert}, {J{\"o}nsson}, \& {Ryde}}]{Guerco2019}
{Guer{\c{c}}o}, R., {Cunha}, K., {Smith}, V.~V., {et~al.} 2019, \apj, 885, 139

\bibitem[{{Guer{\c{c}}o} {et~al.}(2022){Guer{\c{c}}o}, {Ram{\'\i}rez}, {Cunha},
  {Smith}, {Prantzos}, {Sellgren}, \& {Daflon}}]{2022ApJ...929...24G}
{Guer{\c{c}}o}, R., {Ram{\'\i}rez}, S., {Cunha}, K., {et~al.} 2022, \apj, 929,
  24

\bibitem[{{Guggenberger} {et~al.}(2016){Guggenberger}, {Hekker}, {Basu}, \&
  {Bellinger}}]{guggen2016}
{Guggenberger}, E., {Hekker}, S., {Basu}, S., \& {Bellinger}, E. 2016, \mnras,
  460, 4277

\bibitem[{{Gullikson} {et~al.}(2014){Gullikson}, {Dodson-Robinson}, \&
  {Kraus}}]{2014AJ....148...53G}
{Gullikson}, K., {Dodson-Robinson}, S., \& {Kraus}, A. 2014, \aj, 148, 53

\bibitem[{{Hardegree-Ullman} {et~al.}(2020){Hardegree-Ullman}, {Zink},
  {Christiansen}, \& {Dressing}}]{2020ApJS..247...28H}
{Hardegree-Ullman}, K.~K., {Zink}, J.~K., {Christiansen}, J.~L., \& {Dressing},
  C. D. e.~a. 2020, \apjs, 247, 28

\bibitem[{{Hekker} \& {Christensen-Dalsgaard}(2017)}]{hekker2016}
{Hekker}, S. \& {Christensen-Dalsgaard}, J. 2017, \aapr, 25, 1

\bibitem[{{Hinkle} {et~al.}(1995){Hinkle}, {Wallace}, \&
  {Livingston}}]{1995PASP..107.1042H}
{Hinkle}, K., {Wallace}, L., \& {Livingston}, W. 1995, \pasp, 107, 1042

\bibitem[{{Hinkle} {et~al.}(2000){Hinkle}, {Wallace}, {Valenti}, \&
  {Harmer}}]{2000vnia.book.....H}
{Hinkle}, K., {Wallace}, L., {Valenti}, J., \& {Harmer}, D. 2000, {Visible and
  Near Infrared Atlas of the Arcturus Spectrum 3727-9300 A}

\bibitem[{{H{\o}g} {et~al.}(2000){H{\o}g}, {Fabricius}, {Makarov}, {Urban},
  {Corbin}, \& {Wycoff}}]{2000A&A...355L..27H}
{H{\o}g}, E., {Fabricius}, C., {Makarov}, V.~V., {et~al.} 2000, \aap, 355, L27

\bibitem[{{Howell} {et~al.}(2014){Howell}, {Sobeck}, {Haas}, {Still},
  {Barclay}, \& {Mullally}}]{2014PASP..126..398H}
{Howell}, S.~B., {Sobeck}, C., {Haas}, M., {et~al.} 2014, \pasp, 126, 398

\bibitem[{{Huber} {et~al.}(2011){Huber}, {Bedding}, {Stello}, {Hekker},
  {Mathur}, {Mosser}, {Verner}, {Bonanno}, \& {Buzasi}}]{2011ApJ...743..143H}
{Huber}, D., {Bedding}, T.~R., {Stello}, D., {et~al.} 2011, \apj, 743, 143

\bibitem[{{Jofr{\'e}} {et~al.}(2015){Jofr{\'e}}, {Petrucci}, {Saffe}, {Saker},
  {Artur de la Villarmois}, {Chavero}, {G{\'o}mez}, \& {Mauas}}]{teff4}
{Jofr{\'e}}, E., {Petrucci}, R., {Saffe}, C., {et~al.} 2015, \aap, 574, A50

\bibitem[{{J{\"o}nsson} {et~al.}(2020){J{\"o}nsson}, {Holtzman}, {Allende
  Prieto}, {Cunha}, {Garc{\'\i}a-Hern{\'a}ndez}, \&
  {Hasselquist}}]{HD218330HD97716}
{J{\"o}nsson}, H., {Holtzman}, J.~A., {Allende Prieto}, C., {et~al.} 2020, \aj,
  160, 120

\bibitem[{{J{\"o}nsson} {et~al.}(2017){J{\"o}nsson}, {Ryde}, {Spitoni},
  {Matteucci}, {Cunha}, {Smith}, {Hinkle}, \& {Schultheis}}]{Jonsson2017b}
{J{\"o}nsson}, H., {Ryde}, N., {Spitoni}, E., {et~al.} 2017, \apj, 835, 50

\bibitem[{{Kjeldsen} \& {Bedding}(1995)}]{1995A&A...293...87K}
{Kjeldsen}, H. \& {Bedding}, T.~R. 1995, \aap, 293, 87

\bibitem[{{Koleva} \& {Vazdekis}(2012)}]{teff5}
{Koleva}, M. \& {Vazdekis}, A. 2012, \aap, 538, A143

\bibitem[{{Kurucz}(1992)}]{kuruz1992}
{Kurucz}, R.~L. 1992, in The Stellar Populations of Galaxies, ed. B.~{Barbuy}
  \& A.~{Renzini}, Vol. 149, 225

\bibitem[{{Lagarde} {et~al.}(2019){Lagarde}, {Reyl{\'e}}, {Robin},
  {Tautvai{\v{s}}ien{\.{e}}}, \& {Drazdauskas}}]{2019A&A...621A..24L}
{Lagarde}, N., {Reyl{\'e}}, C., {Robin}, A.~C., {Tautvai{\v{s}}ien{\.{e}}}, G.,
  \& {Drazdauskas}, A. e.~a. 2019, \aap, 621, A24

\bibitem[{{Li} {et~al.}(2013){Li}, {Ludwig}, {Caffau}, {Christlieb}, \&
  {Zhao}}]{Li2013}
{Li}, H.~N., {Ludwig}, H.~G., {Caffau}, E., {Christlieb}, N., \& {Zhao}, G.
  2013, \apj, 765, 51

\bibitem[{{Lind} {et~al.}(2009){Lind}, {Asplund}, \&
  {Barklem}}]{2009A&A...503..541L}
{Lind}, K., {Asplund}, M., \& {Barklem}, P.~S. 2009, \aap, 503, 541

\bibitem[{{Massarotti} {et~al.}(2008){Massarotti}, {Latham}, {Stefanik}, \&
  {Fogel}}]{teff7}
{Massarotti}, A., {Latham}, D.~W., {Stefanik}, R.~P., \& {Fogel}, J. 2008, \aj,
  135, 209

\bibitem[{{Mucciarelli} \& {Bellazzini}(2020)}]{2020RNAAS...4...52M}
{Mucciarelli}, A. \& {Bellazzini}, M. 2020, Research Notes of the American
  Astronomical Society, 4, 52

\bibitem[{{Nandakumar} {et~al.}(2023){Nandakumar}, {Ryde}, \&
  {Mace}}]{nanda2023}
{Nandakumar}, G., {Ryde}, N., \& {Mace}, G. 2023, \aap, 676, A79

\bibitem[{{Nault} \& {Pilachowski}(2013)}]{Nault2013}
{Nault}, K.~A. \& {Pilachowski}, C.~A. 2013, \aj, 146, 153

\bibitem[{{Oliva} {et~al.}(2012){Oliva}, {Origlia}, {Maiolino}, {Baffa},
  {Biliotti}, {Bruno}, \& {Falcini}}]{2012SPIE.8446E..3TO}
{Oliva}, E., {Origlia}, L., {Maiolino}, R., {et~al.} 2012, in Society of
  Photo-Optical Instrumentation Engineers (SPIE) Conference Series, Vol. 8446,
  Ground-based and Airborne Instrumentation for Astronomy IV, ed. I.~S.
  {McLean}, S.~K. {Ramsay}, \& H.~{Takami}, 84463T

\bibitem[{{Origlia} {et~al.}(2019){Origlia}, {Dalessandro}, {Sanna},
  {Mucciarelli}, {Oliva}, {Cescutti}, {Rainer}, {Bragaglia}, \&
  {Bono}}]{origlia2019}
{Origlia}, L., {Dalessandro}, E., {Sanna}, N., {et~al.} 2019, \aap, 629, A117

\bibitem[{{Origlia} {et~al.}(2014){Origlia}, {Oliva}, {Baffa}, {Falcini},
  {Giani}, {Massi}, \& {Montegriffo}}]{2014SPIE.9147E..1EO}
{Origlia}, L., {Oliva}, E., {Baffa}, C., {et~al.} 2014, in Society of
  Photo-Optical Instrumentation Engineers (SPIE) Conference Series, Vol. 9147,
  Ground-based and Airborne Instrumentation for Astronomy V, ed. S.~K.
  {Ramsay}, I.~S. {McLean}, \& H.~{Takami}, 91471E

\bibitem[{{Placco} {et~al.}(2021){Placco}, {Sneden}, {Roederer}, {Lawler}, {Den
  Hartog}, {Hejazi}, {Maas}, \& {Bernath}}]{2021RNAAS...5...92P}
{Placco}, V.~M., {Sneden}, C., {Roederer}, I.~U., {et~al.} 2021, Research Notes
  of the American Astronomical Society, 5, 92

\bibitem[{{Pr{\v{s}}a} {et~al.}(2016){Pr{\v{s}}a}, {Harmanec}, {Torres},
  {Mamajek}, {Asplund}, \& et~al.}]{2016AJ....152...41P}
{Pr{\v{s}}a}, A., {Harmanec}, P., {Torres}, G., {et~al.} 2016, \aj, 152, 41

\bibitem[{{Raghavan} {et~al.}(2010){Raghavan}, {McAlister}, {Henry}, {Latham},
  \& {Marcy}}]{2010ApJS..190....1R}
{Raghavan}, D., {McAlister}, H.~A., {Henry}, T.~J., {Latham}, D.~W., \&
  {Marcy}, G. W. e.~a. 2010, \apjs, 190, 1

\bibitem[{{Rainer} {et~al.}(2018){Rainer}, {Harutyunyan}, {Carleo}, {Oliva}, \&
  {Benatti}}]{Rainer2018}
{Rainer}, M., {Harutyunyan}, A., {Carleo}, I., {Oliva}, E., \& {Benatti}, S.
  e.~a. 2018, in Society of Photo-Optical Instrumentation Engineers (SPIE)
  Conference Series, Vol. 10702, Ground-based and Airborne Instrumentation for
  Astronomy VII, ed. C.~J. {Evans}, L.~{Simard}, \& H.~{Takami}, 1070266

\bibitem[{{Ram{\'\i}rez} \& {Allende Prieto}(2011)}]{2011ApJ...743..135R}
{Ram{\'\i}rez}, I. \& {Allende Prieto}, C. 2011, \apj, 743, 135

\bibitem[{{Ram{\'\i}rez} {et~al.}(2014){Ram{\'\i}rez}, {Mel{\'e}ndez}, {Bean},
  {Asplund}, {Bedell}, \& {Monroe}}]{Ramirez2014}
{Ram{\'\i}rez}, I., {Mel{\'e}ndez}, J., {Bean}, J., {et~al.} 2014, \aap, 572,
  A48

\bibitem[{{Reyes} {et~al.}(2022){Reyes}, {Stello}, {Hon}, \&
  {Zinn}}]{2022MNRAS.511.5578R}
{Reyes}, C., {Stello}, D., {Hon}, M., \& {Zinn}, J.~C. 2022, \mnras, 511, 5578

\bibitem[{{Ryde} {et~al.}(2020){Ryde}, {J{\"o}nsson}, {Mace}, {Cunha},
  {Spitoni}, {Af{\c{s}}ar}, {Jaffe}, {Forsberg}, {Kaplan}, {Kidder}, {Lee},
  {Oh}, {Smith}, {Sneden}, {Sokal}, {Strickland}, \& {Thorsbro}}]{Ryde2020}
{Ryde}, N., {J{\"o}nsson}, H., {Mace}, G., {et~al.} 2020, \apj, 893, 37

\bibitem[{{Sneden}(1973)}]{MOOG1973}
{Sneden}, C. 1973, \apj, 184, 839

\bibitem[{{Soderblom}(2010)}]{soderblom2010}
{Soderblom}, D.~R. 2010, \araa, 48, 581

\bibitem[{{Sousa} {et~al.}(2007){Sousa}, {Santos}, {Israelian}, {Mayor}, \&
  {Monteiro}}]{ARES2007}
{Sousa}, S.~G., {Santos}, N.~C., {Israelian}, G., {Mayor}, M., \& {Monteiro},
  M.~J.~P.~F.~G. 2007, \aap, 469, 783

\bibitem[{{Suda} {et~al.}(2008){Suda}, {Katsuta}, {Yamada}, {Suwa}, {Ishizuka},
  {Komiya}, {Sorai}, {Aikawa}, \& {Fujimoto}}]{2008PASJ...60.1159S}
{Suda}, T., {Katsuta}, Y., {Yamada}, S., {et~al.} 2008, \pasj, 60, 1159

\bibitem[{{Ting} {et~al.}(2018){Ting}, {Hawkins}, \&
  {Rix}}]{2018ApJ...858L...7T}
{Ting}, Y.-S., {Hawkins}, K., \& {Rix}, H.-W. 2018, \apjl, 858, L7

\bibitem[{{Ventura} \& {D'Antona}(2008)}]{ventura08}
{Ventura}, P. \& {D'Antona}, F. 2008, \aap, 479, 805

\end{thebibliography}

\begin{appendix}

\section{Comparison of stellar parameters}

\begin{figure*}
   \centering
   \includegraphics[width=1 \linewidth]{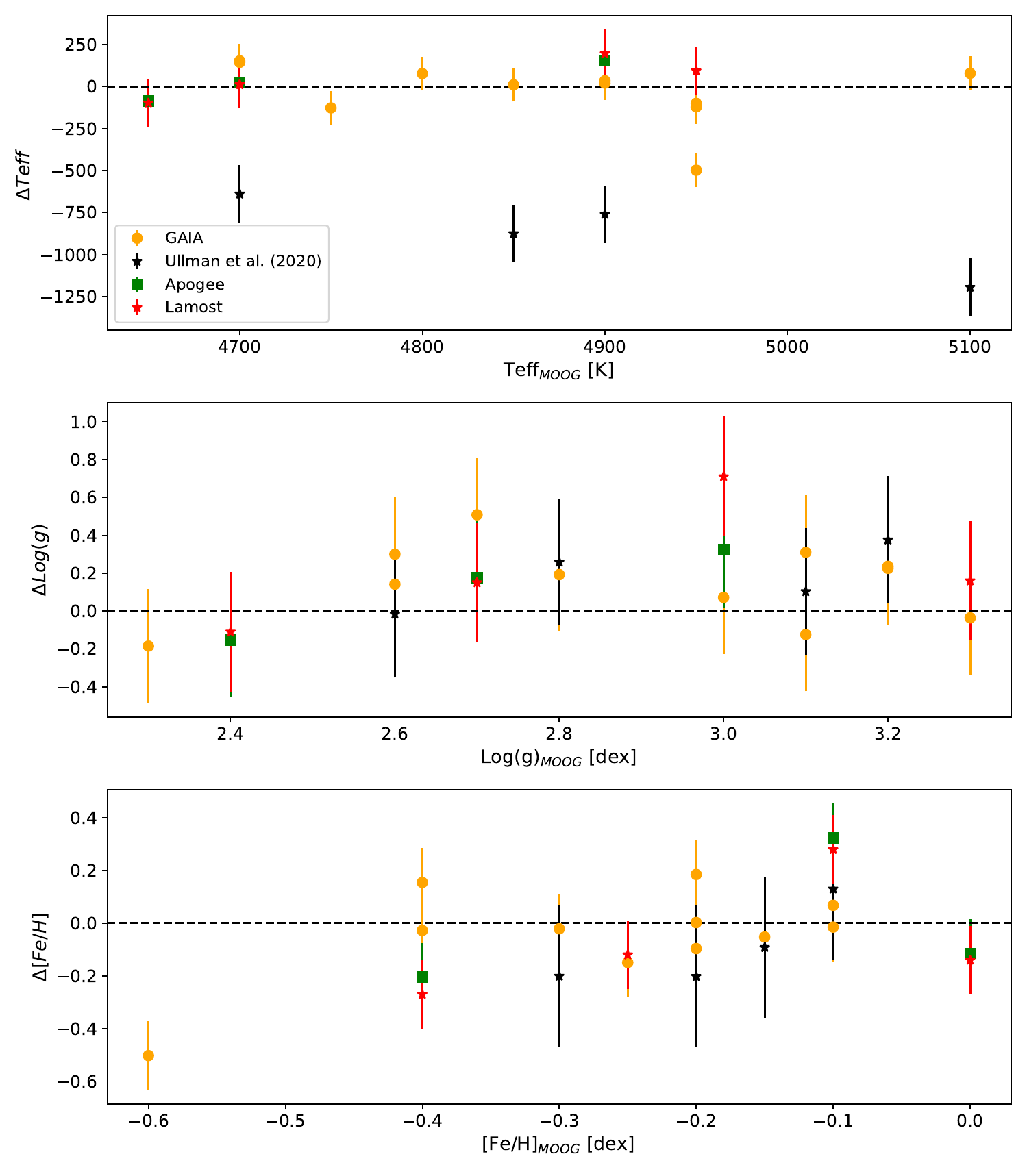}
   \caption{Comparison between stellar parameters obtained from this study and {\em Gaia} \citep{2023A&A...674A..26C}, APOGEE \citep{HD218330HD97716}, Lamost \citep{2018ApJ...858L...7T} and \cite{2020ApJS..247...28H}.}
   \label{fig:MOOGcomparison}
\end{figure*}

\section{Abundances in HD24680}


\begin{table*}[]
\centering
\caption{Elemental abundances of elements affected by mass transfer from an AGB star in HD24680.  }
\label{tab:AGB}
\begin{tabular}{cccccccc}
\hline
Star& {[}Na/Fe{]} & {[}Al/Fe{]} & {[}Sr/Fe{]} & {[}Y/Fe{]} & {[}Zr/Fe{]} & {[}La/Fe{]} & {[}Eu/Fe{]} \\ \hline
HD24680 & 0.56 $\pm$ 0.05 & 0.05 $\pm$ 0.07  & 0.25 $\pm$ 0.16 & 0.3 $\pm$ 0.06 & 0.4 $\pm$ 0.10  & 0.46 $\pm$ 0.14 & 0.15 $\pm$ 0.04\\ \hline
\end{tabular}%
\end{table*}

\section{OH molecular lines in IR region}
\begin{figure*}
   \centering
   \includegraphics[width=1 \linewidth]{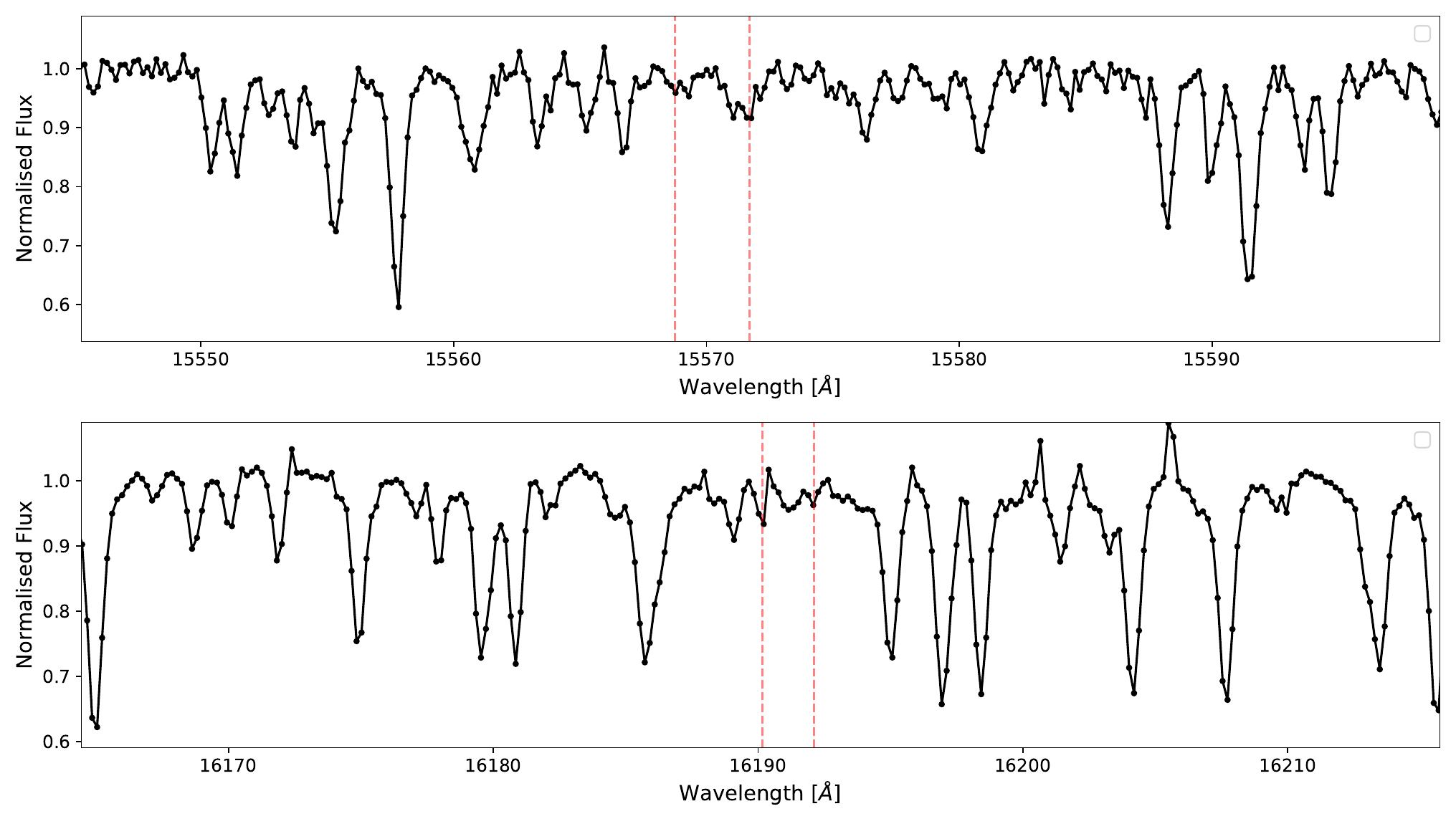}
   \caption{Infrared spectrum of HD 4313 with four OH lines used for analysis marked with red dashed line. This signifies the problem of weak lines and crowding.}
   \label{fig:oxygen_weak}
\end{figure*}

\section{Comparison of CNO abundance}

\begin{figure*}
   \centering
   \includegraphics[width=1 \linewidth]{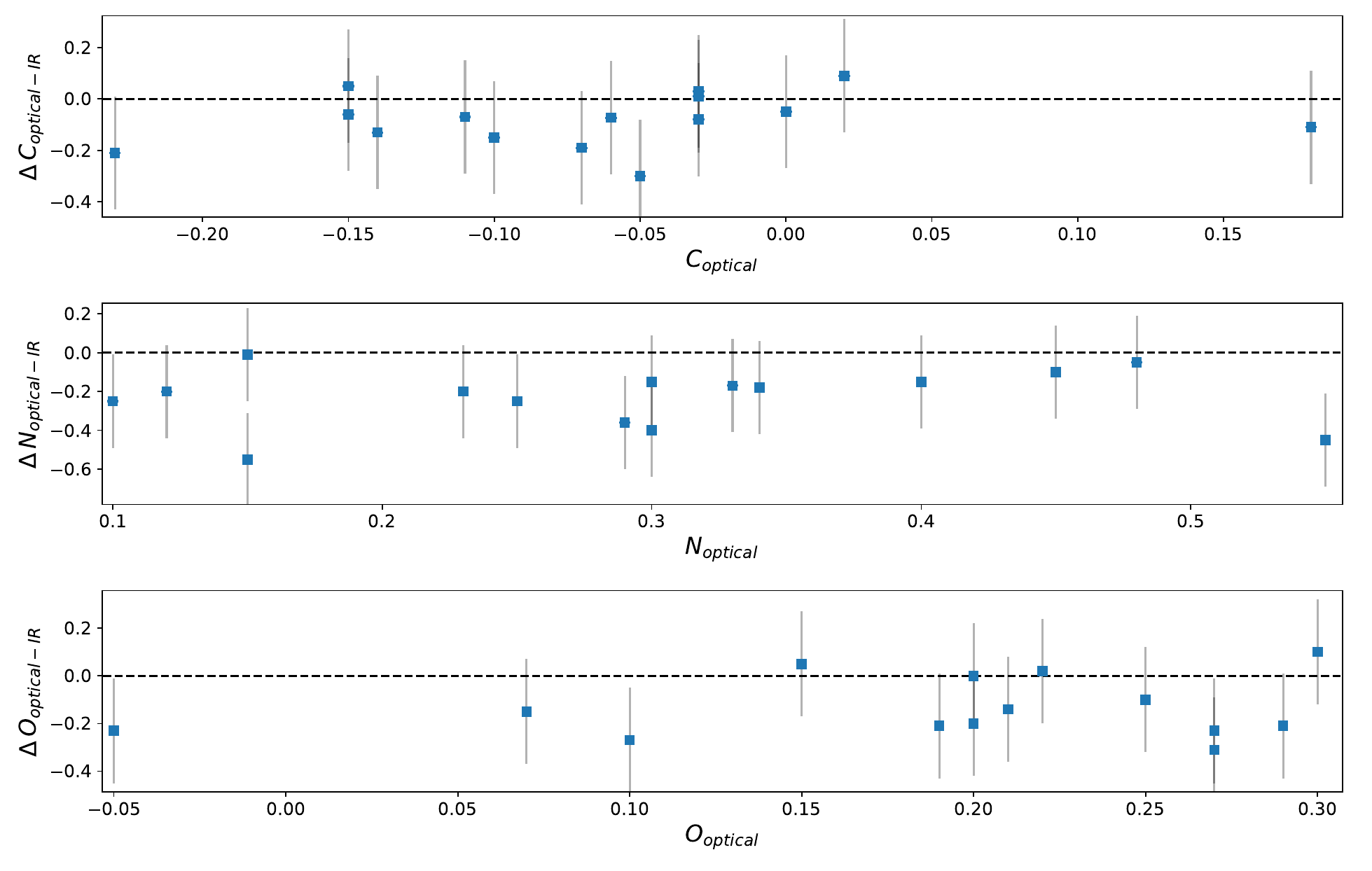}
   \caption{Comparison of abundances of CNO elements obtained from optical and IR spectrum. The y-axis, represents the difference between the two abundances.}
   \label{fig:CNO}
\end{figure*}

\section{Verification of the obtained stellar parameters}

\begin{figure*}
   \centering
   \includegraphics[width=0.5 \paperwidth]{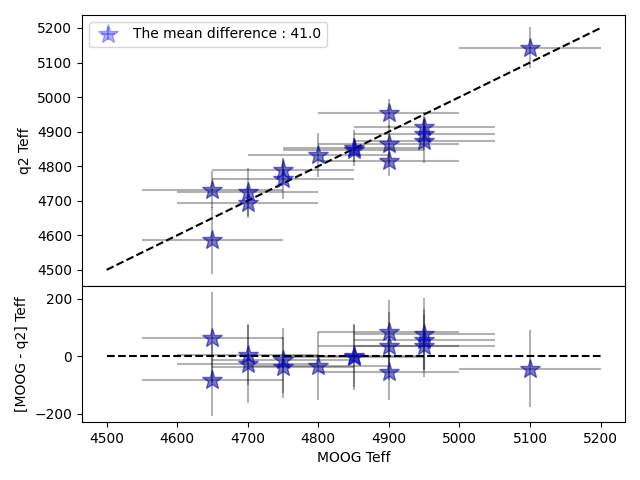}
   \includegraphics[width=0.5 \paperwidth]{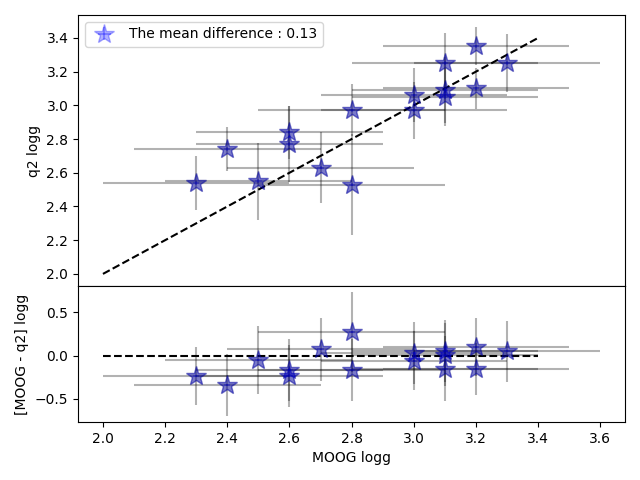}
   \includegraphics[width=0.5 \paperwidth]{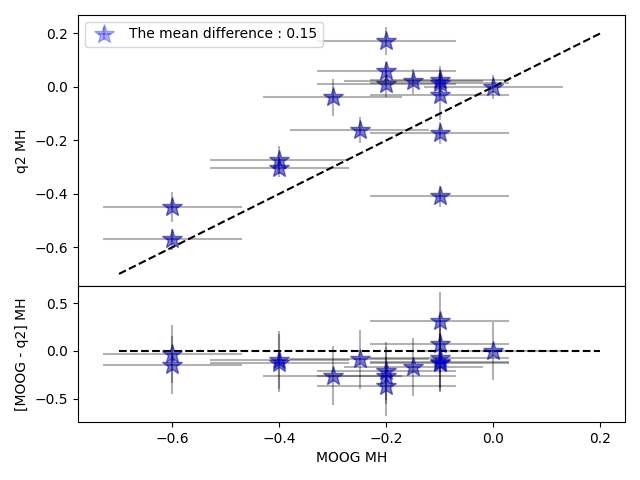}
  \caption{Comparison of stellar parameters obtained from q2 and MOOG. }

   \label{fig:q2}
\end{figure*}

\section{Fitting of molecular bands}

\begin{figure*}
   \centering
   \includegraphics[width=1 \textwidth]{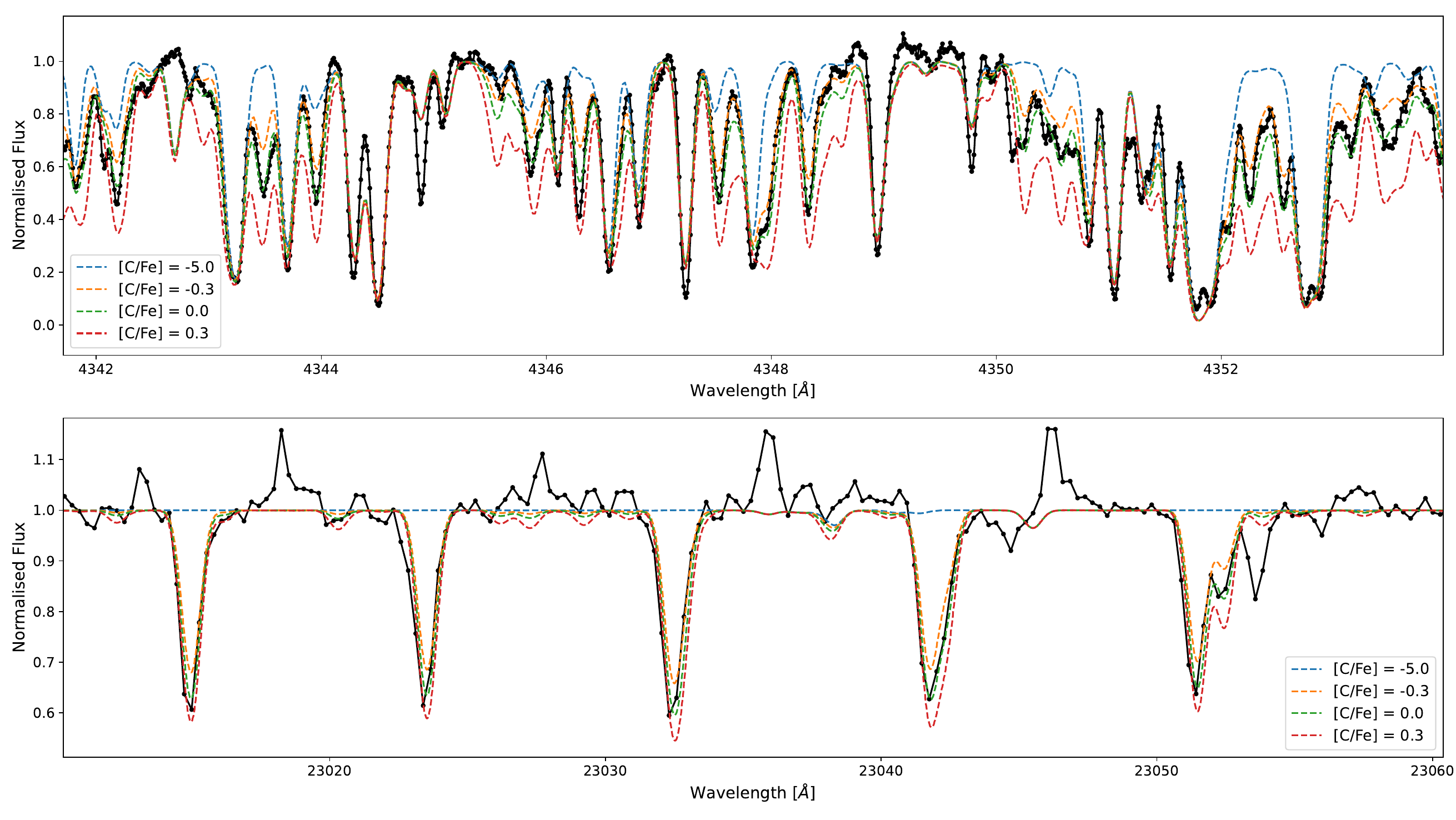}
   \caption{Fitting of the CH- and CO- molecular bands in HD218330. Different synthetic lines represent a different carbon abundance with the blue line ([C/Fe] = -5) showing a spectrum with almost no carbon. Carbon abundance of 0.0 dex is the best fit in both CH- and CO- molecular bands.}
   \label{fig:c}
\end{figure*}

\begin{figure*}
   \centering
   \includegraphics[width=1 \textwidth]{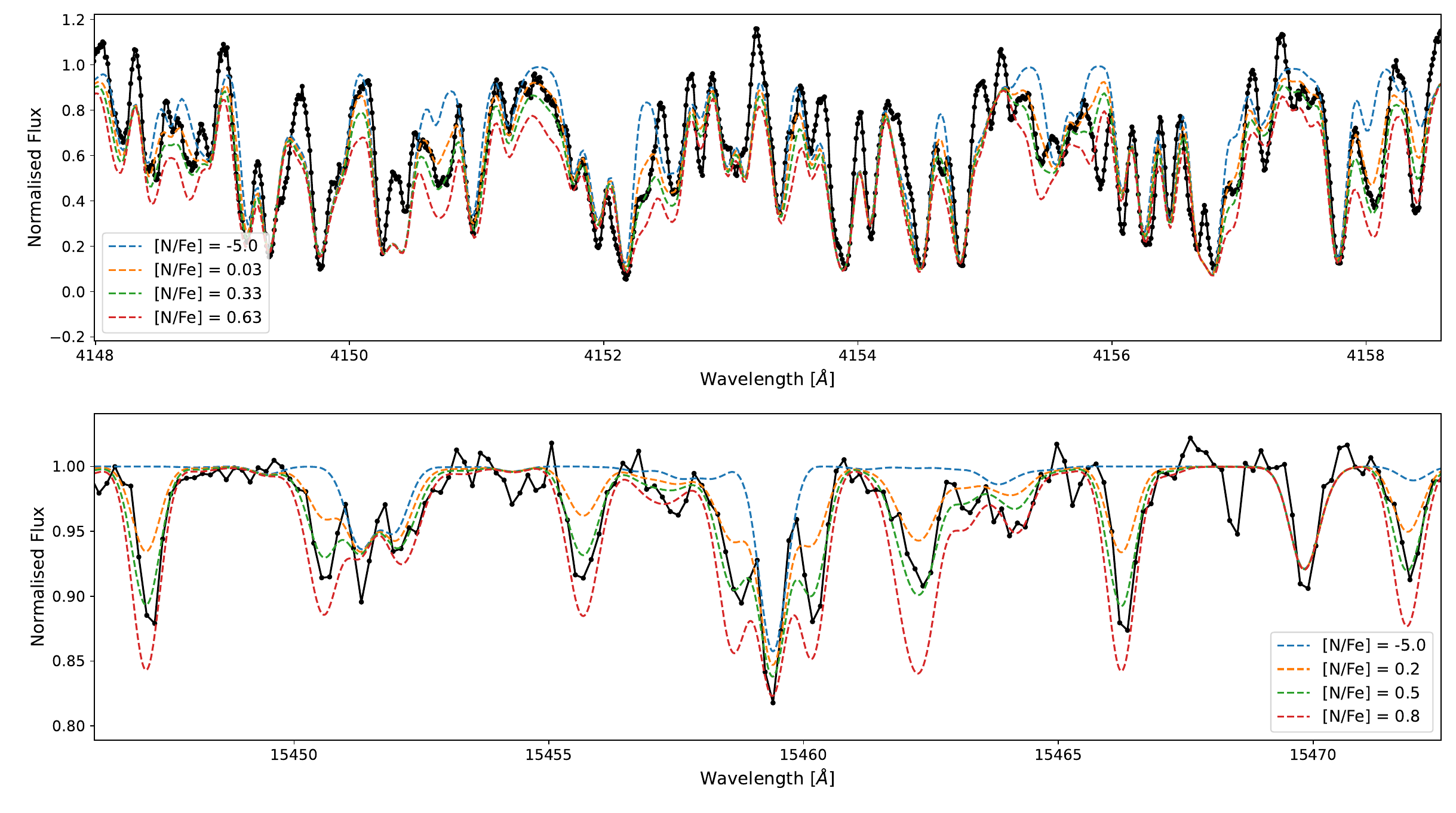}
   \caption{Fitting of the CN- in the optical (top panel) and IR (bottom panel) of HD218330. Different synthetic lines represent a different nitrogen abundance with the blue line ([N/Fe] = -5) showing a spectrum with almost no nitrogen. Nitrogen abundances of 0.33 and 0.5 dex are the best-fit in the optical and IR, respectively.}
   \label{fig:n}
\end{figure*}

\section{Asteroseismic parameters for solar and main sequence scaling relations}
\begin{table*}[]
\centering
\caption{Similar to Table \ref{tab:astero2}. The two scaling relations, i.e. Solar and Main sequence, were taken from \cite{1995A&A...293...87K} and \cite{2019MNRAS.486.4612B}, respectively.}
\label{tab:astero}
\resizebox{\textwidth}{!}{%
\begin{tabular}{lcccccccccccccc}
\hline
\multicolumn{1}{c}{\multirow{2}{*}{Star}} & \multicolumn{1}{c}{\multirow{2}{*}{$\nu_{\rm max}$}} & \multicolumn{1}{c}{\multirow{2}{*}{$\Delta\nu$}} & \multicolumn{6}{c}{Solar scaling relation}  & \multicolumn{6}{c}{Main sequence scaling relation}   \\ \cline{4-15} 
\multicolumn{1}{c}{} & \multicolumn{1}{c}{}  & \multicolumn{1}{c}{}& \multicolumn{1}{c}{Mass} & \multicolumn{1}{c}{$\sigma$} & \multicolumn{1}{c}{Radius} & \multicolumn{1}{c}{$\sigma$} & \multicolumn{1}{c}{log(g)} & \multicolumn{1}{c}{$\sigma$} & \multicolumn{1}{c}{Mass} & \multicolumn{1}{c}{$\sigma$} & \multicolumn{1}{c}{Radius} & \multicolumn{1}{c}{$\sigma$} & \multicolumn{1}{c}{log(g)} & \multicolumn{1}{c}{$\sigma$} \\ \hline
HD 22045  & 123.30 + 7.85   & 10.468 + 0.135  & 1.28  &  0.23   & 5.98   & 0.38&  3.00 &0.03  & 1.34 & 0.09& 6.30   & 0.16& 2.96 & 0.05\\
HD 76445  & 169.70 + 2.44   & 14.140 + 0.109  & 1.02  &  0.04   & 4.53   & 0.06&  3.14 &0.01  & 1.15 & 0.04& 4.89   & 0.06& 3.12 & 0.02\\
HD 99596  & 72.61 + 8.44& 6.776 +0.122& 1.54  &  0.47   & 8.49   & 0.98&  2.77 &0.05 & 1.53 & 0.18& 8.82   & 0.36& 2.73 & 0.05\\
HD 100872 & 28.29 + 2.92& 3.868 + 0.118   & 0.83  &  0.23   & 10.05  & 1.04&  2.36 &0.05 & 1.16 & 0.13& 11.77  & 0.56& 2.36 & 0.03\\
HD 97716  & 33.16 + 2.26& 4.236 + 0.046   & 0.87  &  0.16   & 9.63   & 0.65&  2.42 &0.03 & 1.33 & 0.10& 11.69  & 0.29& 2.42 & 0.02\\
HD 97491  & 83.28 + 3.79& 8.436 + 0.129   & 0.97  &  0.12   & 6.28   & 0.28&  2.83 &0.02 & 1.11 & 0.06& 6.83   & 0.16& 2.81 & 0.02\\
HD 97197  & 55.47 + 1.31& 5.437 + 0.055   & 1.56  &  0.10   & 9.87   & 0.23&  2.64 &0.01 & 1.53 & 0.06& 10.28  & 0.15& 2.60 & 0.03\\ \hline
\end{tabular}%
}
\end{table*}

\section{Telluric modelling}

\begin{figure*}
   \centering
   \includegraphics[width=1 \linewidth]{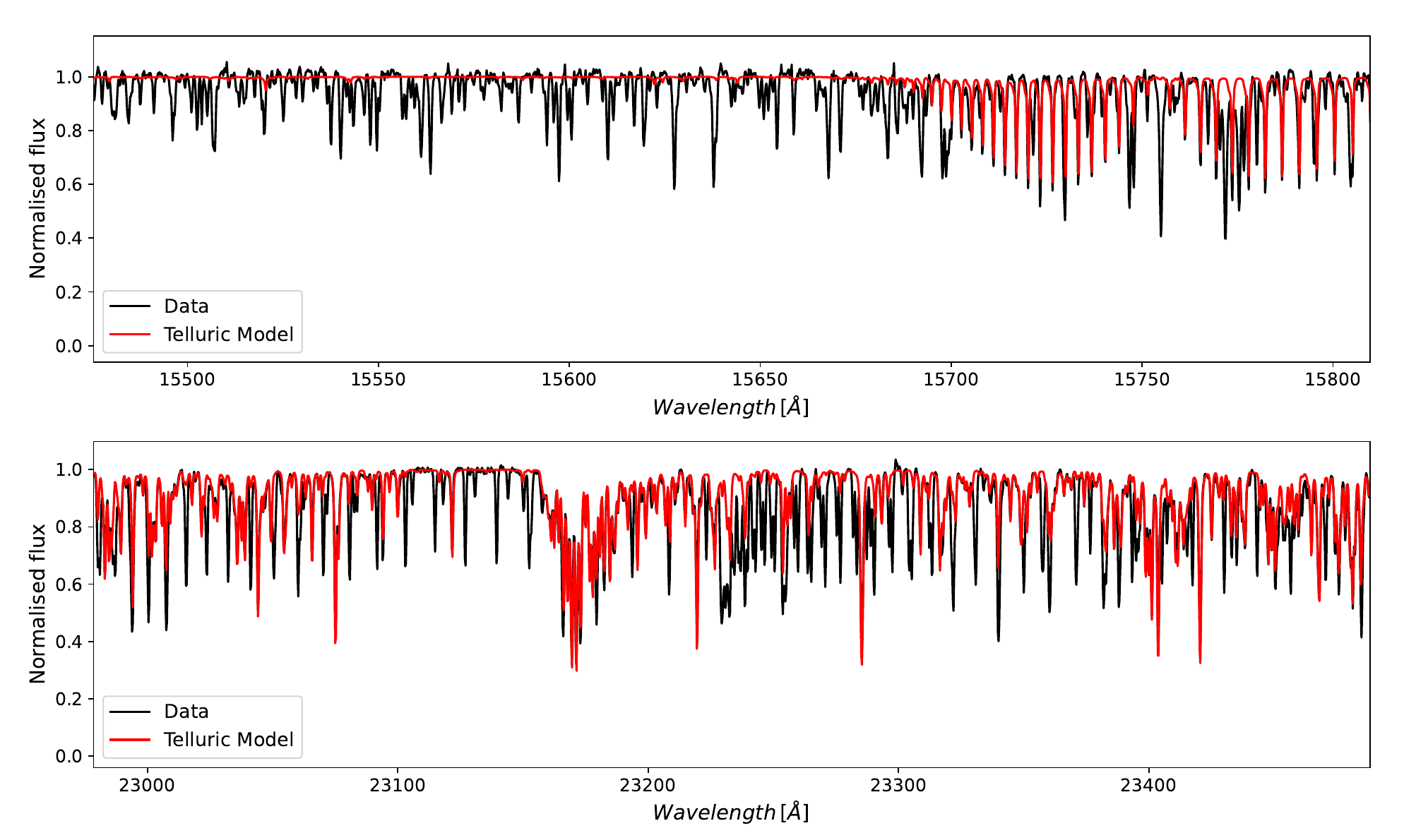}
   \caption{Telluric modelling of IR spectra in HD 5214. }
   \label{fig:telluric}
\end{figure*}

\section{Linelist}

\begin{table*}[]
\centering
\caption{Linelist of FeI and FeII lines used to determine the stellar parameters}
\label{tab:ironelements}
\resizebox{\textwidth}{!}{%
\begin{tabular}{cccccccccc}
\hline
\begin{tabular}[c]{@{}c@{}}Wavelength\\ (\AA)\end{tabular} & \begin{tabular}[c]{@{}c@{}}Excitation\\ potential\\ (eV)\end{tabular} & log(gf) & Element & \begin{tabular}[c]{@{}c@{}}Atomic \\ number\end{tabular} & \begin{tabular} [c]{@{}c@{}}Wavelength\\ (\AA)\end{tabular} & \begin{tabular}[c]{@{}c@{}}Excitation \\ potential\\ (eV)\end{tabular} & log(gf) & Element & \begin{tabular}[c]{@{}c@{}}Atomic\\ number\end{tabular} \\ \hline
4389.25& 0.05   & -4.58   & FeI& 26& 6165.36& 4.14& -1.46   & FeI& 26   \\
4445.47& 0.09   & -5.44   & FeI& 26& 6173.34& 2.22& -2.88   & FeI& 26   \\
4602.01& 1.61   & -3.15   & FeI& 26& 6187.99& 3.94& -1.62   & FeI& 26   \\
4690.14& 3.69   & -1.61   & FeI& 26& 6200.31& 2.61& -2.42   & FeI& 26   \\
4788.76& 3.24   & -1.73   & FeI& 26& 6213.43& 2.22& -2.52   & FeI& 26   \\
4799.41& 3.64   & -2.13   & FeI& 26& 6219.28& 2.21& -2.43   & FeI& 26   \\
4808.15& 3.25   & -2.69   & FeI& 26& 6226.74& 3.88& -2.11   & FeI& 26   \\
4950.11& 3.42   & -1.56   & FeI& 26& 6232.64& 3.65& -1.22   & FeI& 26   \\
4994.13& 0.92   & -3.08   & FeI& 26& 6240.65& 2.22& -3.29   & FeI& 26   \\
5141.74& 2.42   & -2.23   & FeI& 26& 6322.69& 2.59& -2.43   & FeI& 26   \\
5198.71& 2.22   & -2.14   & FeI& 26& 6380.74& 4.19& -1.32   & FeI& 26   \\
5225.53& 0.11   & -4.79   & FeI& 26& 6392.54& 2.28& -4.03   & FeI& 26   \\
5247.05& 0.09   & -4.96   & FeI& 26& 6430.85& 2.18& -2.01   & FeI& 26   \\
5250.21& 0.12   & -4.94   & FeI& 26& 6593.87& 2.43& -2.39   & FeI& 26   \\
5295.31& 4.42   & -1.59   & FeI& 26& 6597.56& 4.81& -0.97   & FeI& 26   \\
5373.71& 4.47   & -0.74   & FeI& 26& 6625.02& 1.01& -5.34   & FeI& 26   \\
5379.57& 3.69   & -1.51   & FeI& 26& 6703.57& 2.76& -3.02   & FeI& 26   \\
5386.33& 4.15   & -1.67   & FeI& 26& 6705.11& 4.61& -0.98   & FeI& 26   \\
5441.34& 4.31   & -1.63   & FeI& 26& 6710.32& 1.49& -4.88   & FeI& 26   \\
5466.41& 4.37   & -0.57   & FeI& 26& 6713.75& 4.81& -1.41   & FeI& 26   \\
5466.99& 3.57   & -2.23   & FeI& 26& 6725.36& 4.11& -2.19   & FeI& 26   \\
5491.83& 4.19   & -2.19   & FeI& 26& 6726.67& 4.61& -1.03   & FeI& 26   \\
5554.89& 4.55   & -0.36   & FeI& 26& 6733.15& 4.64& -1.47   & FeI& 26   \\
5560.21& 4.43   & -1.09   & FeI& 26& 6739.52& 1.56& -4.79   & FeI& 26   \\
5618.63& 4.21   & -1.27   & FeI& 26& 6750.15& 2.42& -2.62   & FeI& 26   \\
5638.26& 4.22   & -0.77   & FeI& 26& 6793.26& 4.08& -2.33   & FeI& 26   \\
5651.47& 4.47   & -1.75   & FeI& 26& 6806.85& 2.73& -3.11   & FeI& 26   \\
5679.02& 4.65   & -0.75   & FeI& 26& 6810.26& 4.61& -0.99   & FeI& 26   \\
5705.46& 4.31   & -1.36   & FeI& 26& 6837.01& 4.59& -1.69   & FeI& 26   \\
5731.76& 4.26   & -1.21   & FeI& 26& 6839.83& 2.56& -3.35   & FeI& 26   \\
5775.08& 4.22   & -1.31   & FeI& 26& 6843.66& 4.55& -0.83   & FeI& 26   \\
5778.45& 2.59   & -3.44   & FeI& 26& 4491.41& 2.86& -2.66   & FeII& 26.1 \\
5784.66& 3.41   & -2.53   & FeI& 26& 4508.29& 2.86& -2.52   & FeII& 26.1 \\
5793.91& 4.22   & -1.62   & FeI& 26& 4576.33& 2.84& -2.95   & FeII& 26.1 \\
5806.73& 4.61   & -0.95   & FeI& 26& 4620.51& 2.83& -3.21   & FeII& 26.1 \\
5852.22& 4.55   & -1.23   & FeI& 26& 4993.34& 2.81& -3.73   & FeII& 26.1 \\
5855.08& 4.61   & -1.48   & FeI& 26& 5197.58& 3.23& -2.22   & FeII& 26.1 \\
5956.69& 0.86   & -4.55   & FeI& 26& 5234.62& 3.22& -2.18   & FeII& 26.1 \\
5987.07& 4.81   & -0.21   & FeI& 26& 5264.81& 3.23& -3.13   & FeII& 26.1 \\
6005.54& 2.59   & -3.43   & FeI& 26& 5414.07& 3.22& -3.58   & FeII& 26.1 \\
6056.01& 4.73   & -0.41   & FeI& 26& 6084.09& 3.21& -3.83   & FeII& 26.1 \\
6065.48& 2.61   & -1.53   & FeI& 26& 6149.24& 3.89& -2.75   & FeII& 26.1 \\
6079.01& 4.65   & -1.02   & FeI& 26& 6247.55& 3.89& -2.38   & FeII& 26.1 \\
6082.71& 2.22   & -3.57   & FeI& 26& 6369.46& 2.89& -4.11   & FeII& 26.1 \\
6093.64& 4.61   & -1.31   & FeI& 26& 6416.92& 3.89& -2.75   & FeII& 26.1 \\
6096.67& 3.98   & -1.81   & FeI& 26& 6432.68& 2.89& -3.57   & FeII& 26.1 \\
6151.62& 2.18   & -3.28   & FeI& 26& 6456.38& 3.91& -2.05   & FeII& 26.1 \\ \hline
\end{tabular}%
}
\end{table*}

\begin{table*}[]
\centering
\caption{Linelist of TiI and TiII lines used to determine the stellar parameters}
\label{tab:Tielements}
\resizebox{\textwidth}{!}{%
\begin{tabular}{cccccccccc}
\hline
\begin{tabular}[c]{@{}c@{}}Wavelength\\ (\AA)\end{tabular} & \begin{tabular}[c]{@{}c@{}}Excitation\\ potential\\ (eV)\end{tabular} & log(gf) & Element & \begin{tabular}[c]{@{}c@{}}Atomic \\ number\end{tabular} & \begin{tabular} [c]{@{}c@{}}Wavelength\\ (\AA)\end{tabular} & \begin{tabular}[c]{@{}c@{}}Excitation \\ potential\\ (eV)\end{tabular} & log(gf) & Element & \begin{tabular}[c]{@{}c@{}}Atomic\\ number\end{tabular} \\ \hline
4186.12 & 1.5 & -0.24 & TiI & 22 & 5514.34 & 1.4 & -0.66 & TiI & 22   \\
4287.40 & 0.8 & -0.37 & TiI & 22 & 5514.53 & 1.4 & -0.50 & TiI & 22   \\
4427.10 & 1.5 & 0.23  & TiI & 22 & 5565.47 & 2.2 & -0.22 & TiI & 22   \\
4453.31 & 1.4 & -0.03 & TiI & 22 & 5739.98 & 2.2 & -0.92 & TiI & 22   \\
4453.70 & 1.8 & 0.10  & TiI & 22 & 5866.45 & 1.0 & -0.79 & TiI & 22   \\
4471.24 & 1.7 & -0.15 & TiI & 22 & 5880.27 & 1.0 & -2.00 & TiI & 22   \\
4518.02 & 0.8 & -0.25 & TiI & 22 & 5922.11 & 1.0 & -1.38 & TiI & 22   \\
4548.76 & 0.8 & -0.28 & TiI & 22 & 5937.81 & 1.0 & -1.94 & TiI & 22   \\
4623.10 & 1.7 & 0.16  & TiI & 22 & 6258.10 & 1.4 & -0.39 & TiI & 22   \\
4722.61 & 1.0 & -1.47 & TiI & 22 & 6261.10 & 1.4 & -0.53 & TiI & 22   \\
4758.90 & 0.8 & -2.17 & TiI & 22 & 6303.76 & 1.4 & -1.58 & TiI & 22   \\
4778.25 & 2.2 & -0.35 & TiI & 22 & 6312.24 & 1.4 & -1.55 & TiI & 22   \\
4781.71 & 0.8 & -1.95 & TiI & 22 & 6554.22 & 1.4 & -1.15 & TiI & 22   \\
4797.98 & 2.3 & -0.63 & TiI & 22 & 4316.79 & 2.0 & -1.62 & TiII & 22.1 \\
4805.41 & 2.3 & 0.07  & TiI & 22 & 4320.95 & 1.1 & -1.88 & TiII & 22.1 \\
4820.41 & 1.5 & -0.38 & TiI & 22 & 4395.83 & 1.2 & -1.93 & TiII & 22.1 \\
4840.87 & 0.8 & -0.43 & TiI & 22 & 4443.80 & 1.0 & -0.71 & TiII & 22.1 \\
4870.12 & 2.2 & 0.44  & TiI & 22 & 4468.49 & 1.1 & -0.63 & TiII & 22.1 \\
4885.08 & 1.8 & 0.41  & TiI & 22 & 4493.52 & 1.0 & -2.78 & TiII & 22.1 \\
4899.91 & 1.8 & 0.31  & TiI & 22 & 4518.33 & 1.0 & -2.56 & TiII & 22.1 \\
4937.73 & 0.8 & -2.08 & TiI & 22 & 4571.97 & 1.5 & -0.31 & TiII & 22.1 \\
4995.07 & 2.2 & -1.00 & TiI & 22 & 4583.40 & 1.1 & -2.84 & TiII & 22.1 \\
5016.16 & 0.8 & -0.48 & TiI & 22 & 4609.26 & 1.1 & -3.32 & TiII & 22.1 \\
5020.03 & 0.8 & -0.33 & TiI & 22 & 4657.20 & 1.2 & -2.29 & TiII & 22.1 \\
5036.46 & 1.4 & 0.14  & TiI & 22 & 4708.66 & 1.2 & -2.35 & TiII & 22.1 \\
5038.40 & 1.4 & 0.02  & TiI & 22 & 4764.52 & 1.2 & -2.69 & TiII & 22.1 \\
5040.61 & 0.8 & -1.67 & TiI & 22 & 4798.53 & 1.0 & -2.66 & TiII & 22.1 \\
5043.58 & 0.8 & -1.59 & TiI & 22 & 4865.61 & 1.1 & -2.70 & TiII & 22.1 \\
5062.10 & 2.1 & -0.39 & TiI & 22 & 4874.00 & 3.0 & -0.86 & TiII & 22.1 \\
5064.65 & 0.0 & -0.94 & TiI & 22 & 4911.19 & 3.1 & -0.64 & TiII & 22.1 \\
5087.06 & 1.4 & -0.88 & TiI & 22 & 5069.09 & 3.1 & -1.62 & TiII & 22.1 \\
5145.46 & 1.4 & -0.54 & TiI & 22 & 5185.90 & 1.8 & -1.41 & TiII & 22.1 \\
5192.97 & 0.0 & -0.95 & TiI & 22 & 5211.53 & 2.5 & -1.41 & TiII & 22.1 \\
5210.38 & 0.0 & -0.82 & TiI & 22 & 5336.78 & 1.5 & -1.60 & TiII & 22.1 \\
5219.70 & 0.0 & -2.22 & TiI & 22 & 5381.02 & 1.5 & -1.97 & TiII & 22.1 \\
5295.78 & 1.0 & -1.59 & TiI & 22 & 5396.24 & 1.5 & -3.18 & TiII & 22.1 \\
5389.17 & 0.8 & -2.35 & TiI & 22 & 5418.76 & 1.5 & -2.13 & TiII & 22.1 \\
5471.19 & 1.4 & -1.42 & TiI & 22 & 6680.13 & 3.0 & -1.89 & TiII & 22.1\\ \hline
\end{tabular}%
}
\end{table*}

\begin{table*}[]
\centering
\caption{Linelist used for abundance analysis of $\alpha$-, Fe-peak elements, Lithium and Yttrium.}
\label{tab:alphaelements}
\resizebox{\textwidth}{!}{%
\begin{tabular}{cccccccccc}
\hline
\begin{tabular}[c]{@{}c@{}}Wavelength\\ (\AA)\end{tabular} & \begin{tabular}[c]{@{}c@{}}Excitation\\ potential\\ (eV)\end{tabular} & log(gf) & Element & \begin{tabular}[c]{@{}c@{}}Atomic\\ number\end{tabular} & \begin{tabular}[c]{@{}c@{}}Wavelength\\ (\AA)\end{tabular} & \begin{tabular}[c]{@{}c@{}}Excitation\\ potential\\ (eV)\end{tabular} & log(gf) & Element & \begin{tabular}[c]{@{}c@{}}Atomic\\ number\end{tabular} \\ \hline
6707.77& 0.00 & 0.00& LiI & 3 & 5219.70   & 0.02  & -2.24   & TiI & 22\\
4730.04 & 4.34 & -2.39   & MgI & 12& 5295.78   & 1.07  & -1.58   & TiI & 22\\
5711.09 & 4.34 & -1.73   & MgI & 12& 5389.17   & 0.80  & -2.35   & TiI & 22\\
6318.72 & 5.11 & -1.95   & MgI & 12& 5471.19   & 1.40  & -1.42   & TiI & 22\\
6319.24 & 5.11 & -2.32   & MgI & 12& 5503.90& 2.50  & -0.05   & TiI & 22\\
7657.61 & 5.11 & -1.28   & MgI & 12& 5514.34   & 1.40  & -0.66   & TiI & 22\\
5684.48 & 4.95 & -1.55   & SiI & 14& 5514.53   & 1.40  & -0.50   & TiI & 22\\
5690.42 & 4.93 & -1.77   & SiI & 14& 5565.47   & 2.20  & -0.22   & TiI & 22\\
5701.10 & 4.93 & -1.95   & SiI & 14& 5739.98   & 2.20  & -0.92   & TiI & 22\\
5772.15 & 5.08 & -1.65   & SiI & 14& 5866.45   & 1.00  & -0.79   & TiI & 22\\
5793.07 & 4.93 & -1.96   & SiI & 14& 5880.27   & 1.00  & -2.00   & TiI & 22\\
4425.44 & 1.88 & -0.36   & CaI & 20& 5922.11   & 1.00  & -1.38   & TiI & 22\\
4435.69 & 1.89 & -0.52   & CaI & 20& 5937.81   & 1.00  & -1.94   & TiI & 22\\
4512.27 & 2.53 & -1.90   & CaI & 20& 6303.76   & 1.44  & -1.51   & TiI & 22\\
4526.93 & 2.71 & -0.55   & CaI & 20& 6312.23   & 1.46  & -1.50   & TiI & 22\\
4578.56 & 2.52 & -0.70   & CaI & 20& 6554.22   & 1.40  & -1.15   & TiI & 22\\
5260.39 & 2.52 & -1.72   & CaI & 20& 4395.83   & 1.20  & -1.93   & TiII& 22.1  \\
5261.71 & 2.52 & -0.58   & CaI & 20& 4583.40   & 1.17  & -2.87   & TiII& 22.1  \\
5349.47 & 2.71 & -0.31   & CaI & 20& 4708.66   & 1.24  & -2.37   & TiII& 22.1  \\
5588.76 & 2.53 & 0.36& CaI & 20& 4798.53   & 1.00  & -2.66   & TiII& 22.1  \\
5590.13 & 2.52 & -0.57   & CaI & 20& 4874.00  & 3.00  & -0.86   & TiII& 22.1  \\
5594.47 & 2.52 & 0.10& CaI & 20& 5069.09   & 3.10   & -1.62   & TiII& 22.1  \\
5857.46 & 2.93 & 0.24& CaI & 20& 5336.78   & 1.58  & -1.63   & TiII& 22.1  \\
5867.57 & 2.93 & -1.57   & CaI & 20& 5418.76   & 1.58  & -2.11   & TiII& 22.1  \\
6161.30 & 2.52 & -1.27   & CaI & 20& 4708.02   & 3.17  & 0.09& Cr1 & 24\\
6163.75 & 2.52 & -1.29   & CaI & 20& 4801.03   & 3.12  & -0.13   & Cr1 & 24\\
6166.44 & 2.52 & -1.14   & CaI & 20& 4936.34   & 3.11  & -0.24   & Cr1 & 24\\
6169.04 & 2.52 & -0.80   & CaI & 20& 5272.00   & 3.45  & -0.42   & Cr1 & 24\\
6169.56 & 2.53 & -0.48   & CaI & 20& 5287.20   & 3.44  & -0.89   & Cr1 & 24\\
6439.08 & 2.53 & 0.39& CaI & 20& 5300.74   & 0.98  & -2.08   & Cr1 & 24\\
6449.82 & 2.52 & -0.50   & CaI & 20& 5304.19   & 3.46  & -0.68   & Cr1 & 24\\
6455.61 & 2.52 & -1.34   & CaI & 20& 5628.65   & 3.42  & -0.76   & Cr1 & 24\\
6462.57 & 2.52 & 0.26& CaI & 20& 5781.19   & 3.01  & -1.00   & Cr1 & 24\\
6471.67 & 2.53 & -0.69   & CaI & 20& 6882.50   & 3.44  & -0.38   & Cr1 & 24\\
6493.79 & 2.52 & -0.11   & CaI & 20& 5157.98   & 3.61  & -1.51   & NiI & 28\\
6499.65 & 2.52 & -0.82   & CaI & 20& 5537.11   & 3.85  & -2.20   & NiI & 28\\
6572.80 & 0.00 & -4.24   & CaI & 20& 6176.80   & 4.09  & -0.26   & NiI & 28\\
6717.69 & 2.71 & -0.52   & CaI & 20& 6204.60   & 4.09  & -0.82   & NiI & 28\\
5581.96 & 2.52 & -0.56   & CaI & 20& 6223.98   & 4.11  & -1.10   & NiI & 28\\
6499.65 & 2.52 & -0.82   & CaI & 20& 6378.26   & 4.15  & -1.05   & NiI & 28\\
6508.85 & 2.53 & -2.10   & CaI & 20& 4823.31   & 0.99  & -1.11   & YII & 39.1  \\
4471.24 & 1.70 & -0.15   & TiI & 22& 4854.87   & 0.99  & -0.38   & YII & 39.1  \\
4758.90 & 0.80 & -2.17   & TiI & 22& 4883.69   & 1.08  & 0.07& YII & 39.1  \\
4778.25 & 2.20 & -0.35   & TiI & 22& 5087.43   & 1.08  & -0.17   & YII & 39.1  \\
4797.98 & 2.30 & -0.63   & TiI & 22& 5123.22   & 0.99  & -0.83   & YII & 39.1  \\
4995.07 & 2.20  & -1.00   & TiI & 22& 5200.41   & 0.99  & -0.57   & YII & 39.1  \\
5020.03& 0.80 & -0.33   & TiI & 22& 5205.73   & 1.03  & -0.34   & YII & 39.1  \\
5036.46 & 1.40 & 0.14& TiI & 22& 5402.78   & 1.84  & -0.64   & YII & 39.1  \\
5040.61 & 0.80 & -1.67   & TiI & 22& 5473.39   & 1.74  & -1.02   & YII & 39.1  \\
5145.46 & 1.40  & -0.54   & TiI & 22& 5544.62   & 1.74  & -1.09   & YII & 39.1  \\ \hline
\end{tabular}%
}
\end{table*}

\end{appendix} 
\end{document}